\documentclass{article}

\usepackage{graphicx} 

\usepackage{geometry}
\usepackage[round]{natbib}

\usepackage{xcolor,colortbl}
\usepackage{latexsym}              
\usepackage{amsmath}               
\usepackage{amssymb}               
\usepackage{amsfonts}              
\usepackage{amsthm}                
\usepackage{dsfont}
\usepackage{bbm}
\usepackage{tikz}                  
\usetikzlibrary{arrows, positioning, shapes}
\usetikzlibrary{bayesnet}
\usepackage{booktabs}
\usepackage{longtable}
\usepackage{algorithm, algorithmic}
\usepackage{subcaption}
\usepackage{multirow}
\usepackage{enumitem}

\definecolor{review}{RGB}{0,0,0}
\usepackage{soul}

\RequirePackage[%
  pdfstartview=FitH,%
  breaklinks=true,%
  bookmarks=true,%
  colorlinks=true,%
  linkcolor= blue,
  anchorcolor=blue,%
  citecolor=blue,
  filecolor=blue,%
  menucolor=blue,%
  urlcolor=blue%
  ]{hyperref}
\usepackage{cleveref}
\Crefname{figure}{Fig}{Figs}
\crefname{figure}{Fig}{Figs}

\allowdisplaybreaks

\definecolor{colP}{RGB}{136,34,85}
\definecolor{colK}{RGB}{51,34,136}
\definecolor{colD}{RGB}{17,119,51}
\definecolor{colN}{RGB}{73,0,146}

\newcommand{\peptp}{\textcolor{colP}{p}}
\newcommand{\samplen}{\textcolor{colN}{n}}
\newcommand{\drawd}{\textcolor{colD}{d}}
\newcommand{\groupk}{\textcolor{colK}{k}}

\newcommand{\peptP}{\textcolor{colP}{P}}
\newcommand{\sampleN}{\textcolor{colN}{N}}
\newcommand{\drawD}{\textcolor{colD}{D}}
\newcommand{\groupK}{\textcolor{colK}{K}}

\newcommand{\yb}{\boldsymbol{y}}

\newcommand{\mub}{\boldsymbol{\mu}}
\newcommand{\Sigmab}{\boldsymbol{\Sigma}}

\newcommand{\dif}{\mathop{}\!\mathrm{d}}
\newcommand{\tr}[1]{\textrm{tr}\left(#1\right)}

\newtheorem{proposition}{\fontfamily{yv1}\selectfont Proposition}
\newtheorem{lemma}{\fontfamily{yv1}\selectfont Lemma}

\title{A Bayesian Framework for Multivariate Differential Analysis}
\author{Marie Chion$^{1,*}$ \& Arthur Leroy$^{2,3}$}
\date{}

\begin{document}

\maketitle

\noindent $^1$ Medical Research Council Biostatistics Unit, University of Cambridge, United Kingdom. \\
$^2$ Université Paris-Saclay, INRAE, AgroParisTech, GABI, France. \\
$^3$ Université Paris-Saclay, AgroParisTech, INRAE, UMR MIA Paris-Saclay, France. \\
$^*$ mc2411@cam.ac.uk

\section*{Abstract}
Differential analysis is a routine procedure in the statistical analysis toolbox across many applied fields, including quantitative proteomics, the main illustration of the present paper. 
The state-of-the-art \texttt{limma} approach uses a hierarchical formulation with moderated-variance estimators for each analyte directly injected into the t-statistic. 
While standard hypothesis testing strategies are recognised for their low computational cost, allowing for quick extraction of the most differential among thousands of elements, they 
generally overlook key aspects such as handling missing values, inter-element correlations, and uncertainty quantification. 
The present paper proposes a fully Bayesian framework for differential analysis, leveraging a conjugate hierarchical formulation for both the mean and the variance.
Inference is performed by computing the posterior distribution of compared experimental conditions and sampling from the distribution of differences.
This approach provides well-calibrated uncertainty quantification at a similar computational cost as hypothesis testing by leveraging closed-form equations. 
Furthermore, a natural extension enables multivariate differential analysis that accounts for possible inter-element correlations.
We also demonstrate that, in this Bayesian treatment, missing \textcolor{review}{at random} data should generally be ignored in univariate settings, and further derive a tailored approximation that handles multiple imputation for the multivariate setting. 
We argue that probabilistic statements in terms of effect size and associated uncertainty are better suited to practical decision-making. Therefore, we finally propose simple and intuitive inference criteria, such as the overlap coefficient, which express group similarity as a probability rather than traditional, and often misleading, p-values.  
The performance of this approach is evaluated through an extensive empirical study using both synthetic and controlled real-world proteomics datasets. 
Overall, we believe that this Bayesian framework for (multivariate) differential analysis provides a valuable and intuitive counterpart to standard methods at a comparable computational cost.

\section*{Author summary}
In many areas of biology, researchers need to decide whether one condition truly changes the levels of proteins or other measured features. Standard statistical tools often reduce this question to a single significance value, which can be hard to interpret and may handle missing measurements poorly. In this work, we developed a probability-based method for comparing groups that focuses on two practical questions: how large is the difference, and how certain are we about it? Using quantitative proteomics as our main example, we show that this approach can deal naturally with uncertainty, make use of relationships between related measurements, and, in simple cases, avoid unnecessary filling-in of missing values. Because the method relies on formulas that can be computed directly, it remains fast enough for large studies while giving results that are more informative than traditional testing alone. Rather than encouraging yes-or-no decisions based on a threshold, our framework helps researchers judge the size and reliability of observed changes. We hope this offers a clearer and more useful way to study biological differences in proteomics and beyond.

\section{Introduction}
\paragraph{Context.}

Differential analysis is a statistical framework used to identify meaningful differences between groups, conditions, or time points within complex datasets. 
By quantifying how measured variables change across predefined situations, it allows researchers to isolate specific effects or patterns of interest. 
Such methods play a central role in biostatistics, where distinguishing the true signal from background variability is essential.
Throughout the paper, we illustrate our methodological proposal in the specific context of quantitative proteomics, although the underlying statistical models can be adapted to many other contexts.

Differential proteomics aims to compare peptide and/or protein expression levels across several biological conditions. 
The amount of data provided by label-free mass spectrometry-based quantitative proteomics experiments requires reliable statistical modelling tools to assess which proteins are differentially abundant. 
In summary, \Cref{tab:biblio} presents the main state-of-the-art routines for differential proteomics analysis. 
They are based on well-known statistical methods, though they face several challenges.
First, while quantitative proteomics data usually contain missing values, they rely on complete datasets.
In label-free quantitative proteomics, the proportion of missing values ranges from 10\% to 50\% \cite{lazarAccountingMultipleNatures2016}.
Imputation remedies this problem by replacing a missing value with a user-defined one.
In particular, multiple imputation \citep{littleStatisticalAnalysisMissing2019} consists of generating several imputed datasets, which are combined to obtain an estimator of the parameter of interest (often a peptide or protein's mean intensity under a given condition) and an estimator of its variability. 
Recent work in \cite{chionAccountingMultipleImputationinduced2022} includes the uncertainty induced by the multiple imputation process in the moderated $t$-testing framework, previously described in \cite{smythLinearModelsEmpirical2004}.
This approach relies on a hierarchical model to deduce the posterior distribution of the variance estimator for each analyte.
The expectation of this distribution is used as a moderated estimation of variance and is substituted into the expression of the $t$-statistic.

Despite such theoretical advances, traditional tools such as $t$-tests and their more recent variants, as presented in \Cref{tab:biblio}, suffer from several limitations that we aim to address.
Inference based on \emph{Null Hypothesis Significance Testing} (NHST) and p-values has been widely questioned over the past decades.
Many authors demonstrated that NHST often leads to underestimated rates of false discoveries, publication bias, and contributes as a major factor to the reproducibility crisis in experimental science \citep{ioannidis2005most,colquhoun2014investigation,wasserstein2019moving}. 
Additionally, NHST does not provide an interpretable distinction between effect sizes and uncertainty quantification, whereas Bayesian statistics offers a valuable alternative in most cases \citep{kruschkeBayesianNewStatistics2018}.
Recently, some authors provided convenient approaches and their implementations \citep{kruschkeBayesianEstimationSupersedes2013} for handling differential analysis problems using Bayesian inference. 
For instance, the \texttt{R} package \texttt{BEST} (standing for Bayesian Estimation Supersedes T-test) has widely contributed to the diffusion of those practices in experimental fields. 
Subsequently, in the proteomics field, 
\cite{obrienEffectsNonignorableMissing2018} suggested a Bayesian selection model to mitigate the problem of missing values.
In the proteomics literature, Bayesian methods have been reviewed by \cite{crookChallengesOpportunitiesBayesian2022}.  
In particular, \cite{theIntegratedIdentificationQuantification2019} implemented a probabilistic model in \texttt{Triqler} that accounts for variability across identification and quantification, as well as differential analysis. 
More recently, \cite{Bollon2025IsoBayes} introduced \texttt{IsoBayes}, a Bayesian framework that propagates uncertainty, including peptide detection errors and ambiguous peptide-to-isoform mappings, when inferring isoform-level abundance and differential expression. 

Although traditional differential analysis routines usually operate on thousands of peptides simultaneously, their computations assume independence across analytes. 
\cite{tumminello2022multivariate} developed a multivariate statistical test for differential expression analysis, restricted to discrete transcriptomics data.
To the best of our knowledge, no Bayesian framework has been proposed so far for conducting \emph{multivariate} differential expression analysis. 
However, the existence of correlations, for instance, between peptides of the same protein, seems like a reasonable assumption.
Modelling and accounting for such structures explicitly could enhance the ability to discover and quantify meaningful differences between groups or conditions.  
In response to the aforementioned methodological issues, we propose a novel framework for differential analysis that accounts for uncertainty quantification and inter-element correlations, with an emphasis on the specific context of quantitative proteomics.

\begin{table}[t]
\centering
\begin{tabular}{|c|c|}
\hline
Method & Software                                                \\ \hline
t-tests                  & \begin{tabular}[c]{@{}c@{}}
                            Perseus \citep{tyanovaPerseusComputationalPlatform2016} \\ 
                            DAPAR \citep{wieczorekDAPARProStaRSoftware2017} \\
                            PANDA-view \citep{changPANDAviewEasytouseTool2018}
                            \end{tabular} \\ \hline
ANOVA                    & \begin{tabular}[c]{@{}c@{}}
                            Perseus \citep{tyanovaPerseusComputationalPlatform2016} \\
                            PANDA-view \citep{changPANDAviewEasytouseTool2018}\\
                            \end{tabular} \\ \hline
Moderated t-test (limma) & \begin{tabular}[c]{@{}c@{}}
                            DAPAR \citep{wieczorekDAPARProStaRSoftware2017} \\ 
                            mi4p \citep{chionAccountingMultipleImputationinduced2022}
                            \end{tabular}    \\ \hline
Linear model             & \begin{tabular}[c]{@{}c@{}}
                            MSstats \citep{choiMSstatsPackageStatistical2014} \\ 
                            proDA \citep{ahlmann-eltzeProDAProbabilisticDropout2020}
                            \end{tabular} \\ \hline
\end{tabular}
\caption{State-of-the-art software for differential proteomics analysis}
\label{tab:biblio}
\end{table}
Leveraging standard results of Bayesian inference with conjugate priors, we derive a fully Bayesian approach that handles missing data and multiple imputation, both commonly encountered in proteomics.
We propose a hierarchical model with prior distributions on both mean and variance parameters to provide a well-calibrated quantification of the uncertainty for subsequent differential analysis. 
The inference is performed by computing the posterior distribution of the difference of means between two experimental conditions. 
In contrast to more flexible models with complex hierarchical structures, our choice of conjugate priors yields analytical expressions that enable direct sampling from posterior distributions without the need for time-consuming Monte Carlo Markov Chain (MCMC) methods. 
This results in a fast inference scheme comparable to classical NHST procedures while providing more interpretable results expressed as probabilistic statements. 

\paragraph{Outline.}
The paper is organised as follows:
\Cref{sec:model_uni} presents well-known results about Bayesian inference for Gaussian-inverse-gamma conjugated priors. 
Following analogous results for the multivariate case, \Cref{sec:model_multi} introduces a general Bayesian framework for evaluating mean differences in differential proteomics contexts. \Cref{sec:uncorr} provides insights on the particular case where the considered analytes are uncorrelated. The proofs of these methodological developments can be found in Text B in S1 Appendix.
\Cref{sec:experiments} evaluates our framework, called ProteoBayes, through an extensive simulation study and comparisons with existing approaches.
We further illustrated the framework with hands-on examples using real proteomics datasets and highlighted its benefits for practitioners.

\section{Modelling}
\subsection{Bayesian inference for Normal-Inverse-Gamma conjugated priors}
\label{sec:model_uni}
Before deriving our complete workflow, let us recall some classical results in Bayesian inference that will further serve the framework with hands-on examples using real proteomics datasets and highlight its benefits owing to expression:
\begin{equation*}
	 y =\mu + \varepsilon,
\end{equation*}
\begin{itemize}
    \item $\mu \mid \sigma^2 \sim \mathcal{N} \left(\mu_0, \dfrac{1}{\lambda_0} \sigma^2 \right)$ is the prior distribution over the mean,
    \item $\varepsilon \sim \mathcal{N} (0, \sigma^2)$ is the error term,
    \item $\sigma^2 \sim \Gamma^{-1} (\alpha_0, \beta_0)$ is the prior distribution over the variance,
\end{itemize}
\noindent with $\{ \mu_0, \lambda_0, \alpha_0, \beta_0 \}$ an arbitrary set of prior hyper-parameters.
In \Cref{graph_model_uni}, we provide an illustration summarising these hypotheses.

\begin{figure}[ht]
\begin{center}
\begin{tikzpicture}

  \node[obs](y) {$y$};
  
  \node[latent, left=of y,  xshift=-0.2cm, yshift= 0.9cm](mu) {$\mu$};
  \node[const, above=of mu, xshift=-0.5cm, yshift= 1cm](m) {$\mu_0$};
  \node[const, above=of mu, xshift= 0.5cm, yshift= 1cm](l) {$\lambda_0$};

  \node[latent, above=of y, yshift= 0.8cm] (s) {$\sigma^2$};
  \node[const, above=of s,  xshift=-0.5cm] (a) {$\alpha_0$};
  \node[const, above=of s,  xshift= 0.5cm] (b) {$\beta_0$};

  \factor[above=of mu] {s-mu} {left:$\mathcal{N}$} {m,s,l} {mu};
  \factor[above=of s] {mu-mu} {left:$\Gamma^{-1}$} {a,b} {s};
  \factor[above=of y] {mu-s} {right:$\mathcal{N}$} {mu,s} {y};  

\end{tikzpicture}
\caption{Graphical model of the hierarchical structure when assuming a Gaussian-inverse-gamma prior, conjugated with a Gaussian likelihood with unknown mean and variance.}
\label{graph_model_uni} 
\end{center}     
\end{figure}

From the previous assumptions, we can deduce the likelihood of the model for a sample of observations $\yb = \{ y_1, \dots, y_N \}$: 
\begin{align*}
\displaystyle
	 p(\yb \mid \mu, \sigma^2) 
	 &= \prod_{n = 1}^{N} p(y_n \mid \mu, \sigma^2) \\
	 &= \prod_{n = 1}^{N} \mathcal{N}\left( y_n; \mu, \sigma^2 \right),
\end{align*}
Let us recall that the proposed prior, known as the Gaussian-inverse-gamma, is conjugate to the Gaussian likelihood with unknown mean $\mu$ and variance $\sigma^2$. 
The probability density function (PDF) of such a prior distribution can be written as follows:
\begin{equation*}
\displaystyle
	 p(\mu, \sigma^2 \mid \mu_0, \lambda_0, \alpha_0, \beta_0) = \frac{\sqrt{\lambda_0}}{\sqrt{2 \pi}} \frac{\beta_0^{\alpha_0}}{\Gamma(\alpha_0)}\left(\frac{1}{\sigma^{2}}\right)^{\alpha_0 + \frac{3}{2}} \exp \left(-\frac{2 \beta_0 +\lambda_0(\mu -\mu_0)^{2}}{2 \sigma^{2}}\right).
\end{equation*}
In this particular case, it is a well-known result that the inference is tractable, and the posterior distribution remains a Gaussian-inverse-gamma \citep{murphyConjugateBayesianAnalysis2007}. We provided an extended proof of this result in Text B.1.
Therefore, the joint posterior distribution can be expressed as: 
\begin{equation}
\label{eq:joint_post_uni}
	\mu, \sigma^2 \mid \yb \sim \mathcal{N} \Gamma^{-1} \left( \mu_{\sampleN}, \lambda_{\sampleN}, \alpha_{\sampleN}, \beta_{\sampleN} \right)
\end{equation}
\noindent with:
\begin{itemize}
	\item $\mu_{\sampleN} = \dfrac{\sampleN \bar{y} + \lambda_0 \mu_0}{\lambda_0 + N}$,
	\item $\lambda_{\sampleN} = \lambda_0 + \sampleN$,
	\item $\alpha_{\sampleN} = \alpha_0 + \dfrac{\sampleN}{2}$,
	\item $\beta_{\sampleN} = \beta_0 + \dfrac{1}{2} \sum\limits_{\samplen = 1}^{\sampleN}(y_{\samplen} - \bar{y})^2 + \dfrac{\lambda_0 \sampleN}{2(\lambda_0 + \sampleN)} (\bar{y} - \mu_0)^2 $.
\end{itemize}

Although these updated expressions for the hyperparameters already yield valuable results, we shall see in the sequel that we are more interested in the marginal distribution over the mean parameter $\mu$ for comparison purposes.
Computing this marginal from the joint posterior in \Cref{eq:joint_post_uni} remains tractable as well by integrating over $\sigma^2$:
\begin{align*}
\displaystyle
	p(\mu \mid \yb) 
	&= \int p(\mu, \sigma^2 \mid \yb) \dif \sigma^2 \\
	&= \frac{\sqrt{\lambda_{\sampleN}}}{\sqrt{2 \pi}} \frac{\beta_{\sampleN}^{\alpha_{\sampleN}}}{\Gamma(\alpha_{\sampleN})}  \int \left(\frac{1}{\sigma^{2}}\right)^{\alpha_{\sampleN} + \frac{3}{2}} \exp \left(-\frac{2 \beta_{\sampleN} +\lambda_{\sampleN}(\mu -\mu_{\sampleN})^{2}}{2 \sigma^{2}}\right) \dif \sigma^2 \\ 
	&= \frac{\Gamma(\frac{\nu + 1}{2})}{\Gamma(\frac{\nu}{2})} \frac{1}{\sqrt{\pi \nu \hat{\sigma}^2}} ( 1 +\frac{1}{\nu } \frac{(\mu -\mu_{\sampleN})^{2}}{\hat{\sigma}^2})^{- \frac{\nu + 1}{2}}\\
    &= T_{\nu}(\mu; \ \mu_{\sampleN}, \hat{\sigma}^2),
\end{align*}
\noindent with:
\begin{itemize}
	\item $\nu = 2\alpha_{\sampleN}$,
	\item $\hat{\sigma}^2 = \dfrac{\beta_{\sampleN}}{\alpha_{\sampleN} \lambda_{\sampleN}}$.
\end{itemize}
The marginal posterior distribution over $\mu$ can thus be expressed as a non-standardised Student's $t$-distribution that we express below in terms of the initial hyper-parameters:
\begin{equation}
\label{eq:t_dist_uni}
	\mu \mid \yb \sim T_{2\alpha_0 + \sampleN} \left(\dfrac{\sampleN \bar{y} + \lambda_0 \mu_0}{\lambda_0 + \sampleN} , \dfrac{\beta_0 + \dfrac{1}{2} \sum\limits_{\samplen = 1}^{\sampleN}(y_n - \bar{y})^2 + \dfrac{\lambda_0 \sampleN}{2(\lambda_0 + \sampleN)} (\bar{y} - \mu_0)^2}{(\alpha_0 + \frac{\sampleN}{2})( \lambda_0 + \sampleN)} \right).
\end{equation} 
We shall see in the next section how to leverage this approach to introduce a novel comparison-of-means methodology based on such analytical posterior computations.

\subsection{General Bayesian framework for evaluating mean differences}
\label{sec:model_multi}

Recalling our differential proteomics context that assesses the differences in mean intensity values for $\peptP$ peptides or proteins quantified in $\sampleN$ samples divided into $\groupK$ groups (also called \emph{conditions}).
As before, \Cref{graph_model_multi} illustrates the hierarchical generative structure assumed for each group $\groupk = 1, \dots, \groupK$.
\begin{figure}[ht]
\begin{center}
\begin{tikzpicture}
  \node[obs](y) {$\yb_{\groupk}$};
  
  \node[latent, left=of y,  xshift=-0.2cm, yshift= 0.9cm](mu) {$\mub_{\groupk}$};
  \node[const, above=of mu, xshift=-0.5cm, yshift= 1cm] (m) {$\mub_0$};
  \node[const, above=of mu, xshift= 0.5cm, yshift= 1cm] (l) {$\lambda_0$};

  \node[latent, above=of y, yshift= 0.8cm] (s) {$\Sigmab_{\groupk}$};
  \node[const, above=of s,  xshift=-0.5cm] (a) {$\Sigmab_0$};
  \node[const, above=of s,  xshift= 0.5cm] (b) {$\nu_0$};

  \factor[above=of mu] {s-mu} {left:$\mathcal{N}$} {m,s,l} {mu};
  \factor[above=of s] {mu-mu} {left:$\mathcal{W}^{-1}$} {a,b} {s};
  \factor[above=of y] {mu-s} {right:$\mathcal{N}$} {mu,s} {y};  
  
  \plate {} {(mu)(y)(s)} {$\forall \groupk = 1, \dots, \groupK$} ;

\end{tikzpicture}
\caption{Graphical model of the hierarchical structure of the generative model for the vector $\yb_{\groupk}$ of peptide intensities in $\groupK$ groups of biological samples, \textit{i.e.} $\groupK$ experimental conditions.}
\label{graph_model_multi} 
\end{center}     
\end{figure}

Maintaining the notation analogous to previous ones, the generative model for $\yb_{\groupk} \in \mathbb{R}^{\peptP}$, can be written as:
\begin{equation*}
	 \yb_{\groupk} = \mub_{\groupk} + \boldsymbol{\varepsilon}_{\groupk}, \ \forall \groupk = 1, \dots,  \groupK,
\end{equation*}
\noindent where:
\begin{itemize}
    \item $\mub_{\groupk} \mid \Sigmab_{\groupk} \sim \mathcal{N} \left(\mub_0, \dfrac{1}{\lambda_0}\Sigmab_{\groupk} \right)$ is the prior mean intensities vector of the $\groupk$-th group,
    \item $\boldsymbol{\varepsilon}_{\groupk} \sim \mathcal{N} (0, \Sigmab_{\groupk})$ is the error term of the $\groupk$-th group,
    \item $\Sigmab_{\groupk} \sim \mathcal{W}^{-1} (\Sigmab_0, \nu_0)$ is the prior variance-covariance matrix of the $\groupk$-th group,
\end{itemize}
\noindent with $\{ \mub_0, \lambda_0, \Sigmab_0, \nu_0 \}$ a set of hyper-parameters that needs to be chosen as modelling hypotheses and $\mathcal{W}^{-1}$ represents the inverse-Wishart distribution, used as the conjugate prior for an unknown covariance matrix of a multivariate Gaussian distribution \citep{bishopPatternRecognitionMachine2006}.

Traditionally, in Bayesian inference, those quantities must be carefully chosen to achieve the most accurate estimation, particularly with small sample sizes.
Incorporating expert or prior knowledge into the model would also come from appropriately setting these hyperparameters. 
We discuss in more detail the choice and influence of those prior hyperparameters in \Cref{sec:hyperparam}.
However, this article's ultimate purpose is not to estimate but to compare group means (i.e., differential analysis).
Interestingly, providing a perfect estimation of the posterior distributions over $\{\mub_{\groupk} \}_{\groupk = 1, \dots, \groupK}$ does not appear as the main concern here, as the posterior difference of means (i.e. $p(\mub_{\groupk} - \mub_{\groupk^{\prime}} \mid \yb_{\groupk}, \yb_{{\groupk}^{\prime}}))$ represents the actual quantity of interest.
Although providing meaningful prior hyperparameters leads to more accurate uncertainty quantification, we shall mainly set those quantities equal across all groups to ensure an unbiased comparison.
\textcolor{review}{This sharing applies only to the prior specification; the covariance matrices themselves remain group-specific through $\Sigmab_{\groupk}$ and are updated separately from the observed data in each condition.}

\textcolor{review}{The multivariate formulation should be viewed as the natural extension of the univariate conjugate Gaussian model to vector-valued observations. Its assumptions are therefore analogous in nature, but the model is more demanding in practice because it requires estimation of covariance structures, which is substantially harder than estimating independent variances from limited replication. Accordingly, we primarily intend this framework for moderate-dimensional groups of correlated analytes, such as peptides within a protein, where explicit covariance modelling is both biologically meaningful and statistically feasible.}

The present framework aims to estimate a posterior distribution for each mean parameter vector $\mub_{\groupk}$, using the same prior assumptions across groups.
The comparison between the means of all groups would then rely solely on the ability to sample directly from these distributions and compute empirical posteriors for the difference in means.
As a bonus, this framework remains compatible with multiple imputation strategies previously introduced to handle missing data that frequently arise in applicative contexts \citep{chionAccountingMultipleImputationinduced2022}.
From the previous hypotheses, we can deduce the likelihood of the model for an i.i.d. sample $\{ \yb_{\groupk,\textcolor{colN}{1}}, \dots, \yb_{\groupk,\sampleN_{\groupk}} \}$: 
\begin{align*}
\displaystyle
	 p(\yb_{\groupk,\textcolor{colN}{1}}, \dots, \yb_{\groupk,\sampleN_{\groupk}} \mid \mub_{\groupk}, \Sigmab_{\groupk}) 
	 &= \prod_{\samplen = 1}^{\sampleN_{\groupk}} p(\yb_{\groupk,\samplen} \mid \mub_{\groupk}, \Sigmab_{\groupk}) \\
	 &= \prod_{\samplen = 1}^{\sampleN_{\groupk}} \mathcal{N} \left( \yb_{\groupk,\samplen}; \ \mub_{\groupk}, \Sigmab_{\groupk} \right),
\end{align*}

However, as previously noted, such datasets often contain missing data, and we shall introduce a consistent notation here. Assume $\mathcal{H}$ to be the set of all observed data, we additionally define:
\begin{itemize}
	\item $\yb_{\groupk}^{(0)} = \{y_{k,\samplen}^{\peptp} \in \mathcal{H}, \ n = 1, \dots N_{\groupk}, \ p = 1, \dots, P\}$, the set of elements that are observed in the $\groupk$-th group,
	\item $\yb_{\groupk}^{(1)} = \{y_{k,\samplen}^{\peptp} \notin \mathcal{H}, \ n = 1, \dots N_{\groupk}, \ p = 1, \dots, P\}$, the set of elements that are missing the $\groupk$-th group.
\end{itemize}

Moreover, as we remain in the context of multiple imputation, we define $\{ \tilde{\yb}_{\groupk}^{(1),\textcolor{colD}{1}}, \dots, \tilde{\yb}_{\groupk}^{(1),\drawD}\}$ as the set of $\drawD$ draws of an imputation process applied on missing data in the $\groupk$-th group.
In such a context, a closed-form approximation for the multiple-imputed posterior distribution of $\mub_{\groupk}$ can be derived for each group as stated in \Cref{prop:multi}.

\begin{proposition}
\label{prop:multi}
For all $\groupk = 1, \dots, \groupK$, the posterior distribution of $\mub_{\groupk}$ can be approximated by a mixture of multiple-imputed multivariate $t$-distributions, such as:
\begin{align*}
	p(\mub_{\groupk} \mid \yb_{\groupk}^{(0)}) \simeq \dfrac{1}{\drawD} \sum\limits_{\drawd = 1}^{\drawD} T_{\nu_{\groupk}}\left( \mub; \tilde{\mub}_{\groupk}^{(\drawd)},  \tilde{\Sigmab}_{\groupk}^{(\drawd)} \right)
\end{align*}
with: 
\begin{itemize}
	\item $\nu_{\groupk} = \nu_0 + \sampleN_{\groupk} - \peptP + 1$,
	\item $\tilde{\mub}_{\groupk}^{(\drawd)} = \dfrac{\lambda_0 \mub_0 + \sampleN_{\groupk} \bar{\yb}_{\groupk}^{(\drawd)} }{\lambda_0 + \sampleN_{\groupk}}$ ,
	\item $ \tilde{\Sigmab}_{\groupk}^{(\drawd)} = \dfrac{\Sigmab_0 + \sum\limits_{\samplen = 1}^{\sampleN_{\groupk}} (\tilde{\yb}_{\groupk,\samplen}^{(\drawd)} - \bar{\yb}_{\groupk}^{(\drawd)})(\tilde{\yb}_{\groupk,\samplen}^{(\drawd)} - \bar{\yb}_{\groupk}^{(\drawd)})^{\intercal} + \dfrac{\lambda_0 \sampleN_{\groupk}}{(\lambda_0 + \sampleN_{\groupk})} (\bar{\yb}_{\groupk}^{(\drawd)} - \mub_0)(\bar{\yb}_{\groupk}^{(\drawd)} - \mub_0)^{\intercal}}{(\nu_0 + \sampleN_{\groupk} - \peptP + 1)( \lambda_0 + \sampleN_{\groupk})}$,
\end{itemize}
where we introduced the shorthand $\tilde{\yb}_{\groupk,\samplen}^{(\drawd)} = \begin{bmatrix}
\yb_{\groupk,\samplen}^{(0)} \\
\tilde{\yb}_{\groupk,\samplen}^{(1),\drawd}
\end{bmatrix}$ to represent the $\drawd$-th imputed vector of observed data, and the corresponding average vector $\bar{\yb}_{\groupk}^{(d)} =  \dfrac{1}{\sampleN_{\groupk}} \sum\limits_{\samplen = 1}^{\sampleN_{\groupk}} \tilde{\yb}_{\groupk,\samplen}^{(\drawd)}$.
\end{proposition}
The proof of \Cref{prop:multi} can be found in Text B.2.
This analytical formulation is particularly convenient for approximating the posterior distribution of the mean vector for each group using multiple-imputed datasets. 
Although such a linear combination of multivariate $t$-distributions is not a known specific distribution in itself, it is now straightforward to generate realisations of posterior samples by simply drawing from the $\drawD$ multivariate $t$-distributions, each being specific to an imputed dataset, and then computing the mean of the $\drawD$ vectors.
Therefore, the empirical distribution resulting from a large number of samples generated by this procedure would be easy to visualise and compare. 
Generating the empirical distribution of the mean's difference between two groups $\groupk$ and $\groupk^{\prime}$ comes directly by computing the difference between each couple of samples drawn from both posterior distributions $p(\mub_{\groupk} \mid \yb_{\groupk}^{(0)})$ and $p(\mub_{\groupk}^{\prime} \mid \yb_{k^{\prime}}^{(0)})$.
In Bayesian statistics, relying on empirical distributions drawn from the posterior is common practice in the context of Markov chain Monte Carlo (MCMC) algorithms, but often comes at a high computational cost. 
In our framework, we maintained analytical distributions from model hypotheses to enable probabilistic inference with adequate uncertainty quantification, while remaining tractable and avoiding MCMC procedures.
Therefore, the computational cost of the method roughly remains as low as its frequentist counterparts, as inference merely requires updating hyper-parameter values and drawing from corresponding $t$-distributions.
Empirical evidence of this claim is provided in the further simulation study and summarised in \Cref{tab:running_time}.

As usual, when it comes to comparing the means between two groups, we still need to assess if the posterior distribution of the difference appears, in a sense, to be sufficiently away from zero. 
This practical inference choice is not specific to our context and remains highly dependent on the study's context. 
Moreover, because the present model is multidimensional, we may also question the metric used to compute vector differences. 
In a sense, our posterior distribution of means' differences offers an elegant solution to the traditional problem of multiple testing often encountered in applied science and calls for tailored definitions of what could be called a \emph{meaningful} result (\emph{significant} does not appear as an appropriate term anymore in this more general context).
For example, displaying the distribution of squared differences would penalise large differences in the elements of the mean vector. 
In contrast, the absolute difference would give a more balanced conception of the average divergence between the two groups. 
Clearly, as any marginal of a multivariate $t$-distribution remains a (multivariate) $t$-distribution, comparing specific elements of the mean vectors merely by restricting to the appropriate dimension is also straightforward. 
In particular, comparing two groups in the univariate case would be a particular case of \Cref{prop:multi} with $\peptP = 1$. 
Recalling our proteomics context, we could still compare the mean peptide intensities between groups, one peptide at a time, or compare all peptides at once, accounting for possible correlations within each group.
However, an appropriate way to account for those correlations could be to group peptides by their reference protein.
Let us provide in \Cref{alg:algo} a summary of the overall procedure for comparing mean vectors of two different experimental conditions (i.e. Bayesian multivariate differential analysis). 
The algorithm first updates the parameters of the posterior distribution of the group means from the observed (and possibly imputed) data. Then, an imputation index is randomly chosen (optional), and samples of the group means are repeatedly
drawn from the associated posteriors. Finally, the sampled means are compared across conditions
to obtain an empirical posterior distribution of the mean difference.
\begin{algorithm}
\caption{Posterior distribution of the vector of means' difference}
\label{alg:algo}
\begin{algorithmic}
    \STATE Initialise the hyper-posteriors $\mub_0^{\groupk} = \mub_0^{{\groupk}^{\prime}}$, $\lambda_0^{\groupk} = \lambda_0^{{\groupk}^{\prime}}$, $\Sigmab_0^{\groupk} = \Sigmab_0^{{\groupk}^{\prime}}$, $\nu_0^{\groupk} = \nu_0^{{\groupk}^{\prime}}$
    \newline
    \FOR{$ \drawd = 1, \dots, \drawD$}
    \vspace{0.3cm}
    \STATE - Compute $\{ \mub_{\sampleN}^{{\groupk},(\drawd)}, \lambda_{\sampleN}^{\groupk}, \Sigmab_{\sampleN}^{{\groupk},(\drawd)}, \nu_{\sampleN}^{\groupk} \}$ and $\{ \mub_{\sampleN}^{{\groupk}^{\prime},(\drawd)}, \lambda_{\sampleN}^{{\groupk}^{\prime}}, \Sigmab_{\sampleN}^{{\groupk}^{\prime},(\drawd)}, \nu_{\sampleN}^{{\groupk}^{\prime}} \}$ from hyper-posteriors and data
    \ENDFOR
    \newline
    \FOR{$ r = 1, \dots, R$}
    \vspace{0.3cm}
    \STATE - Draw a random imputation index $d_r \sim Uniform\{1, \dots, D\}$ 
    \STATE - Draw realisations $\hat{\mub}_{{\groupk}}^{[r]} \sim T_{\nu_{\sampleN}^{\groupk}}\left(\mub_{\sampleN}^{{\groupk},(d_r)}, \dfrac{\Sigmab_{\sampleN}^{{\groupk},(d_r)}}{ \lambda_{\sampleN}^{\groupk} \nu_{\sampleN}^{\groupk} } \right)$ and $\ \hat{\mub}_{{\groupk}^{\prime}}^{[r]} \sim T_{\nu_{\sampleN}^{{\groupk}^{\prime}}}\left(\mub_{\sampleN}^{{\groupk}^{\prime},(d_r)}, \dfrac{\Sigmab_{\sampleN}^{{\groupk}^{\prime},(d_r)}}{ \lambda_{\sampleN}^{{\groupk}^{\prime}} \nu_{\sampleN}^{{\groupk}^{\prime}} } \right)$ 
    \STATE - Compute a realisation $\hat{\mub}_{\Delta}^{[r]} = \hat{\mub}_{{\groupk}}^{[r]} - \hat{\mub}_{{\groupk}^{\prime}}^{[r]}$  from the difference's distribution
    \ENDFOR 
    \newline
    \RETURN $\{ \hat{\mub}_{\Delta}^{[1]}, \dots, \hat{\mub}_{\Delta}^{[R]} \}$,  an R-sample drawn from the posterior distribution of the mean's difference
\end{algorithmic}
\end{algorithm}

\subsection{The uncorrelated case: no more multiple testing nor imputation}
\label{sec:uncorr}

Let us note that modelling covariances across all variables, as in \Cref{prop:multi}, often poses a challenge, is computationally expensive in high dimensions, and is not always well-adapted. 
However, we detailed in \Cref{sec:model_uni} results that, although classical in Bayesian statistics, remain too rarely exploited in applied science.
In particular, we can leverage these results to adapt \Cref{alg:algo} to the univariate case for handling the same problem as in \cite{chionAccountingMultipleImputationinduced2022} with a probabilistic flavour. 
In the classical setting of the absence of correlations between peptides (\textit{i.e.} $\Sigmab$ being diagonal), the problem reduces to the analysis of $\peptP$ independent inference problems (as $\mub$ is supposed Gaussian) and the posterior distributions can be derived in closed-form, as we recalled in \Cref{eq:joint_post_uni}.
Moreover, let us highlight a pleasant property that arises from relaxing this assumption: (multiple-)imputation is no longer needed in this context. 
Using the same notation as before and the uncorrelated assumption (and thus the induced independence between analytes for $\peptp \neq {\peptp}^{\prime}$), we can write:
\begin{align}
\label{eq:factorise}
	p\left( \mub_{\groupk} \mid \yb_{\groupk}^{(0)} \right) 
	&=  \int p\left( \mub_{\groupk}, \yb_{\groupk}^{(1)} \mid \yb_{\groupk}^{(0)}\right)  \dif \yb_{\groupk}^{(1)}  \\
	&=  \int p\left( \mub_{\groupk} \mid \yb_{\groupk}^{(0)}, \yb_{\groupk}^{(1)}\right) p\left( \yb_{\groupk}^{(1)} \mid \yb_{\groupk}^{(0)} \right) \dif \yb_{\groupk}^{(1)}  \\
	&= \int \prod\limits_{\peptp=1}^{\peptP} \left\{  p \left( \mu_{\groupk}^{\peptp} \mid y_{\groupk}^{p,(0)}, y_{\groupk}^{\peptp,(1)} \right) p\left( y_{\groupk}^{{\peptp},(1)} \mid y_{\groupk}^{{\peptp},(0)} \right) \right\} \dif \yb_{\groupk}^{(1)} \\
	&= \prod\limits_{{\peptp}=1}^{\peptP} \int \left\{  p \left( \mu_{\groupk}^{\peptp} \mid y_{\groupk}^{{\peptp},(0)}, y_{\groupk}^{{\peptp},(1)} \right) p\left( y_{\groupk}^{{\peptp},(1)} \mid y_{\groupk}^{{\peptp},(0)} \right)\dif y_{\groupk}^{{\peptp},(1)}\right\} \\
	&= \prod\limits_{{\peptp}=1}^{\peptP} p \left( \mu_{\groupk}^{\peptp} \mid y_{\groupk}^{{\peptp},(0)} \right) \\
	&= \prod\limits_{{\peptp}=1}^{\peptP} T_{2\alpha_0^{\peptp} + N_{\groupk}^{\peptp}} \left(\mu_{\groupk}^{\peptp}; \ \mu_{k,N}^{\peptp} , \ \hat{\sigma^{\peptp}_{\groupk}}^2 \right),
\end{align} 
\noindent with:
\begin{itemize}
	\item $\mu_{k,N}^{\peptp} = \dfrac{N_{\groupk}^{\peptp} \bar{y}_{\groupk}^{{\peptp},(0)} + \lambda_0^{\peptp} \mu_0^{\peptp}}{\lambda_0^{\peptp} + N_{\groupk}^{\peptp}}$,
	\item $\hat{\sigma^{\peptp}_{\groupk}}^2 = \dfrac{\beta_0^p + \dfrac{1}{2} \sum\limits_{n = 1}^{N_{\groupk}^{\peptp}}(y_{k,n}^{{\peptp},(0)} - \bar{y}_{\groupk}^{{\peptp},(0)})^2 + \dfrac{\lambda_0 N_{\groupk}^{\peptp}}{2(\lambda_0^{\peptp} + N_{\groupk}^{\peptp})} (\bar{y}_{\groupk}^{{\peptp},(0)} - \mu_0^{\peptp})^2}{(\alpha_0^{\peptp} + \frac{N_{\groupk}^{\peptp}}{2})( \lambda_0^{\peptp} + N_{\groupk}^{\peptp})}$.
\end{itemize}

It can be noticed that $p\left( \mub_{\groupk} \mid \yb_{\groupk}^{(0)} \right)$ factorises naturally over $\peptp = 1, \dots, \peptP$, and thus only depends upon the data that have actually been observed for each peptide. 
We observe that integrating over missing data $\yb_{\groupk}^{(1)}$ is straightforward in this framework, and neither Rubin's approximation nor imputation (whether multiple or not) appears necessary.
The observed data $\yb_{\groupk}^{(0)}$ already bear all relevant information as if each unobserved value could merely be ignored without effect on the posterior distribution. 

Let us emphasise that this property of factorisation and tractable integration over missing data comes directly from the covariance structure as a diagonal matrix and thus only constitutes a particular case of the previous model, though convenient. 
It should also be noted that this result applies only to values that are Missing At Random (MAR). 
Under MAR, the missingness mechanism is ignorable for inference on the means, and the observed-data posterior is sufficient. This does not extend to Missing Not At Random (MNAR) settings, where missingness depends on the unobserved abundance itself, for instance, through censoring below a detection limit. In such cases, missingness is informative and should be modelled explicitly, as in selection or Heckman-type models \citep{heckmanSampleSelectionBias1979, orwollProteomicAssessmentSerum2020} and related probabilistic approaches such as \emph{proDA} \citep{ahlmann-eltzeProDAProbabilisticDropout2020, ahlmann-eltzeProDADifferentialAbundance2023} or \emph{Triqler} \citep{theTriqlerMaxQuantEnhancing2021}. Therefore, our result should be interpreted as a no-imputation result for the univariate MAR setting, not as a general recommendation for all proteomics missing-data problems.

To conclude, whereas the analytical derivation of posterior distributions with Gaussian-inverse-gamma constitutes a well-known result, our proposition to define such probabilistic means' comparison procedure provides, under the standard uncorrelated-peptides assumption, an elegant and handy alternative to classical techniques that alleviates both imputation and multiple testing issues. 
Let us provide in \Cref{alg:algo_uni} the pseudo-code summarising the univariate inference procedure. 
\textcolor{review}{The only difference with the fully-correlated case comes from the absence of imputed datasets, and thus multiple associated posteriors}:
\begin{algorithm}
\caption{Posterior distribution of the means' difference}
\label{alg:algo_uni}
\begin{algorithmic}
    \FOR{$ {\peptp} = 1, \dots, {\peptP}$}
    \STATE - Initialise the hyper-posteriors $\mu_0^{{\groupk},{\peptp}} = \mu_0^{{\groupk}^{\prime}, {\peptp}}$, $\lambda_0^{{\groupk},{\peptp}} = \lambda_0^{{\groupk}^{\prime}, {\peptp}}$, $\alpha_0^{{\groupk},{\peptp}} = \alpha_0^{{\groupk}^{\prime}, {\peptp}}$, $\beta_0^{{\groupk},{\peptp}} = \beta_0^{{\groupk}^{\prime}, {\peptp}}$
    \newline
    \STATE - Compute $\{ \mu_{\sampleN}^{{\groupk},{\peptp}}, \lambda_{\sampleN}^{{\groupk},{\peptp}}, \alpha_{\sampleN}^{{\groupk},{\peptp}}, \beta_{\sampleN}^{{\groupk},{\peptp}} \}$ and $\{ \mu_{\sampleN}^{{\groupk}^{\prime}, {\peptp}}, \lambda_{\sampleN}^{{\groupk}^{\prime}, p}, \alpha_{\sampleN}^{{\groupk}^{\prime}, {\peptp}}, \beta_{\sampleN}^{{\groupk}^{\prime}, {\peptp}} \}$ from hyper-posteriors and data
    \newline
    \STATE - Draw $R$ realisations $\hat{\mu}_{{\groupk}}^{{\peptp},[r]} \sim T_{\alpha_{\sampleN}^{{\groupk},{\peptp}}}\left(\mu_{\sampleN}^{{\groupk},{\peptp}}, \dfrac{\beta_{\sampleN}^{{\groupk},{\peptp}}}{ \lambda_{\sampleN}^{{\groupk},{\peptp}} \alpha_{\sampleN}^{{\groupk},{\peptp}}} \right)$, $\hat{\mu}_{{\groupk}^{\prime}}^{{\peptp},[r]} \sim T_{\alpha_{\sampleN}^{{\groupk}^{\prime}, {\peptp}}}\left(\mu_{\sampleN}^{{\groupk}^{\prime}, {\peptp}}, \dfrac{\beta_{\sampleN}^{{\groupk}^{\prime}, \peptp}}{ \lambda_{\sampleN}^{{\groupk}^{\prime}, \peptp} \alpha_{\sampleN}^{{\groupk}^{\prime}, {\peptp}} } \right)$
    \newline
    \FOR{$ r = 1, \dots, R$}
    \STATE - Generate a realisation $\hat{\mu}_{\Delta}^{{\peptp},[r]} = \hat{\mu}_{{\groupk}}^{{\peptp},[r]} - \hat{\mu}_{{\groupk}^{\prime}}^{{\peptp},[r]}$ from the difference's distribution
    \ENDFOR 
    \ENDFOR
    \newline
    \RETURN $\{ \hat{\mub}_{\Delta}^{[1]}, \dots, \hat{\mub}_{\Delta}^{[R]} \}$,  an R-sample drawn from the posterior distribution of the mean's difference
\end{algorithmic}
\end{algorithm}

\section{Experiments}
\label{sec:experiments}

In this section, we assess the performance of the ProteoBayes framework using both simulated datasets and well-calibrated quantitative proteomics data. Where applicable—that is, in the context of the univariate approach—we compare its results to those obtained using the limma framework, as implemented in the DAPAR R package \citep{wieczorekDAPARProStaRSoftware2017}. 
We also compared our method to proDA \citep{ahlmann-eltzeProDAProbabilisticDropout2020}, which constructs a probabilistic dropout model to avoid imputing missing values, and MSqRob \citep{goeminne2018experimental}, a robust regression framework that performs peptide-level inference with empirical variance stabilisation.

\subsection{Synthetic datasets}
\label{subsec:simdata}

\paragraph{Univariate datasets}
To generate simulated datasets to evaluate the performance of our method, called ProteoBayes, we used the generative model presented in \Cref{graph_model_uni}. 
A Gaussian distribution $\mathcal{N}(0, 1)$ is taken as a baseline reference. 
To compute mean differences between groups, we generated samples from various distributions $\mathcal{N}(m, \sigma^2)$ where $m$ and $\sigma^2$ will vary depending on the context. 
Unless otherwise stated, each experiment is repeated 1000 times, and the results are averaged using the computed mean and standard deviation of the metrics. 
In each group, we observe 5 distinct samples.

\paragraph{Multivariate datasets}

Similarly to the univariate setting, multivariate datasets are simulated from the generative model proposed in \Cref{graph_model_multi}. Each experiment is repeated 1000 times to compute performance metrics. 
In each group, we observe 5 distinct samples. To emulate different contexts of correlations between peptides that remain intuitive for illustration purposes, we use as a baseline reference a 3-dimensional Gaussian distribution defined as:

\begin{equation}
\label{eq:ref_3d_gaussian}
 \mathcal{N}\left(
\begin{pmatrix}
0 \\
0 \\
0
\end{pmatrix}, 
\Sigma_{ref} =
\begin{bmatrix}
1 & 0.7 & 0.2\\
0.7 & 1 & 0.5 \\
0.2 & 0.5 & 1
\end{bmatrix}\right).
\end{equation}

Several distributions with different mean vectors and covariance matrices are used for comparison purposes and reported accordingly in the results. 

\subsection{Real datasets}
\label{subsec:realdata}

\paragraph{Description of all datasets}

To further evaluate our methodology on real datasets, we used four well-calibrated proteomics experiments, cited in previous methodological works \citep{chionAccountingMultipleImputationinduced2022, etourneauNewTakeMissing2023}. These experiments use a "spike-in" design, which helps us determine which peptides are expected to show differences in expression.
Hence, they provide a diverse and robust framework for benchmarking our method under various experimental conditions. 

\begin{itemize}
    \item The \textbf{Muller2016} dataset refers to the experiment from \cite{mullerBenchmarkingSamplePreparation2016}, where a mixture of UPS1 proteins has been spiked in increasing amounts (0.5, 1, 2.5, 5, 10, and 25 fmol) in a constant background of \textit{Saccharomyces cerevisiae} lysate (yeast), with each condition analysed in triplicate using a data-dependent acquisition method. This dataset is available on the ProteomeXchange website using the PXD003841 identifier.
    \item The \textbf{Bouyssie2020} dataset from \cite{bouyssieProlineEfficientUserfriendly2020} is similar to Muller\_2016 but expands the range of UPS1 spike-in concentrations to include ten levels (0.01, 0.05, 0.1, 0.25, 0.5, 1, 5, 10, 25, and 50 fmol), with each condition analysed in quadruplicate. The dataset is available on ProteomeXchange using the PXD009815 identifier.
    \item The \textbf{Huang2020} dataset from \cite{huangCombiningPrecursorFragment2020} features UPS2 proteins spiked at five concentrations (0.75, 0.83, 1.07, 2.04, and 7.54 amol) into $1 \mu g$ of mouse cerebellum lysate, analysed in pentaplicate using a data-independent acquisition (DIA) method. The dataset is available on the ProteomeXchange repository using the PXD016647 identifier.
    \item The \textbf{Chion2022} dataset refers to the ARATH dataset from \cite{chionAccountingMultipleImputationinduced2022}, where a mixture of UPS1 proteins spiked at seven increasing concentrations (0.05, 0.25, 0.5, 1.25, 2.5, 5, and 10 fmol) into a constant background of Arabidopsis thaliana lysate, with triplicate analyses performed for each condition using a DDA method. The dataset is available on ProteomeXchange using the PXD027800 identifier.
\end{itemize}

For each experiment, a normalisation step on the $log_2$-intensities was performed before analysis using the \verb|normalize.quantiles| function of the \verb|preprocessCore| R package \citep{bolstadPreprocessCoreCollectionPreprocessing2024}.

\paragraph{Illustration dataset}
Additionally, we illustrate our arguments using the Chion2022 experiment, namely the UPS-spiked \textit{Arabidopsis thaliana} dataset. 
Briefly, let us recall that UPS proteins were spiked into a constant background of \textit{Arabidopsis thaliana} (ARATH) protein lysate at increasing concentrations.  
Hence, UPS proteins are differentially expressed, and ARATH proteins are not.
For illustration purposes, we arbitrarily focused the examples on the \texttt{P12081ups$\mid$SYHC\_HUMAN\_UPS} and the \texttt{sp$\mid$F4I893$\mid$ILA\_ARATH} proteins. 
Note that both proteins have nine quantified peptides. Unless otherwise stated, we used the examples of the \texttt{AALEELVK} UPS peptide and the \texttt{VLPLIIPILSK} ARATH peptide, and set the same values for the prior hyperparameters as for synthetic data.

Additionally, let us recall that in our real datasets, the constants have the following values:
\begin{itemize}
	\item $\forall k = 1, \dots, K, \ N_k = 3$ data points, in the absence of missing data,
	\item $P = 9$ peptides, when using the multivariate model,
	\item $D = 7$ draws of imputation,
	\item $R = 10^4$ sample points from the posterior distributions.
\end{itemize}

In this context, where the number $N_k$ of observed biological samples is extremely low, notably when data are missing, we should expect a perceptible influence of the prior hyperparameters and of inherent uncertainty in the posteriors.
However, this influence has been reduced to a minimum in all subsequent graphs for clarity and to ensure a clear understanding of the methodology's underlying properties. 
The high number $R$ of sample points drawn from the posteriors ensures the empirical distribution is smoothly displayed on the graph. 
However, one should note that sampling is really quick in practice and that this number can be easily increased if necessary.

\subsection{Performance metrics}
\label{subsec-perf}
We compared the performance of our method with simple t-tests and with the limma framework implemented in the ProStaR software via the DAPAR R package \cite{wieczorekDAPARProStaRSoftware2017}.
However, due to the intrinsic difference in paradigm, limma being a frequentist tool and ProteoBayes a probabilistic one, we could only compare them in terms of mean difference recovery. 
To evaluate ProteoBayes as a probabilistic tool, we used other metrics, such as credible intervals, the root mean square error (RMSE), and credible interval coverage, to assess the quality of estimation and uncertainty calibration. 

\begin{itemize}
    \item \textbf{Mean difference:} For each peptide, we computed the difference between the mean intensity in the two groups compared. The common practice in proteomics is to use log2-intensities rather than raw intensities. Therefore, the mean difference is similar to the log2-fold change. 
    $$\mu_{diff} = \hat{\mu}_1 - \hat{\mu}_2$$ 
    
    \item \textbf{95\% Credible Interval Width (CI$\mathbf{_{95}}$ width):} This indicator reflects the uncertainty in the posterior distribution of the mean. A smaller CI$_{width}$ indicates greater confidence in the estimated mean intensity. For each peptide, we computed the range of the 95\% credible interval. 
    $$CI_{width} = max(CI_{95}) - min(CI_{95})$$

    \item \textbf{Root Mean Square Error (RMSE):} This indicator describes the average error for all peptides between the posterior mean intensity and the reconstructed reference mean intensity $\mu_{p}^{true}$ (see next paragraph). 
    $$RMSE = \sqrt{ \dfrac{1}{P} \sum_{p = 1}^{P} (\hat{\mu}_{p} - \mu_{p}^{true})^2}$$
    
    \item \textbf{95\% Credible Interval Coverage (CIC$\mathbf{_{95}}$):} This indicator shows how well our method is calibrated. Empirical values should be as close as possible to the theoretical 95\%. This measure is computed as the proportion of peptides for which the reference mean $\mu_{p}^{true}$ falls within the 95\% credible interval bounds. 
    $$CIC_{95} = 100 \times \dfrac{1}{P} \sum\limits_{p = 1}^{P} \ \mathds{1}_{ \{ \mu_{p}^{true} \in \ CI_{95} \} }$$ 
\end{itemize}

Both the RMSE and CIC$_{95}$ indicators rely on a reference mean. 
Ideally, and for synthetic datasets, we would know the true mean intensity for each peptide within a group and be able to compute the metrics exactly.
However, in real-data experiments, this value is unknown, but can be approximated in a carefully controlled design.
More specifically, the spike-in experimental design revolves around known theoretical abundances. 
In proteomics, global quantification assumes that peptide intensity is proportional to peptide abundance, scaled by its response factor. 
This means that while we may not know the absolute mean intensity, we do know the true difference in mean intensity between two groups.
For each group $k$ and each peptide $p$, we thus reconstructed the reference intensity mean $\mu_{p,k}^{true}$ as follows:
\begin{enumerate}
    \item For each peptide, we adjusted its observed intensity by adding the log2-fold change between its group and a designated reference group (in the real data experiments, the highest point of the spike-in range). This created a reconstructed sample of peptide intensities for the reference group.
    \item We then averaged these reconstructed values to obtain the reference mean intensity for the reference group.
    \item Finally, for each peptide in any other group, we derived its reference mean intensity by subtracting the log2-fold change from the reference mean of the reference group.
\end{enumerate}

\subsection{The overlap coefficient and the total variation distance}

\begin{figure}[ht]
        \centering
        \includegraphics[width=\textwidth]{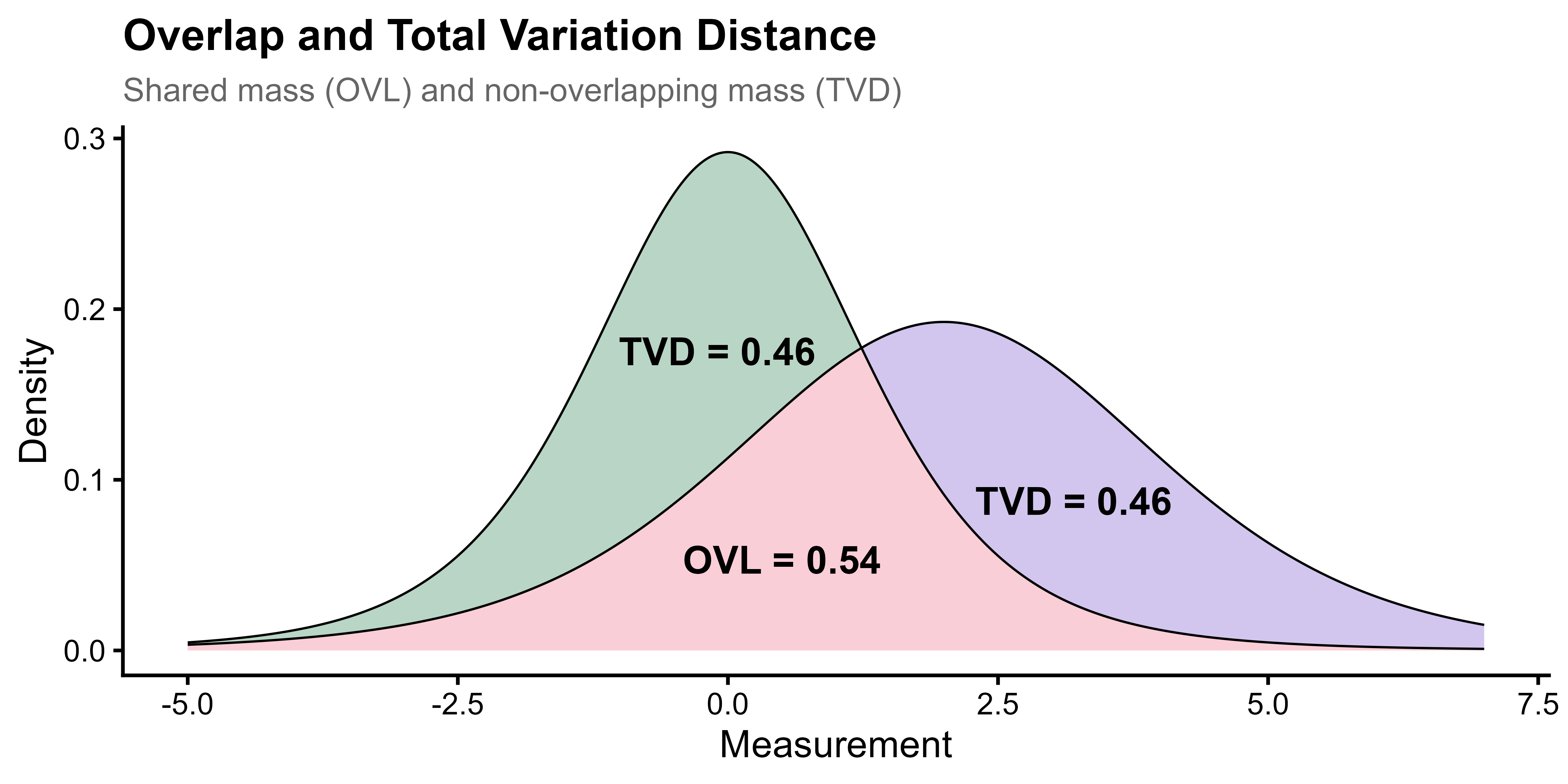}
        \caption{Illustration of the overlap (OVL) and the total variation distance (TVD) between two probability distributions.}
        \label{fig:illu_ovl}
\end{figure}

In addition to the previous standard metrics for assessing estimation performance and uncertainty calibration, let us introduce a particularly relevant quantity provided by the probabilistic treatment of differential analysis, namely the \emph{overlap coefficient} \citep{InmanBradley1989}. 
The overlap coefficient (OVL) is a measure of similarity between two distributions, defined as the shared probability mass. 
It ranges from 0 (disjoint supports) to 1 (identical distributions).
Formally, for two arbitrary probability distributions $P$ and $Q$ on $\mathcal{X}$, it can be computed as: 
$$
\mathrm{OVL}(P, Q) = \int_{\mathcal{X}} \min\bigl(p(x), q(x)\bigr)\, dx
$$
In addition, a related quantity often used in modern machine learning literature is known as the Total Variation Distance (TVD) \citep{LeCam1960}, such that:
$$
\mathrm{TVD}(P, Q) = \frac{1}{2} \int_{\mathcal{X}} \lvert p(x) - q(x) \rvert \, dx = \int_{\mathcal{X}} \max\bigl(p(x) - q(x), 0\bigr)\, dx
$$
Both quantities are therefore linked by the simple relation $OVL(P,Q) = 1 - TVD(P,Q)$, as illustrated in \Cref{fig:illu_ovl}.
Even though one should always be careful when summarising both effect size and uncertainty with a single number, TVD and OVL provide intuitive measures of distributional difference based on the observed data. 
They constitute valuable counterparts to traditional p-values as decision-making tools for assessing differential status, based on genuine probabilities rather than a somewhat arbitrary and too often misunderstood significance threshold. 
Therefore, TVD can be interpreted as the probability that two random variables drawn from the respective distributions differ, while OVL measures their overlap.
In practice, it should be interpreted more as a continuous measure of evidence rather than as a universal binary decision rule. 
Any reporting threshold should therefore be chosen in light of the scientific objective and a practically relevant effect size, in a way that is conceptually related to ROPE-style decision rules \citep{kruschkeRejectingAcceptingParameter2018}, although we do not use it here as a strict binary testing device.
For clarity, we chose to report TVD in the following experiments, as it aligns more closely with the traditional \emph{rejection} viewpoint of hypothesis testing and is likely more straightforward for readers less familiar with the Bayesian paradigm.

\subsection{Choice of hyperparameters}
\label{sec:hyperparam}

Unless otherwise stated, we used the following values for prior hyperparameters throughout the experiment section: 
\begin{itemize}
	\item $\mu_0 = \bar{y}$,
	\item $\lambda = 10^{-10}$,
	\item $\alpha_0 = 0.01$,
	\item $\beta_0 = 0.3$,
	\item $\Sigmab_0 = I_P$
\end{itemize}

where $\bar{y}$ represent the average of observed values computed over all groups.
These values correspond to practical insights from empirical sanity checks, while remaining relatively vague. 
In particular, the hyperparameter $\lambda_0$ corresponds to the confidence one holds in the prior $\mu_0$ value to be the correct mean. 
As in our differential analysis context, this prior mean is shared across all groups/conditions, and does not bear much interest in itself (we are interested in the difference of means, not in their actual values), we purposefully set $\lambda_0$ to a really low value, therefore cancelling most of $\mu_0$ influence in posterior distributions to facilitate fair comparison with competing methods (in particular with non-Bayesian ones).  
$\Sigma_0$ is considered diagonal a priori, which corresponds to a prior assumption of no correlations between peptides, and $\nu_0$ is always chosen as low as possible while satisfying the dimensional constraint $\nu_0 > P - N_K - 1$ for the posterior.
To further examine robustness to prior misspecification, we conducted an extensive sensitivity analysis of all key hyperparameters in Text C and Fig A-E in S1 Appendix.
Although we can argue that those priors could be improved, for instance, with an expert's knowledge, we believe that those choices are sensible, as we remove potential biases and confounding factors in the empirical results in our simulation study that could come from prior specification.
As previously stated, identical values in all groups are essential to ensure a fair and unbiased comparison. 
More generally, prior specification is a central question in Bayesian statistics that has been thoroughly explored, in particular for conjugate models, and extended discussions can be found in \citep{gelman2006prior}.

\subsection{Illustration and interpretation of posterior distributions}

First, let us illustrate the univariate framework described in \Cref{sec:uncorr}, using the Chion2022 dataset. 
In this experiment, we compared the intensity means at the lowest (0.05 fmol UPS1) and highest (10 fmol UPS1) points of the UPS1 spike range.
Remember that our univariate algorithm does not rely on imputation and should be applied directly to raw data.
For illustrative purposes, the chosen peptides were observed in all three biological samples for both experimental conditions.

\begin{figure}[ht]
     \makebox[\textwidth][c]{
     \includegraphics[width=0.95\textwidth]{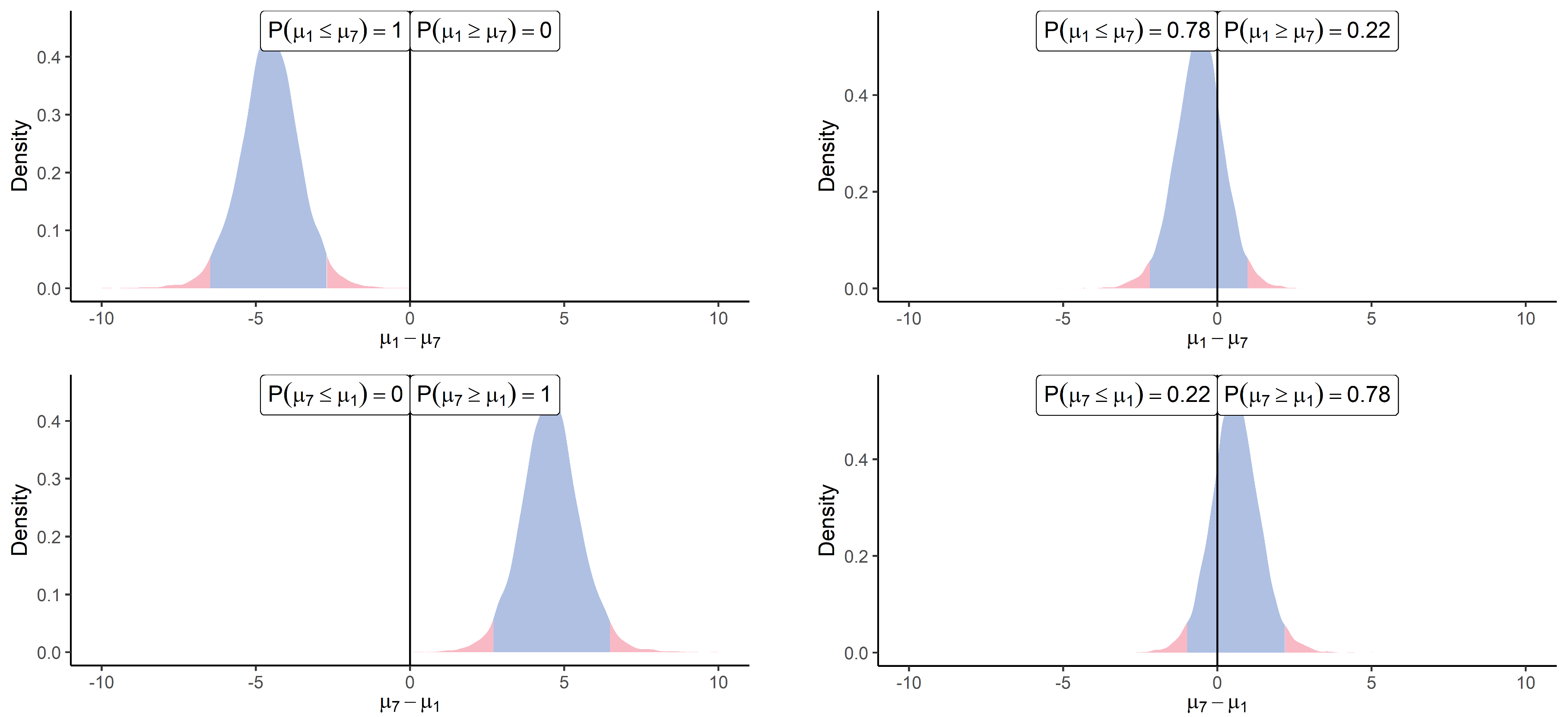}}
        \caption{Posterior distributions of the difference of means between the 0.05 fmol UPS spike condition ($\mu_1$) and the 10 fmol UPS spike condition ($\mu_7$). The blue central region indicates the 95\% credible interval. Left: \texttt{AALEELVK} peptide from the \texttt{P12081ups$\mid$SYHC\_HUMAN\_UPS} protein. Right: \texttt{VLPLIIPILSK} peptide from the \texttt{sp$\mid$F4I893$\mid$ILA\_ARATH} protein.}
        \label{fig:graph1}
\end{figure}

As a result of the application of our univariate algorithm, posterior distributions of the mean difference for both peptides are represented on \Cref{fig:graph1}. 
As the analysis compares conditions, the value 0 has been highlighted on the x-axis to assess both the direction and the magnitude of the difference.
The blue area under the distribution curve corresponds to the 95\% credible interval, meaning that there is a 95\% probability that the true mean difference lies within it. 
The distance to zero of the distributions indicates whether the peptide is differentially expressed or not. 
In particular, the left panel shows the posterior distribution of the means' difference for the UPS peptide.
Its location, far from zero, indicates a high probability (almost surely in this case) that the mean intensity of this peptide differs between the two groups.
Conversely, the posterior distribution of the difference of means for the ARATH peptide (right panel) indicates, as expected, with high probability, that the groups are not so different. 
Those conclusions support the raw data summaries depicted on the bottom panel of \Cref{fig:graph1}.
Moreover, the posterior distribution provides additional insights into whether a peptide is under-expressed or over-expressed in a condition compared to another.
For example, looking back to the UPS peptide, the left panel suggests an over-expression of the \texttt{AALEELVK} peptide in the seventh group (being the condition with the highest amount of UPS spike) compared to the first group (being the condition with the lowest amount of UPS spike), which is consistent with the experimental design.
Furthermore, the middle panel merely highlights that the posterior distribution of the difference $\mu_1 - \mu_7$ is symmetric to that of $\mu_7 - \mu_1$, so the direction of the comparison remains an aesthetic choice. 

We believe this inference procedure, based on the probability that one group has a larger mean than another, is particularly intuitive to practitioners. 
Instead of following automatic, and somewhat arbitrary, decision rules based on a \emph{significance} threshold that can be modified from one experiment to another, probabilistic reasoning emphasises that statistical inference inherently carries a degree of uncertainty.
However, if this uncertainty is carefully quantified and explicitly presented in statistical software, it returns decision-making power to the scientists, the actual experts in the specific field being studied \citep{betenskyPValueRequiresContext2019}.
We argue that scientists should be the ones knowledgeably assessing whether a certain effect size, and its associated uncertainty, can be considered \emph{meaningful} in this specific context \citep{sullivanUsingEffectSize2012}.

\subsection{Univariate Bayesian inference for differential analysis}

In this subsection, we evaluate the univariate framework described in \Cref{sec:uncorr} using the performance indicators defined in \Cref{subsec-perf} on both simulated (see \cref{subsec:simdata}) and real controlled datasets (see \cref{subsec:realdata}).

\subsubsection{Running time comparison}

A drawback often associated with Bayesian methods is the greater computational burden compared to frequentist counterparts.
However, by leveraging conjugate priors in our model and sampling from analytical distributions for inference, we maintained a (univariate) algorithm that was as quick as reference non-probabilist methods in practice, as illustrated in \Cref{tab:running_time}.
As expected, the multivariate version generally suffers from a scaling issue when modelling more than $10^3$ peptides jointly, as we need to estimate large covariance matrices.
However, this benchmark is mainly theoretical, as typical use cases in proteomics would consider numbers of correlated peptides, for instance, within the same protein, way below than $10^3$. 
Therefore, scaling remains roughly as effective as univariate methods when peptides are allocated to subgroups of reasonable size. 
Overall, ProteoBayes provides a competitive alternative to the main differential analysis tools (such as limma, proDA, MSqRob), offering additional benefits of probabilistic and multivariate inference at a negligible computational cost. 

\begin{table}[ht]
\begin{tabular}{c|cc|c|c|c|c|}
\cline{2-7}
 & \multicolumn{2}{c|}{\textbf{ProteoBayes}}   & \multirow{2}{*}{\textbf{t-test}} & \multirow{2}{*}{\textbf{limma}} & \multirow{2}{*}{\textcolor{review}{\textbf{proDA}}} & \multirow{2}{*}{\textcolor{review}{\textbf{MSqRob}}} \\
                                 & \textbf{Univariate} & \textbf{Multivariate} &                                  &                                 &                                 &                                  \\ \hline
\multicolumn{1}{|c|}{$P = 10$}   & 0.12 (0.04)         & 0.22 (0.13)           & 0.02 (0.01)                      & 0.03 (0.02)                     & 0.26 (0.07)                     & 0.06 (0.02)                      \\
\multicolumn{1}{|c|}{$P = 10^2$} & 0.20 (0.07)         & 0.20 (0.08)           & 0.07 (0.08)                      & 0.04 (0.02)                     & 1.16 (0.42)                     & 0.20 (0.07)                      \\
\multicolumn{1}{|c|}{$P = 10^3$} & 1.16 (0.72)         & 0.95 (0.42)           & 0.24 (0.06)                      & 0.09 (0.03)                     & 10.84 (3.20)                    & 1.64 (0.60)                      \\
\multicolumn{1}{|c|}{$P = 10^4$} & 7.75 (3.03)         & 249.17 (27.51)        & 2.64 (0.79)                      & 9.10 (6.34)                     & 93.51 (21.96)                   & 16.23 (3.73)                     \\ \hline
\end{tabular}
\centering
\caption{Running times (in seconds) of univariate and multivariate ProteoBayes compared with t-test, limma, proDA, and MSqRob differential analysis procedures, for an increasing number of peptides. All results are averaged over 10 repetitions of the experiments and reported using the format \emph{Mean (Sd)}.}
    \label{tab:running_time}
\end{table}

\subsubsection{Acknowledging the effect size and uncertainty quantification}

\begin{figure}[ht]
 	\makebox[\textwidth][c]{\includegraphics[width = .7\textwidth]{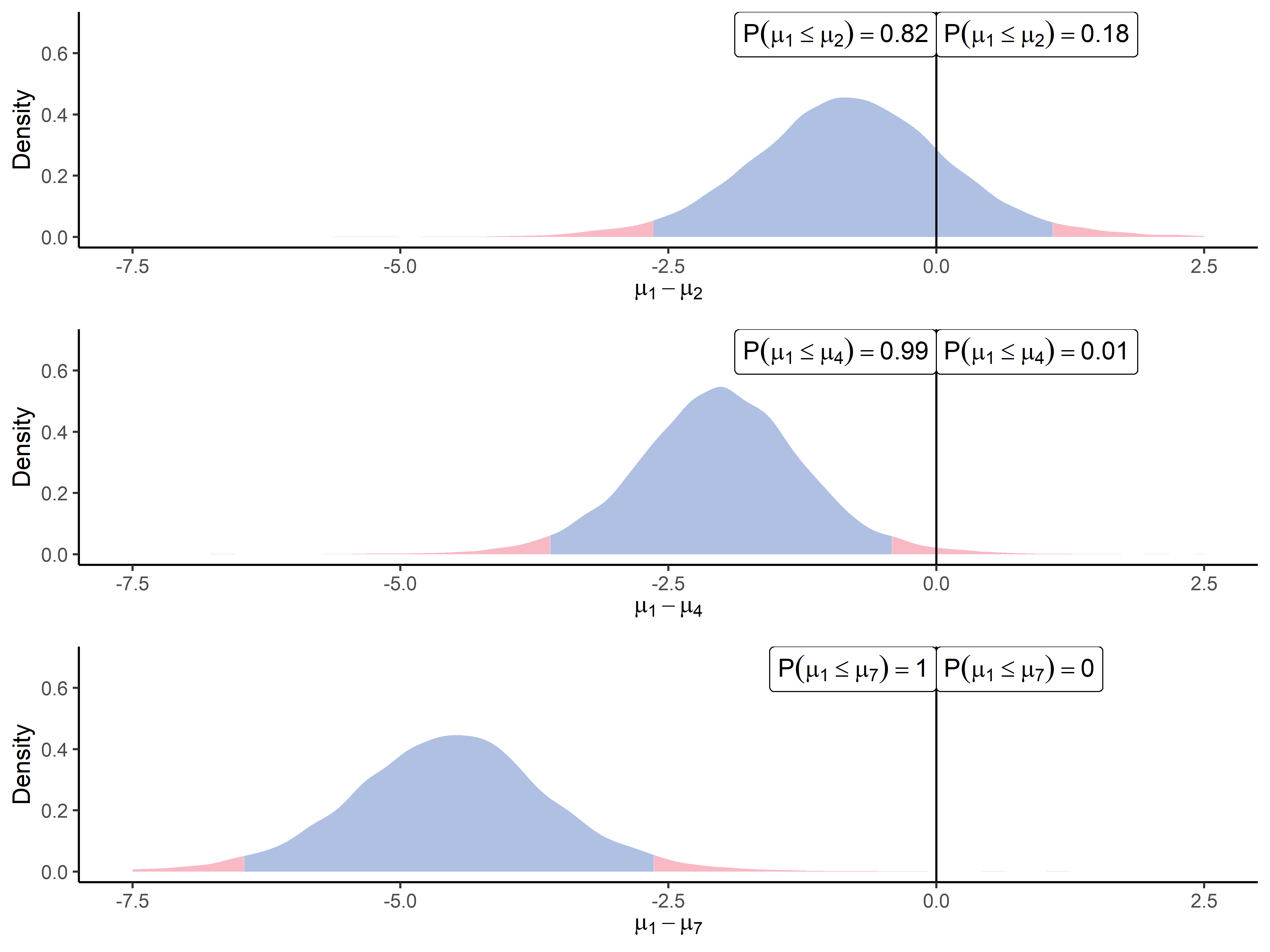}}
    \caption{Posterior distributions of the mean differences $\mu_1 - \mu_2$, $\mu_1 - \mu_4$ and $\mu_1 - \mu_7$ for the \texttt{AALEELVK} peptide from the \texttt{P12081ups$\mid$SYHC\_HUMAN\_UPS} protein.}
    \label{fig:graph4}
\end{figure}

\begin{table}
\makebox[\textwidth][c]{
\begin{tabular}{cc|ccc|cc|c|c|}
\cline{3-9}
\multicolumn{1}{l}{}                                                                          &                      & \multicolumn{3}{c|}{\textbf{ProteoBayes}}                                                                                               & \multicolumn{2}{c|}{\textbf{Quality of estimation}} & \textbf{limma}   & \textcolor{review}{\textbf{MSqRob}}  \\
\multicolumn{1}{l}{}                                                                          &                      & \textcolor{review}{\textbf{TVD}} & \textbf{\begin{tabular}[c]{@{}c@{}}Mean\\  difference*\end{tabular}} & $\mathbf{CI_{95}}$ \textbf{width} & \textbf{RMSE}         & $\mathbf{CIC_{95}}$         & \textbf{p-value} & \textcolor{review}{\textbf{p-value}} \\ \hline
\multicolumn{1}{|c|}{\multirow{6}{*}{\begin{tabular}[c]{@{}c@{}}5\\ samples\end{tabular}}}    & $\mathcal{N}(1, 1)$  & 0.34 (0.24)  & 1.02 (0.62)                                                         & 2.09 (0.63)                                        & 0.45 (0.53)           & 95.10 (21.60)               & 0.22 (0.26)      & 0.65 (0.17)      \\
\multicolumn{1}{|c|}{}                                                                        & $\mathcal{N}(5, 1)$  & 0 (0)        & 5.07 (0.63)                                                         & 2.11 (0.62)                                        & 0.46 (0.54)           & 94.2 (23.39)                & 0 (0)            & 0.04 (0.03)      \\
\multicolumn{1}{|c|}{}                                                                        & $\mathcal{N}(10, 1)$ & 0 (0)        & 10.05 (0.61)                                                        & 2.15 (0.65)                                        & 0.42 (0.50)           & 96.6 (18.34)                & 0 (0)            & 0 (0)            \\
\multicolumn{1}{|c|}{}                                                                        & $\mathcal{N}(1, 5)$  & 0.28 (0.16)  & 1.03 (2.34)                                                         & 9.52 (3.48)                                        & 2.30 (2.88)           & 91.8 (27.45)                & 0.75 (0.18)      & 0.49 (0.28)      \\
\multicolumn{1}{|c|}{}                                                                        & $\mathcal{N}(1, 10)$ & 0.17 (0.10)  & 0.96 (4.59)                                                         & 19.25 (6.62)                                       & 4.57 (5.38)           & 91.6 (27.75)                & 0.58 (0.26)      & 0.33 (0.32)      \\
\multicolumn{1}{|c|}{}                                                                        & $\mathcal{N}(1, 20)$ & 0.10 (0.06)  & 0.75 (8.96)                                                         & 38.58 (13.98)                                      & 8.95 (10.95)          & 93.0 (10.95)                & 0.40 (0.31)      & 0.23 (0.29)      \\ \hline
\multicolumn{1}{|c|}{\multirow{6}{*}{\begin{tabular}[c]{@{}c@{}}1000\\ samples\end{tabular}}} & $\mathcal{N}(1, 1)$  & 0 (0)        & 1 (0.04)                                                            & 0.12 (0.003)                                       & 0.03 (0.04)           & 95.7 (20.3)                 & 0 (0)            & 0 (0)            \\
\multicolumn{1}{|c|}{}                                                                        & $\mathcal{N}(5, 1)$  & 0 (0)        & 4.99 (0.04)                                                         & 0.12 (0.003)                                       & 0.03 (0.04)           & 94.6 (22.61)                & 0 (0)            & 0 (0)            \\
\multicolumn{1}{|c|}{}                                                                        & $\mathcal{N}(10, 1)$ & 0 (0)        & 9.99 (0.04)                                                         & 0.13 (0.003)                                       & 0.03 (0.04)           & 95.9 (19.84)                & 0 (0)            & 0 (0)            \\
\multicolumn{1}{|c|}{}                                                                        & $\mathcal{N}(1, 5)$  & 0 (0)        & 1 (0.16)                                                            & 0.6 (0.01)                                         & 0.16 (0.19)           & 95.5 (20.74)                & 0.04 (0.04)      & 0 (0)            \\
\multicolumn{1}{|c|}{}                                                                        & $\mathcal{N}(1, 10)$ & 0.01 (0.03)  & 0.99 (0.31)                                                         & 1.2 (0.02)                                         & 0.31 (0.37)           & 95.0 (21.81)                & 0.08 (0.12)      & 0.01 (0.05)      \\
\multicolumn{1}{|c|}{}                                                                        & $\mathcal{N}(1, 20)$ & 0.05 (0.04)  & 1.04 (0.58)                                                         & 2.4 (0.06)                                         & 0.62 (0.75)           & 95.2 (21.39)                & 0.15 (0.24)      & 0.09 (0.22)      \\ \hline
\end{tabular}}
\centering
\caption{\textcolor{review}{Simulation study reporting performances of univariate ProteoBayes compared to limma and MSqRob for differential analysis inference. All distributions are compared with the univariate Gaussian baseline $\mathcal{N}(0, 1)$. All results are averaged over 1000 repetitions of the experiments and reported using the format \emph{Mean (Sd)}. (*) Mean differences are identical across all methods and thus only reported once for concision.}}
\label{tab:effect_size_variance}
\end{table}

A key feature of ProteoBayes is to directly perform inference on effect sizes (referred to as \emph{fold change} in proteomics), i.e., the estimated difference in means within posterior distributions, leveraging explicit uncertainty quantification to provide probabilistic statements.
\Cref{fig:graph4} illustrates increasing mean differences with the amount of UPS proteins spiked, across various conditions, consistent with the experimental design. \citep{chionAccountingMultipleImputationinduced2022}.
In addition, \Cref{fig:graph4} highlights the importance of considering the effect size, which is crucial when studying the underlying biological phenomenon. 
To dive into the extensive evaluation of ProteoBayes on synthetic data, we provided in \Cref{tab:effect_size_variance} a thorough analysis of the computation of mean differences for various effect sizes and variance combinations. 
We recover empirical values that are close to the expected mean difference on average, even with only 5 samples, and are almost exact with 1000 observed samples.
Increased variance results in wider credible intervals, as the posterior distributions reflect higher uncertainty.
\textcolor{review}{Empirical p-values reported by limma and MSqRob remain difficult to interpret as they conflate effect size, variance, and sample size (which is why volcano plots are generally displayed in practice), although they remain convenient for ranking analytes.
In this sense, TVD provides an actual probability that two measurements are observed under indistinguishable conditions, which exhibits the expected empirical behaviour in the present experiments.}
In addition to inference metrics for all competing methods, we provided in \Cref{tab:effect_size_variance}, and all subsequent result tables, sanity-check metrics regarding the quality of estimation, including Root Mean Squared Error (RMSE) and the empirical Coverage of the $95\%$ Credible Interval ($CIC_{95}$).
We observe consistent behaviour, while calibration of uncertainty quantification remains remarkably stable, even in low-sample-size regimes. 
Those measures constitute solid empirical evidence that our proposed method recovers accurate posterior distributions. 

We further evaluated our approach on real, controlled datasets from proteomics experiments. 
In the main text, we report the highest-quality experiment across our panel in \Cref{tab:Muller2016}, whereas results for additional datasets are available in Tables B-D in S1 Appendix.
We displayed mean-difference metrics for both limma and ProteoBayes, which are always equal across all experiments.
This behaviour is theoretically expected as we have set a purposefully low value for the prior hyperparameter $\lambda_0 = 10^{-10}$ that entirely cancels all influence of the prior mean $\mu_0$. 
Our empirical results confirm that the fold change in limma constitutes a special case of ProteoBayes, in which we ignore prior information.
Uncertainty remains well-calibrated, though yeast conditions (non-differential) are more challenging and lead to a slight but consistent overestimation of variability.

\begin{figure}[ht]
    \makebox[\textwidth][c]{
        \includegraphics[width=0.95\textwidth]{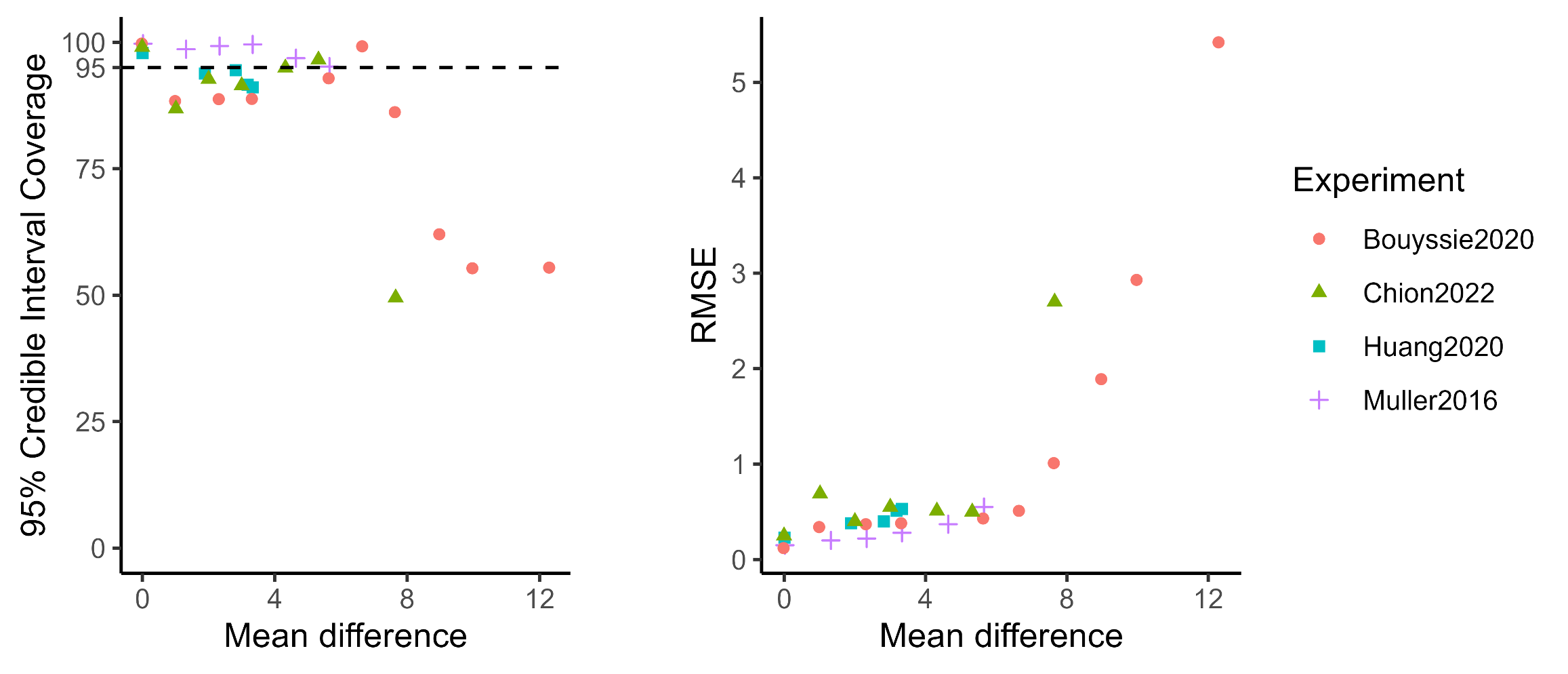}}
    \caption{Graphical summary of the quality of estimation for all real datasets. RMSE and $CIC_{95}$ values are reported with respect to the true mean difference computed in different experimental settings. For $CIC_{95}$, values should be as close as possible to the theoretical threshold $95$. For RMSE, the lower the value, the better.}
    \label{fig:CIC_RMSE_real_datasets}
\end{figure}

While all these real-world controlled experiments yield overall coherent results, we observed some noticeable differences compared to simulations. 
As the true mean difference increases, sanity-check metrics decrease consistently, as displayed in \Cref{fig:CIC_RMSE_real_datasets}.
Across all datasets, we observe that reasonable mean differences remain well estimated. In contrast, both errors and uncertainty calibration deteriorate sharply at higher values in the Bouyssie2020 and Chion2022 experiments (Tables B and D in S1 Appendix). 
Although we previously demonstrated correct calibration in simulations, the largest effect sizes are not well recovered by either limma or ProteoBayes on real datasets.
This could challenge the hypothesis of proportionality between protein quantities and their measured intensities.

In label-free data-dependent acquisition (DDA) proteomics, relative quantification using extracted ion chromatograms (XICs) relies on the assumption of a linear relationship between peptide signal intensity—typically expressed as the integrated chromatographic peak area—and peptide abundance across samples \citep{matzkeComparativeAnalysisComputational2013}. 
This assumption is strong, as it requires stable ionisation efficiency, detector response, and chromatographic performance across a wide dynamic range. 
In practice, however, the intrinsic stochasticity of DDA, combined with noisy signals, particularly for low-abundance peptides, can compromise this proportionality and lead to inaccurate peptide quantification \citep{coxAccurateProteomewideLabelfree2014, rozanovaQuantitativeMassSpectrometryBased2021}.

\begin{table}
\makebox[\textwidth][c]{
\begin{tabular}{|c|c|c|cc|cccc|}
\hline
\multirow{2}{*}{\textbf{Truth}} & \multirow{2}{*}{\textbf{\begin{tabular}[c]{@{}c@{}}Vs.\\ 25 fmol\end{tabular}}} & \multirow{2}{*}{\textbf{\begin{tabular}[c]{@{}c@{}}Nb of\\ peptides\end{tabular}}} & \multicolumn{2}{c|}{\textbf{Mean difference*}}            & \multicolumn{4}{c|}{\textbf{ProteoBayes}}                                                                                                    \\ \cline{4-9} 
                                &                                                                                 &                                                                                    & \multicolumn{1}{c|}{\textbf{True}} & \textbf{ProteoBayes} & \multicolumn{1}{c|}{\textcolor{review}{\textbf{TVD}}} & \multicolumn{1}{c|}{\textbf{CI$_{95}$ width}} & \multicolumn{1}{c|}{\textbf{RMSE}} & \textbf{CIC$_{95}$} \\ \hline
\multirow{5}{*}{\rotatebox[origin=c]{90}{\textbf{UPS}}}           & 0.5 fmol                                                                        & 229                                                                                & \multicolumn{1}{c|}{5.64}          & 5.01 (1.20)          & \multicolumn{1}{c|}{0.09 (0.16)}  & \multicolumn{1}{c|}{7.72 (7.58)}              & \multicolumn{1}{c|}{0.92 (1.52)}   & 95.20 (21.43)       \\
                                & 1 fmol                                                                          & 350                                                                                & \multicolumn{1}{c|}{4.64}          & 4.31 (0.86)          & \multicolumn{1}{c|}{0.11 (0.17)}  & \multicolumn{1}{c|}{6.08 (6.89)}              & \multicolumn{1}{c|}{0.57 (0.91)}   & 96.86 (17.47)       \\
                                & 2.5 fmol                                                                        & 478                                                                                & \multicolumn{1}{c|}{3.32}          & 3.09 (0.71)          & \multicolumn{1}{c|}{0.11 (0.15)}  & \multicolumn{1}{c|}{5.06 (6.14)}              & \multicolumn{1}{c|}{0.47 (0.83)}   & 99.58 (6.46)        \\
                                & 5 fmol                                                                          & 538                                                                                & \multicolumn{1}{c|}{2.32}          & 2.18 (0.58)          & \multicolumn{1}{c|}{0.11 (0.14)}  & \multicolumn{1}{c|}{4.18 (5.45)}              & \multicolumn{1}{c|}{0.39 (0.87)}   & 99.26 (8.60)        \\
                                & 10 fmol                                                                         & 585                                                                                & \multicolumn{1}{c|}{1.32}          & 1.20 (0.39)          & \multicolumn{1}{c|}{0.09 (0.16)}  & \multicolumn{1}{c|}{2.94 (3.83)}              & \multicolumn{1}{c|}{0.32 (0.59)}   & 98.63 (11.62)       \\ \hline
\multirow{5}{*}{\rotatebox[origin=c]{90}{\textbf{YEAST}}}          & 0.5 fmol                                                                        & 19856                                                                              & \multicolumn{1}{c|}{0}             & 0.09 (0.45)          & \multicolumn{1}{c|}{0.76 (0.18)}  & \multicolumn{1}{c|}{3.14 (4.01)}              & \multicolumn{1}{c|}{0.31 (0.74)}   & 99.74 (5.11)        \\
                                & 1 fmol                                                                          & 19784                                                                              & \multicolumn{1}{c|}{0}             & 0.11 (0.43)          & \multicolumn{1}{c|}{0.78 (0.17)}  & \multicolumn{1}{c|}{3.01 (3.98)}              & \multicolumn{1}{c|}{0.30 (1.04)}   & 99.53 (6.80)        \\
                                & 2.5 fmol                                                                        & 19835                                                                              & \multicolumn{1}{c|}{0}             & 0.10 (0.40)          & \multicolumn{1}{c|}{0.79 (0.16)}  & \multicolumn{1}{c|}{3.20 (4.11)}              & \multicolumn{1}{c|}{0.27 (0.67)}   & 99.83 (4.08)        \\
                                & 5 fmol                                                                          & 19740                                                                              & \multicolumn{1}{c|}{0}             & 0.07 (0.38)          & \multicolumn{1}{c|}{0.80 (0.16)}  & \multicolumn{1}{c|}{3.09 (4.08)}              & \multicolumn{1}{c|}{0.26 (0.66)}   & 99.82 (4.21)        \\
                                & 10 fmol                                                                         & 19776                                                                              & \multicolumn{1}{c|}{0}             & 0.04 (0.39)          & \multicolumn{1}{c|}{0.78 (0.16)}  & \multicolumn{1}{c|}{3.17 (4.23)}              & \multicolumn{1}{c|}{0.28 (0.72)}   & 99.70 (5.50)        \\ \hline
\end{tabular}}
\caption{Results table for the differential analysis of the Muller2016 dataset. All results are averaged over all peptides in each group and reported using the format \textit{Mean (Sd)}. \textcolor{review}{(*) Mean differences are identical across all competing methods (limma, MSqRob) thus only reported for ProteoBayes for concision.}}
\label{tab:Muller2016}
\end{table}

\subsubsection{The mirage of MAR imputed data}

After discussing the advantages and the valuable interpretative properties of our methods, let us mention a pitfall that one should avoid for the inferences to remain valid. 
In the case of univariate analysis, we noted in \Cref{eq:factorise} that all useful information is contained in the observed data, and no imputation is needed since we have already integrated out missing at random data. 
Our claim is not that imputation is always inappropriate in proteomics, nor that all missingness mechanisms are ignorable. Rather, in the univariate MAR setting considered here, replacing missing values with artificial observations can create the illusion of increased information, thereby distorting uncertainty quantification.

\begin{figure}[ht]
 	\makebox[\textwidth][c]{\includegraphics[width =.7\textwidth]{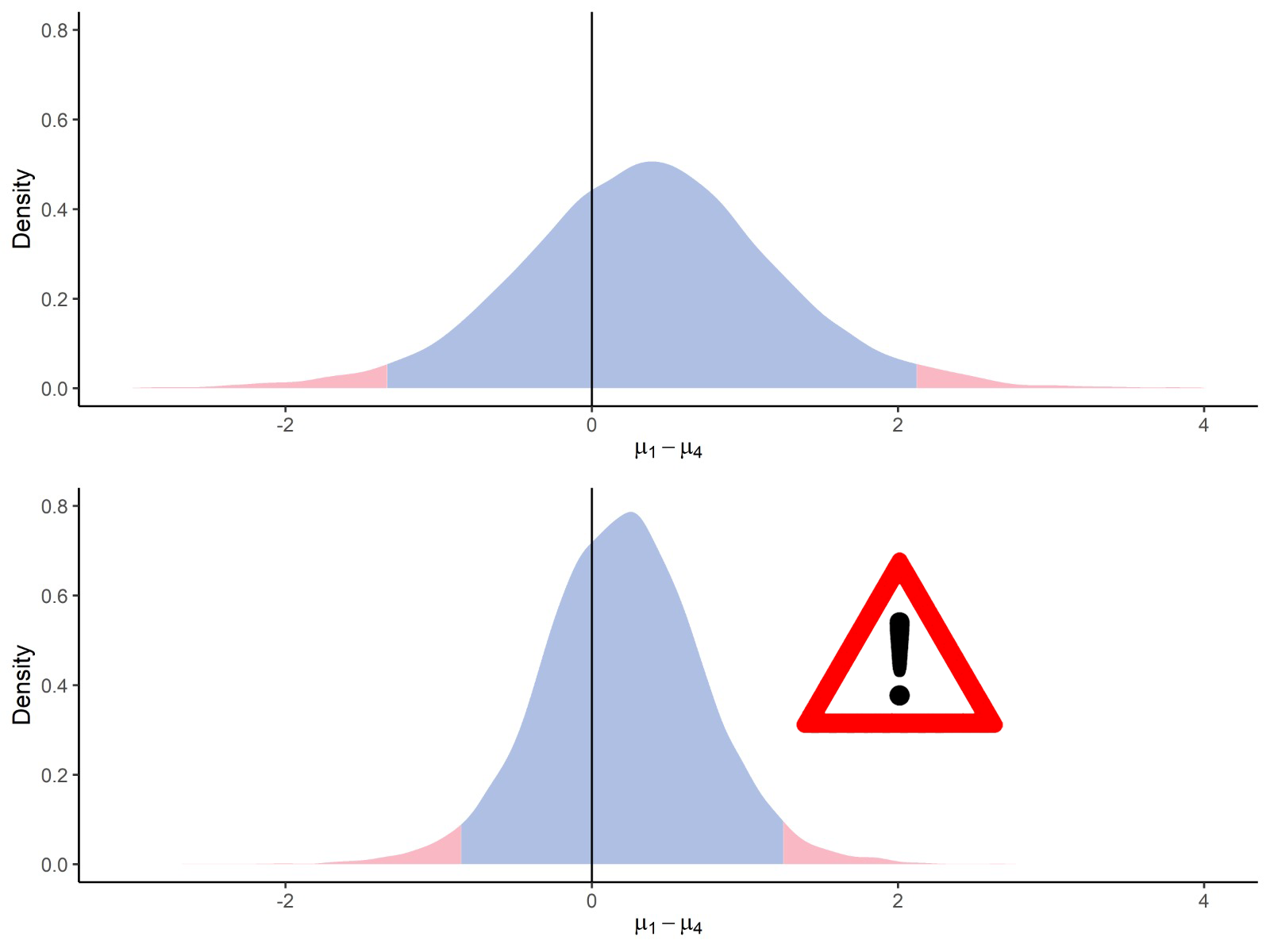}}
    \caption{Posterior distributions of the mean difference $\mu_1 - \mu_4$ for the \texttt{EVQELAQEAAER} peptide from the \texttt{sp$\mid$F4I893$\mid$ILA\_ARATH} protein using the observed dataset (top) and the imputed dataset (bottom)}
    \label{fig:mirage_imp}
\end{figure}

For illustration, we displayed in \Cref{fig:mirage_imp} an example of our univariate algorithm applied to a real dataset (top panel) with 2 replicates and to the same dataset with 1 additional imputed replicate (bottom panel). 
In this context, we observe lower variance in the imputed dataset, though this reduction constitutes an artefact of the imputation process.

To explore this behaviour more systematically, \Cref{tab:tab_mirage_imputation} reports performance metrics similar to those before, although we deliberately introduced varying levels of \textcolor{review}{MAR} data to conduct analyses with and without imputation before applying all competing methods.
As expected, we observe that imputation degrades the calibration of uncertainty quantification in ProteoBayes and should thus be avoided, since the posterior distributions naturally adapt to the amount of data collected (as indicated by the larger credible intervals), even at high rates of missing data. 
For test-based inference methods, we observe two problematic yet expected behaviours that depend on the context. 
In the absence of imputation, we can see p-values increasing, even though we did not change the underlying mean difference. 
This is expected, as p-values somewhat collapse information from both the effect size and its associated uncertainty into a single number, which is often hard to interpret for this very reason. 
Conversely, if we perform imputation before testing, we can see that p-values artificially decrease as the missing-data ratio increases, leading to spurious inferences solely due to overestimating the amount of observed information.
\textcolor{review}{Even the \emph{proDA} algorithm, specifically designed to deal with missing data and avoid imputation, struggles to provide meaningful results in our experiments, and often failed to converge in really sparse regimes (as highlighted by pathological results for $80\%$ missing data).
While our claim about ignoring missing data holds only under MAR hypotheses, we conducted additional experiments to evaluate empirical robustness to MNAR. 
In Table E in S1 Appendix, instead of randomly sampling missing data, we systematically removed the lowest observed values (to emulate the detection threshold often encountered in proteomics), while still increasing the missingness ratio in both imputation and no-imputation contexts.
We can observe that calibration metrics deteriorate, as expected, with increased RMSE and damaged credible intervals. However, this behaviour seems much worse when imputing missing data, leading to strong overconfidence. 
In terms of mean difference and total variation distance, ProteoBayes remains reasonably robust and cautious, whereas both limma and proDA seem to be consistently overconfident.} 

Those pathological behaviours in the context of missing data once more highlight the counterintuitive phenomena that arise when conducting inference using simplistic and overly sensitive measures such as p-values.
Let us note that this imputation issue is not specific to the present framework and, more generally, applies to Rubin's rules as well. 
One should keep in mind that these approximations hold only for a reasonable level of missing data.
Otherwise, one may consider adapting the method, for example, by penalising the degree of freedom in the relevant $t$-distributions.
More generally, while imputation is sometimes necessary for the methods to run, one should keep in mind that it always introduces a bias (even when controlled) that should be accounted for.

\begin{table}
\makebox[\textwidth][c]{
\begin{tabular}{c|c|clccl|c|c|}
\cline{2-9}
\multirow{2}{*}{\textbf{}}                   & \multirow{2}{*}{\textbf{\begin{tabular}[c]{@{}c@{}}Missing \\ ratio\end{tabular}}} & \multicolumn{5}{c|}{\textbf{ProteoBayes}}                                                                                                                                                                                                                                                                              & \multirow{2}{*}{\textbf{\begin{tabular}[c]{@{}c@{}}limma \\ p-value\end{tabular}}} & \multirow{2}{*}{\textbf{\begin{tabular}[c]{@{}c@{}} \textcolor{review}{proDA} \\ \textcolor{review}{p-value}\end{tabular}}} \\ \cline{3-7}
                                             &                                                                                    & \textbf{\begin{tabular}[c]{@{}c@{}}Mean \\ difference\end{tabular}} & \textcolor{review}{\textbf{RMSE}}                    & $\mathbf{CIC_{95}}$                & \textbf{\begin{tabular}[c]{@{}c@{}}$\mathbf{CI_{95}}$ \\ width\end{tabular}} & \multicolumn{1}{c|}{\textcolor{review}{\textbf{TVD} }} &                                                                                    &                                                                                                                                                                  \\ \hline
\multicolumn{1}{|c|}{\textbf{Complete data}} & 0\%                                                                                & \multicolumn{1}{c|}{1 (0.44)}                                       & \multicolumn{1}{l|}{0.29 (0.31)} & \multicolumn{1}{c|}{94.94 (21.92)} & \multicolumn{1}{c|}{1.35 (0.30)}                                             & 0.19 (0.19)                                                                               & 0.11 (0.18)                                                                        & 0.13 (0.19)                                                                                                                                                      \\ \hline
\multicolumn{1}{|c|}{\textbf{}}              & 20\%                                                                               & \multicolumn{1}{c|}{1 (0.51)}                                       & \multicolumn{1}{l|}{0.37 (0.47)} & \multicolumn{1}{c|}{94.74 (22.32)} & \multicolumn{1}{c|}{1.57 (0.44)}                                             & 0.25 (0.23)                                                                               & 0.14 (0.22)                                                                        & 0.17 (0.22)                                                                                                                                                      \\
\multicolumn{1}{|c|}{\textbf{No imputation}} & 50\%                                                                               & \multicolumn{1}{c|}{0.99 (0.67)}                                    & \multicolumn{1}{l|}{0.52 (0.57)} & \multicolumn{1}{c|}{95.86 (19.91)} & \multicolumn{1}{c|}{2.32 (1.10)}                                             & 0.35 (0.26)                                                                               & 0.24 (0.27)                                                                        & 0.29 (0.28)                                                                                                                                                      \\
\multicolumn{1}{|c|}{}                       & 80\%                                                                               & \multicolumn{1}{c|}{1 (0.91)}                                       & \multicolumn{1}{l|}{0.74 (1.01)} & \multicolumn{1}{c|}{97.56 (15.42)} & \multicolumn{1}{c|}{3.91 (1.88)}                                             & 0.50 (0.24)                                                                               & 0.33 (0.30)                                                                        & 1.00 (0.00)                                                                                                                                                      \\ \hline
\multicolumn{1}{|c|}{\textbf{}}              & 20\%                                                                               & \multicolumn{1}{c|}{1 (0.48)}                                       & \multicolumn{1}{l|}{0.33 (0.38)} & \multicolumn{1}{c|}{88.96 (31.34)} & \multicolumn{1}{c|}{1.19 (0.31)}                                             & 0.19 (0.23)                                                                               & 0.10 (0.19)                                                                        & 0.13 (0.22)                                                                                                                                                      \\
\multicolumn{1}{|c|}{\textbf{Imputation}}    & 50\%                                                                               & \multicolumn{1}{c|}{1.01 (0.49)}                                    & \multicolumn{1}{l|}{0.33 (0.38)} & \multicolumn{1}{c|}{78.00 (41.43)} & \multicolumn{1}{c|}{0.93 (0.31)}                                             & 0.11 (0.18)                                                                               & 0.07 (0.17)                                                                        & 0.06 (0.15)                                                                                                                                                      \\
\multicolumn{1}{|c|}{}                       & 80\%                                                                               & \multicolumn{1}{c|}{1 (0.48)}                                       & \multicolumn{1}{l|}{0.32 (0.34)} & \multicolumn{1}{c|}{61.34 (48.70)} & \multicolumn{1}{c|}{0.62 (0.25)}                                             & 0.08 (0.16)                                                                               & 0.03 (0.12)                                                                        & 0.03 (0.10)                                                                                                                                                      \\ \hline
\end{tabular}}
\caption{Performance metrics of ProteoBayes, limma, \textcolor{review}{and proDA}, for different scenarios of missing data ratios. Missing data are randomly removed, and imputation is performed by replacing missing values with the average of the observed ones. All results are averaged over 1000 repetitions of the experiments with 10 samples per peptide and reported using the format Mean (Sd).}
\label{tab:tab_mirage_imputation}
\end{table}

\subsection{Multivariate Bayesian inference}

\subsubsection{Comparing multivariate distributions}

\begin{figure}[ht]
    \includegraphics[width = \textwidth]{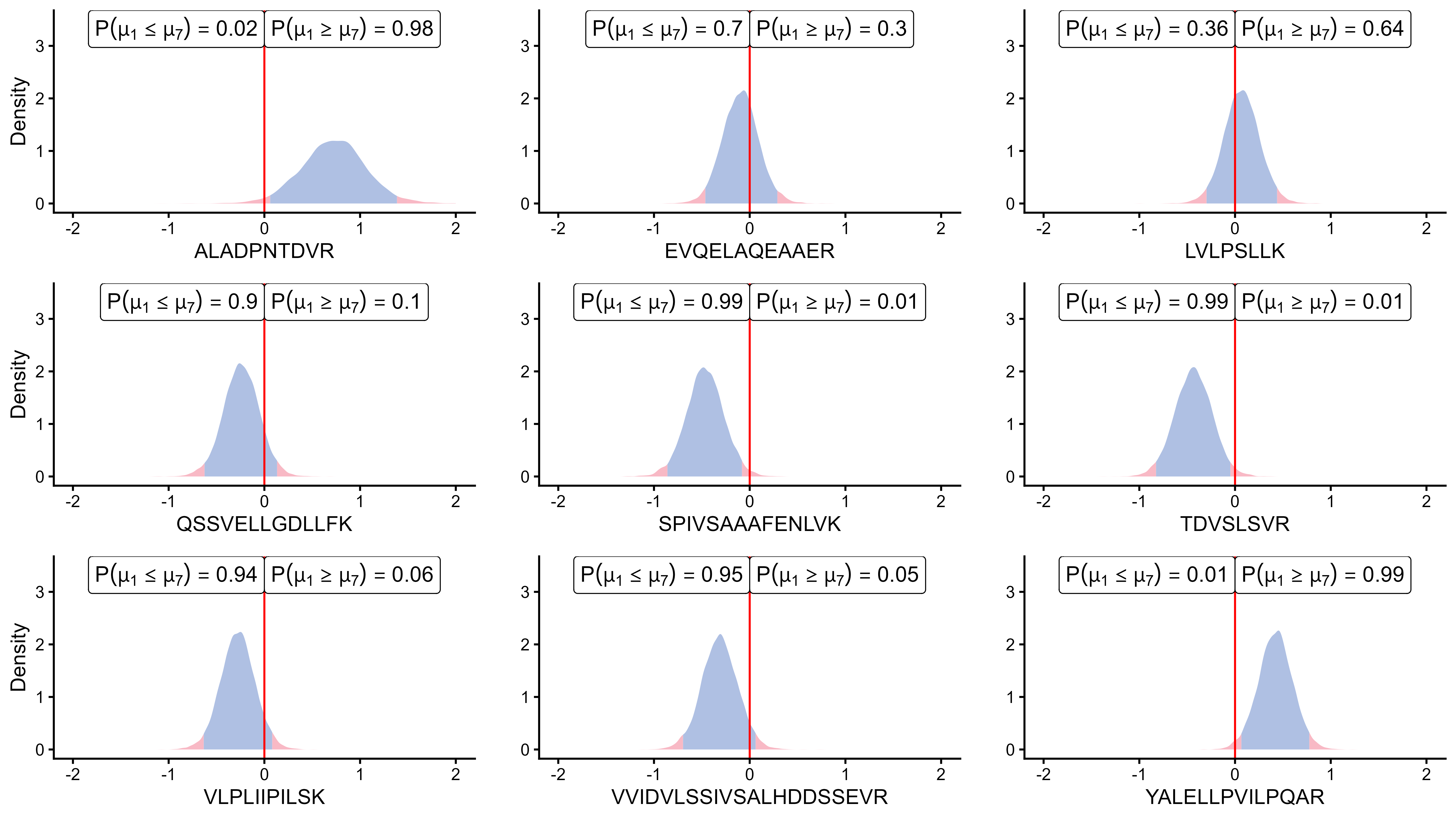}
    \caption{Posterior marginal distributions of mean differences $\mu_1 - \mu_7$ for the nine peptides from the sp\textbar F4I893\textbar ILA\_ARATH protein using multivariate ProteoBayes.}
    \label{fig:multi-marginals}
\end{figure}

\begin{figure}[ht]
    \centering
 	\includegraphics[width = 0.6\textwidth]{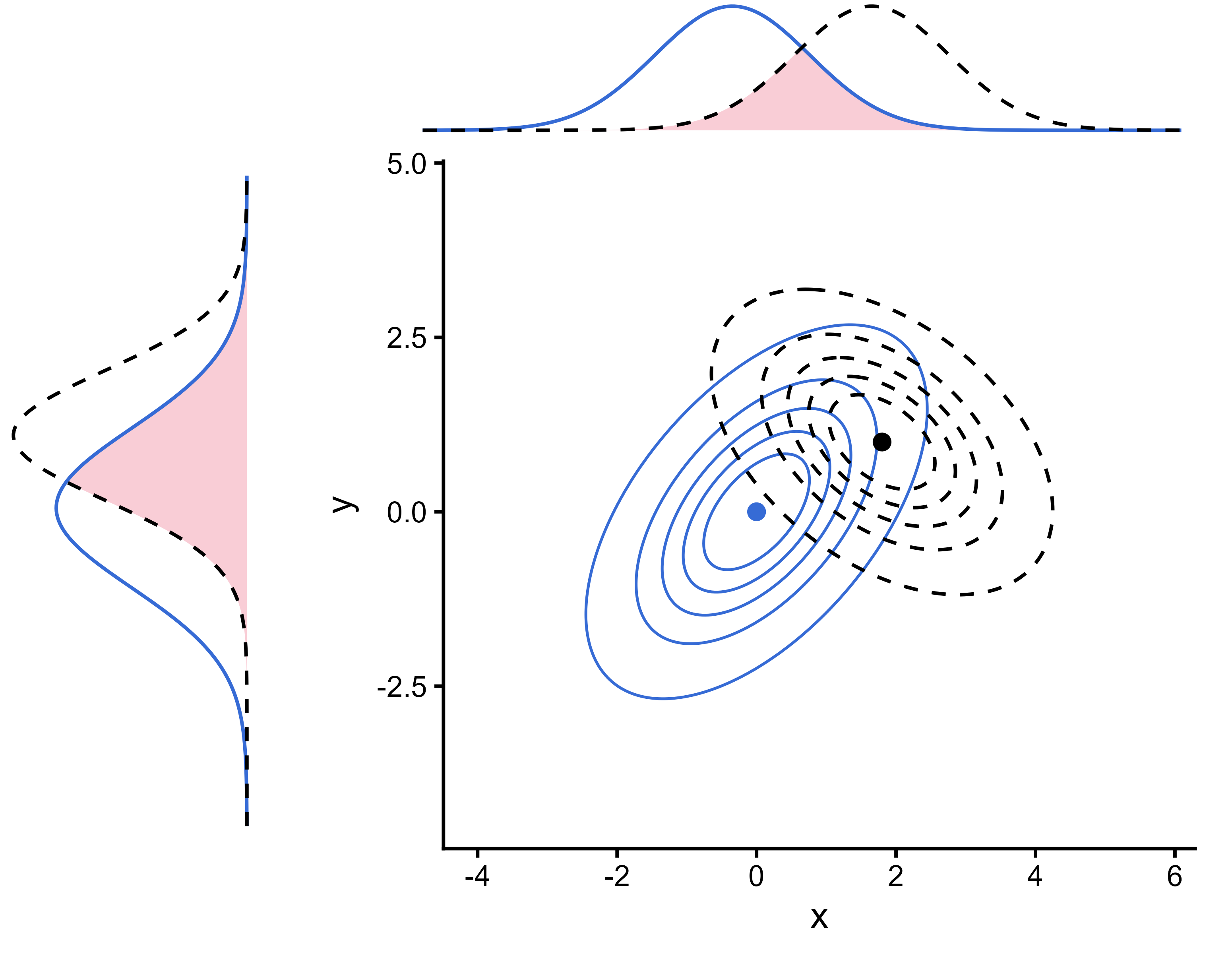}
    \caption{Illustration of two 2-dimensional probability distributions along with their marginals (respectively in blue plain lines and black dashed lines). The pink region depicts the \emph{overlap coefficient}, which measures the similarity between two univariate distributions and can be interpreted as a probability. This graph highlights the difficulty of differential analysis in a multivariate setting, as univariate intuitions quickly become irrelevant when comparing distributions in higher dimensions.}
    \label{fig:2d-gaussian}
\end{figure}

To the best of our knowledge, the present paper is among the first attempts to tackle the problem of multivariate differential analysis (e.g., protein inference rather than peptide-wise univariate comparisons), especially in a Bayesian setting.
The vast majority of routine methods operate in univariate settings, often performing thousands of independent tests across all studied elements to identify a subset of the most differentially expressed elements between groups/conditions. 
However, this approach largely ignores joint structures that are likely to exist (for instance, between peptides of the same protein) and would influence statistical results if carefully accounted for. 
Therefore, our aim in this section is to highlight the ability of the proposed method to capture such correlations and leverage them to perform genuine protein inference (in contrast with post-hoc inference, where decisions about proteins rely on arbitrary aggregation of peptide-wise differential statuses). 
We illustrate the difficulty of deriving reliable decision tools at the protein level from univariate analysis in \Cref{fig:multi-marginals}.
In this illustrative example, which we know to be non-differential, we observe 9 peptides from the same proteins, whose marginal distributions of the difference between the 2 groups appear slightly over- or under-abundant for some peptides, but are roughly similar overall. 
Making a decision based on aggregation measures is always somewhat arbitrary (e.g., average, majority, or the simple presence of a single differential peptide have been proposed in the literature, with no clear consensus). In particular, when the magnitude and the orientation of the effect size and the uncertainty are not adequately taken into account, as in traditional null-hypothesis testing frameworks. 

Considering our quantities of interest as multivariate probability distributions, we now need to propose relevant inference measures and decision tools in this novel context. 
To illustrate the difficulties arising in such multivariate settings (which often relate to the well-known \emph{curse of dimensionality}), we displayed in \Cref{fig:2d-gaussian} an example of 2-dimensional Gaussian densities, along their respective marginals, and overlapping regions. 
This figure illustrates that inference can be more subtle in multivariate settings and may quickly lead to intractable regimes. 
First, it is essential to recall that properties of the marginals differ from those of the joint distribution, in the sense that correlations play a crucial role in the \emph{shape} of each distribution (i.e., on the central graph, the contour lines are almost orthogonal, indicating opposite signs in their respective covariance matrices).
Additionally, criteria based on distances tend to become irrelevant as the dimension grows (intuitively, all objects are "far" from each other in high dimensions), and computations required to recover a reliable empirical distribution of the mean differences between two groups/conditions quickly become intractable (the number of necessary samples increases as $\mathcal{O}(2^D)$, with $D$ being the number of joint peptides).

Fortunately, we argue that the full probability distribution is, most of the time, unnecessary for answering practitioners' queries about a protein's differential status. 
There exists another quantity of interest appearing sufficient to conduct inference while remaining remarkably trivial to compute, even in high dimension (i.e. $\geq 10^4$ peptides). 
Therefore, we propose to compute the probability distribution of $N^+$, which we define as the number of peptides with higher values in one group/condition compared with the other. 
Symmetrically, if we denote $N^-$ the number of lower values, this can be deduced as $N^- = D - N^+$.
By being agnostic to peptide permutations, the probability distribution of $D^+$ is straightforward and quick to estimate from samples, and it provides a valuable measure of uncertainty for the multivariate inference procedure.
\textcolor{review}{In practice, we recommend summarising the posterior of $N^+$ through probabilities of the form $Pr(N^+ \geq k \mid \mathrm{data})$ or $Pr(N^+ \leq D - k \mid \mathrm{data})$, where $k$ encodes the minimum number of concordant peptides regarded as biologically meaningful. Large values provide strong evidence for a directional protein-level difference. In this sense, $N^+$ should be viewed as a continuous measure of evidence. The appropriate reporting threshold depends on the study objective, with confirmatory analyses generally requiring stronger evidence than exploratory ones. We also stress that $N^+$ is a pragmatic summary of coordinated peptide-level directionality within a protein group. Because it is invariant to peptide permutations, it does not explicitly model isoform-specific peptide subsets or ambiguous peptide-to-isoform assignments.}
This overall multivariate inference strategy is illustrated in \Cref{fig:multi_inference} on real data, and described in the following section.

\subsubsection{Protein inference: effect size and uncertainty across peptides}

In this section, we consider the comparison of intensity means in a multivariate setting.
As an example, we deliberately considered groups of 9 peptides from the Chion2022 dataset, whose intensities should be correlated to some degree, for both a known differential (P12081ups\textbar SYHC\_HUMAN\_UPS) and a non-differential (sp\textbar F4I893\textbar ILA\_ARATH) protein.
The posterior differences of the mean vector between pairs of conditions have been computed.

\begin{figure}[ht]
    \makebox[\textwidth][c]{
    \includegraphics[width =\textwidth]{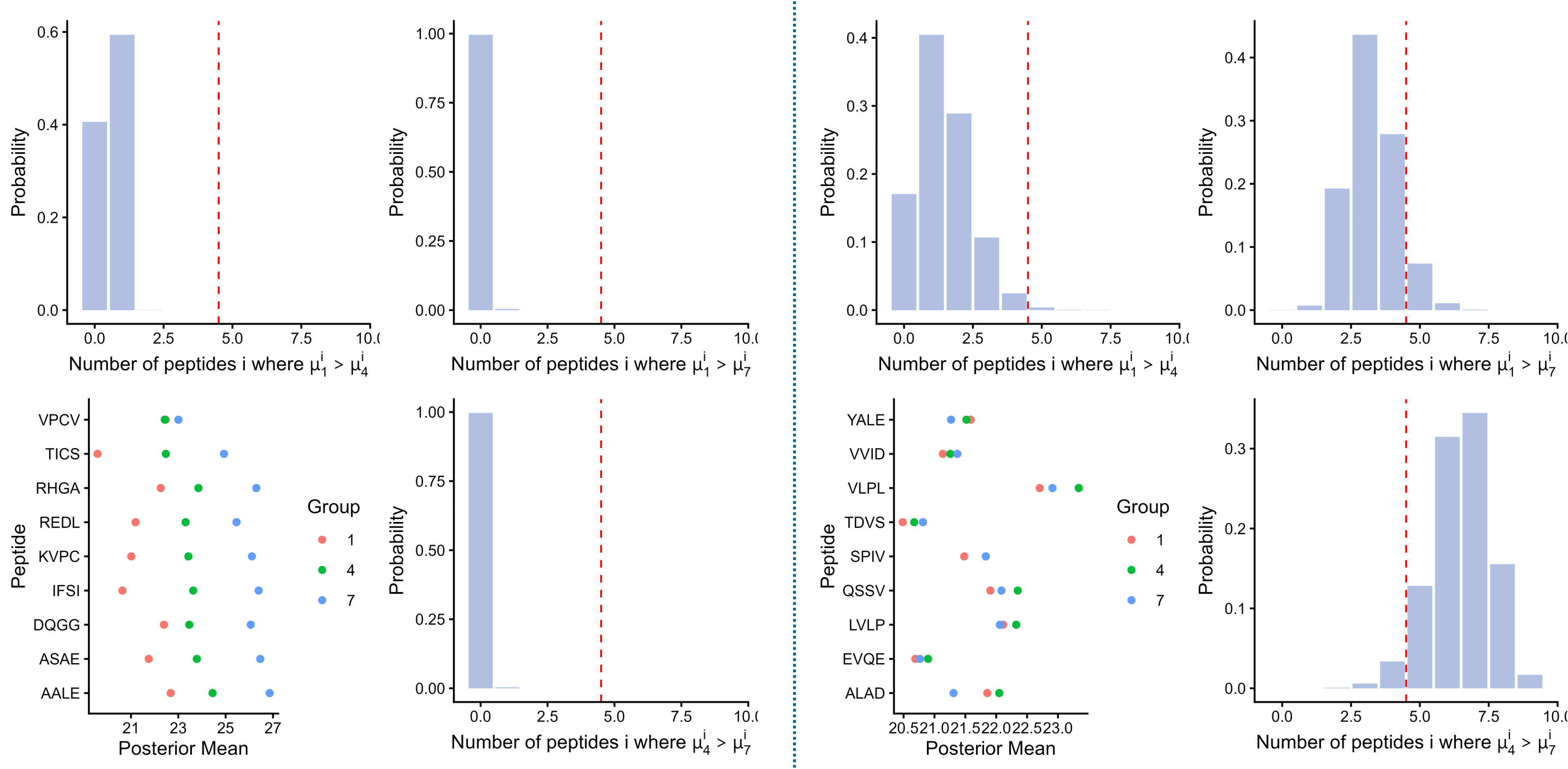}}
    \caption{Illustration of multivariate ProteoBayes inference on differential (left) and non-differential (right) proteins (resp. P12081ups\textbar SYHC\_HUMAN\_UPS and sp\textbar F4I893\textbar ILA\_ARATH), based on 9 peptides between three conditions (i.e Groups 1, 4 and 7).}
    \label{fig:multi_inference}
\end{figure}

As illustrated in \Cref{fig:multi_inference}, both panels depict differences computed between 3 distinct groups (denoted as group 1, 4, and 7), which are increasingly differential in the left panel and non-differential in the right panel. 
At the bottom left, the effect size of mean differences in peptide-wise marginals remains directly interpretable, even in high dimensions, and we can observe large effect sizes as expected for the differential dataset.
The distribution of $N^+$ allows practitioners to assess whether two groups appear fairly similar (i.e. a distribution close to the central red dashed line, corresponding to half of the number of peptides), or clearly distinct (i.e. a distribution concentrated on one side, indicating that most peptides are highly likely differential, in one direction).
In the most extreme case, when comparing groups 1 and 7, at the top right of the left panel, we are almost certain ($\mathbb{P}(N^+ = 0) \approx 1$) that the intensity of all peptides is higher in group 7 than in group 1, as indicated by the distribution concentrated on the value $0$.  

To support visual intuition, we provided in \Cref{tab:tab_multi} an empirical comparison of performance on synthetic datasets between the univariate and multivariate versions of the method.
These results, across various situations with different effect sizes and inter-peptide correlation structures, demonstrate that our approach correctly recovers the mean differences in all conditions.
One can observe that when accounting for correlations, errors remain consistently lower than in an univariate setting. 
In particular, when variance increases (on the diagonal of the covariance matrix), the sharp rise in errors observed with univariate ProteoBayes does not occur with the multivariate version, which leverages inter-peptide correlations to remain accurate. 
These errors remain low, even with only 5 samples per group, and decrease further as we increase the number of samples per group.
Regarding uncertainty quantification, we observe that the multivariate version of the method appears well-calibrated (i.e. $CI_{95}$ coverage close to the expected theoretical value of 95).
It is crucial for the credible intervals themselves to be multivariate, as we highlight in the \emph{univariate} $CIC_{95}$ column, where those computed from marginals are poorly calibrated and consistently overestimate genuine uncertainties. 
Overall, those results highlight both the benefits of our probabilistic approach in providing interpretable, well-calibrated inference tools and the robustness and accuracy achievable by adequately modelling the underlying correlations among peptides. 

\begin{table}
\centering
\makebox[\textwidth][c]{
\begin{tabular}{cc|c|cc|cc|}
\cline{3-7}
   &
   &
  \multicolumn{1}{c|}{} &
  \multicolumn{2}{c|}{\textbf{Multivariate}} &
  \multicolumn{2}{c|}{\textbf{Univariate}} \\

    \multicolumn{1}{l}{} &
    \multicolumn{1}{l|}{} &
  \textbf{Mean difference} &
  \textbf{RMSE} &
  $\mathbf{CIC_{95}}$ & \textbf{RMSE} &
  $\mathbf{CIC_{95}}$  \\ \hline
\multicolumn{1}{|c|}{} &
  \multicolumn{1}{|l|}{
$\mathcal{N}\left(
\begin{bmatrix}
1 \\
1 \\
1
\end{bmatrix}, 
\begin{bmatrix}
1 & 0 & 0\\
0 & 1 & 0 \\
0 & 0 & 1
\end{bmatrix}\right)$}             & 0.99 (0.44)              & 0.44 (0.45)   & 94.5 (22.8) & 0.64 (0.62)   & 100 (0)         \\
\multicolumn{1}{|c|}{\multirow{4}{*}{\begin{tabular}[c]{@{}c@{}}5\\ samples\end{tabular}}} & \multicolumn{1}{|l|}{
$\mathcal{N}\left(
\begin{bmatrix}
1 \\
1 \\
1
\end{bmatrix}, 
\begin{bmatrix}
1 & 0.7 & 0.2\\
0.7 & 1 & 0.5 \\
0.2 & 0.5 & 1
\end{bmatrix}\right)$}   & 1.01 (0.50)    & 0.44 (0.43)   & 93.3 (25.0) & 0.62 (0.58)   & 100 (0)        \\ 
\multicolumn{1}{|c|}{} & \multicolumn{1}{|l|}{
$\mathcal{N}\left(
\begin{bmatrix}
1 \\
1 \\
1
\end{bmatrix}, 
\begin{bmatrix}
10 & 0.7 & 0.2\\
0.7 & 10 & 0.5 \\
0.2 & 0.5 & 10
\end{bmatrix}\right)$} & 1.00 (0.96)              & 0.46 (0.47)   & 92.8 (25.8) & 1.46 (1.33)   & 99.1 (9.5)         \\ 
\multicolumn{1}{|l|}{} & \multicolumn{1}{|l|}{
$\mathcal{N}\left(
\begin{bmatrix}
10 \\
10 \\
10
\end{bmatrix}, 
\begin{bmatrix}
1 & 0.7 & 0.2\\
0.7 & 1 & 0.5 \\
0.2 & 0.5 & 1
\end{bmatrix}\right)$}      & 9.97  (0.50)             & 0.44 (0.45)   & 92.5 (26.4) & 0.60 (0.59)   & 100 (0)         \\ \hline
\multicolumn{1}{|c|}{} &
    \multicolumn{1}{|l|}{
$\mathcal{N}\left(
\begin{bmatrix}
1 \\
1 \\
1
\end{bmatrix}, 
\begin{bmatrix}
1 & 0 & 0\\
0 & 1 & 0 \\
0 & 0 & 1
\end{bmatrix}\right)$}             & 1.00 (0.11)              & 0.10 (0.13)   & 96.2 (19.1) & 0.14 (0.12)   & 100 (0)         \\
\multicolumn{1}{|c|}{\multirow{4}{*}{\begin{tabular}[c]{@{}c@{}}100\\ samples\end{tabular}}} & \multicolumn{1}{|l|}{
$\mathcal{N}\left(
\begin{bmatrix}
1 \\
1 \\
1
\end{bmatrix}, 
\begin{bmatrix}
1 & 0.7 & 0.2\\
0.7 & 1 & 0.5 \\
0.2 & 0.5 & 1
\end{bmatrix}\right)$}   & 1.00 (0.11)    & 0.10 (0.11)   & 94.3 (23.2) & 0.14 (0.13)   & 100 (0)         \\ 
\multicolumn{1}{|c|}{} & \multicolumn{1}{|l|}{
$\mathcal{N}\left(
\begin{bmatrix}
1 \\
1 \\
1
\end{bmatrix}, 
\begin{bmatrix}
10 & 0.7 & 0.2\\
0.7 & 10 & 0.5 \\
0.2 & 0.5 & 10
\end{bmatrix}\right)$} & 0.99 (0.23)    & 0.10 (0.11)   & 94.8 (22.2) & 0.34 (0.29)   & 100 (0)         \\
\multicolumn{1}{|l|}{} & \multicolumn{1}{|l|}{
$\mathcal{N}\left(
\begin{bmatrix}
10 \\
10 \\
10
\end{bmatrix}, 
\begin{bmatrix}
1 & 0.7 & 0.2\\
0.7 & 1 & 0.5 \\
0.2 & 0.5 & 1
\end{bmatrix}\right)$}      & 9.99  (0.13)             & 0.10 (0.10)   & 93.4 (24.9) & 0.14 (0.14)   & 100 (0)         \\ \hline
\end{tabular}}
\caption{Simulation study reporting empirical performances and uncertainty quantification metrics of the univariate and multivariate versions of ProteoBayes. The reported distributions are compared with the Gaussian baseline defined in \Cref{eq:ref_3d_gaussian}. All results are averaged over 100 repetitions of the experiments, and reported using the format \emph{Mean (Sd)}.}
\label{tab:tab_multi}
\end{table}

\section{Conclusion and perspectives}

This article presents a Bayesian inference framework for differential analysis, providing a fully probabilistic perspective that is often limited in traditional approaches based on moderated variance, such as limma. 
Furthermore, we leveraged and adapted well-established results from conjugate Bayesian inference to propose a coherent, computationally efficient strategy for tackling both univariate and multivariate contexts while accounting for missing data.
In particular, multivariate differential analysis is rarely considered in the literature. 
However, we argue that quantitative proteomics constitutes a natural illustration in which correlations across multiple elements (e.g., inter-peptide correlations within the same protein) yield more accurate and robust inference of differential status between experimental conditions.
We also explored, both theoretically and empirically, the recurring question of missing data in proteomics datasets, highlighting the problems caused by systematic imputation. 
We showed that, in the univariate setting with independent peptides and under MAR assumption, missing values can be integrated out analytically, so that imputation is unnecessary and may even degrade uncertainty quantification.
Through various illustrations and simulation studies, we proposed a probabilistic inference framework that we expect will be more interpretable for practitioners by focusing on notions of effect size and uncertainty quantification rather than traditional null hypothesis testing.
The primary interest of this framework, in contrast with other tools in the Bayesian toolbox (which is growing in many applied fields), is its remarkable computational efficiency, as closed-form posteriors and further sampling keep running times comparable to those of frequentist tests. 
Therefore, practitioners can still perform hundreds of thousands of differential analyses (even in multivariate settings if covariance structures include less than $\approx 10^3$ peptides at a time) in a couple of seconds and still benefit from intuitive probabilistic insights to determine, not only whether two conditions are differential or not, but more importantly, \emph{how much} they differ and \emph{how certain} are we? 
With an appropriate decision rule and a suitable correlation structure, Bayesian inference can also be used in large-scale proteomics experiments, such as label-free global quantification strategies. 
Furthermore, such experiments used in biomarker research could greatly benefit from quantifying uncertainty and assessing effect sizes.

While we believe this Bayesian framework provides a new perspective on differential analysis and its practical implementation, we should also mention that the current model has intrinsic limitations. 
The quick computations come at the cost of limited flexibility in the model hypotheses, to preserve conjugacy and closed-form equations. 
For instance, the current formulation assumes a Gaussian likelihood and is not well-suited to count data, which is common in other omics measurements. 
Nonetheless, several approximate modern strategies, such as Laplace Matching \citep{hobbhahn2021laplace} or Variational Inference \citep{blei2017variational}, can be used to perform efficient inference for latent structures with non-Gaussian likelihoods. 
Another limitation could come from the difficulty in estimating high-dimensional covariance structures from a limited number of samples (generally a handful in omics studies). 
On this matter, a possible avenue is to leverage covariance kernels \citep{duvenaud2014automatic}, which are widely used in machine learning nowadays to learn expressive correlation structures from a limited number of hyperparameters and to share information across multiple data sources, thereby enhancing the robustness of estimation.
Finally, while we proposed novel inference strategies that we believe are sound for comparing multivariate distributions in high dimensions, we concede that these proposals could probably be improved, as the curse of dimensionality is a pervasive problem across many fields of statistics. 
Many researchers have proposed sensible methods to mitigate this issue, and future adaptations to our framework could help maintain intuitive, interpretable results, even in this new high-dimensional differential analysis paradigm.

\section*{Supporting information}

\textbf{S1 Appendix.} Supporting information including the following sections: (A) Notations and abbreviations, (B) Proofs, (C) Sensitivity analysis for hyperparameters, (D) Additional real-data experiments results and (E) Simulation experiment on MNAR data.

\section*{Code availability}
The work described in the present article was implemented as an R package called \emph{ProteoBayes}, available on CRAN, while a development version can be found on GitHub (\url{https://github.com/mariechion/ProteoBayes}). A companion web app can also be accessed at \url{https://arthurleroy.shinyapps.io/ProteoBayes/}.

\section*{Data availability}
All datasets and the code for reproducing experiments are available on GitHub (\url{https://github.com/mariechion/ProteoBayes-paper}). All real datasets are also publicly accessible on the ProteomeXchange website using the following identifiers: PXD003841, PXD009815, PXD016647, and PXD027800.

\bibliography{ProteoBayes}

\newpage
\section{Notations and abbreviations}
\begin{longtable}{>{\raggedright\arraybackslash}p{0.25\textwidth} >{\raggedright\arraybackslash}p{0.70\textwidth}}
\toprule
\textbf{Symbol} & \textbf{Meaning} \\
\midrule
\endfirsthead

\toprule
\textbf{Symbol} & \textbf{Meaning} \\
\midrule
\endhead

$y$ & Observation in the univariate model. \\

$y_n$ & $n$-th observed value in a univariate sample. \\

$\mathbf{y}_k$ or $y_k \in \mathbb{R}^P$ & Observation vector for group/condition $k$ in the multivariate model. \\

$y_{k,n}$ & $n$-th sample in group $k$. \\

$y_{k,n}^p$ & Value for peptide/feature $p$, sample $n$, group $k$. \\

$\mu$ & Mean parameter in the univariate model. \\

$\mu_k$ & Mean vector for group $k$. \\

$\mu_k^p$ & Mean of peptide/feature $p$ in group $k$. \\

$\mu_0$ & Prior mean hyperparameter. \\

$\lambda_0$ & Prior precision-like hyperparameter controlling confidence in $\mu_0$. \\

$\sigma^2$ & Variance in the univariate model. \\

$\Sigma_k$ & Covariance matrix for group $k$. \\

$\Sigma_0$ & Prior covariance-scale hyperparameter in the multivariate model. \\

$\alpha_0, \beta_0$ & Inverse-gamma prior hyperparameters for variance. \\

$\nu_0$ & Inverse-Wishart prior degrees-of-freedom hyperparameter. \\

$\varepsilon$ & Univariate error term. \\

$\varepsilon_k$ & Multivariate error term for group $k$. \\

$N$ & Number of observations in the univariate derivations. \\

$N_k$ & Number of samples in group $k$. \\

$N_k^p$ & Number of observed values for peptide $p$ in group $k$. \\

$K$ & Number of groups/conditions. \\

$P$ & Number of peptides/features jointly modelled. \\

$\bar{y}$ & Sample mean in the univariate model. \\

$\bar{y}_k^{(d)}$ & Mean vector of the $d$-th imputed dataset in group $k$. \\

$\bar{y}_{k}^{p,(0)}$ & Observed-data mean for peptide $p$ in group $k$. \\

$\mu_N, \lambda_N, \alpha_N, \beta_N$ & Updated posterior hyperparameters in the univariate conjugate model. \\

$\mu_{k,N}^{(d)}, \lambda_{k,N}, \Sigma_{k,N}^{(d)}, \nu_{k,N}$ & Updated posterior hyperparameters for group $k$ in the multivariate algorithm. \\

$\hat{\sigma}^2$ & Scale parameter of the marginal posterior Student distribution for $\mu$. \\

$\hat{\sigma}_{k}^{p\,2}$ & Univariate posterior scale for peptide $p$ in group $k$. \\

$T_{\nu}(\cdot)$ & Student $t$ distribution with $\nu$ degrees of freedom. \\

$W^{-1}(\Sigma_0, \nu_0)$ & Inverse-Wishart prior for covariance matrices. \\

$\Gamma^{-1}(\alpha_0, \beta_0)$ & Inverse-gamma prior for variance. \\

$H$ & Set of observed data. \\

$y_k^{(0)}$ & Observed entries in group $k$. \\

$y_k^{(1)}$ & Missing entries in group $k$. \\

$\tilde{y}_k^{(1),d}$ & $d$-th imputed version of the missing data in group $k$. \\

$\tilde{y}_{k,n}^{(d)}$ & Complete vector for sample $n$, group $k$, after $d$-th imputation. \\

$D$ & Number of imputed datasets / imputation draws. \\

$R$ & Number of posterior samples drawn for inference. \\

$d_r$ & Random imputation index drawn in Algorithm 1. \\

$\hat{\mu}_k^{[r]}$ & $r$-th posterior draw of the mean for group $k$. \\

$\hat{\mu}_{\Delta}^{[r]}$ & $r$-th posterior draw of the difference between two groups. \\

$p(\mu_k - \mu_{k'} \mid y_k, y_{k'})$ & Posterior distribution of the difference of means between groups $k$ and $k'$. \\

$\mu_{\mathrm{diff}}$ & Estimated mean difference, defined as $\hat{\mu}_1 - \hat{\mu}_2$. \\

$\mathrm{CI_{95}}$ & 95\% credible interval. \\

$\mathrm{CI_{95}\ width}$ & Width of the 95\% credible interval. \\

$\mathrm{RMSE}$ & Root mean square error. \\

$\mathrm{CIC_{95}}$ & 95\% credible interval coverage. \\

$\mu_{p,k}^{\mathrm{true}}$ & Reconstructed reference mean for peptide $p$ in group $k$. \\

$\Sigma_{\mathrm{ref}}$ & Baseline covariance matrix used in multivariate simulations. \\

$N^{+}$ & Number of peptides with larger values in one group than the other in multivariate inference. \\

$N^{-}$ & Number of peptides with lower values in one group than the other. \\

MAR & Missing At Random. \\

MNAR & Missing Not At Random. \\

\bottomrule
\caption{Notation and abbreviation table.}
\label{tab:notation}
\end{longtable}

\section{Proofs}
\label{sec:proof}
\subsection{Proof of Bayesian inference for Normal-Inverse-Gamma conjugated priors}
\label{sec:model_uni:proof}
Let us recall below the complete development of this derivation by identification of the analytical form (we ignore conditioning over the hyperparameters for convenience):
\begin{align*}
\displaystyle
	p(\mu, \sigma^2 \mid \yb) 
	& \propto p(\yb \mid \mu, \sigma^2) \times p(\mu, \sigma^2) \\
	&= \left( \dfrac{1}{2 \pi \sigma^2} \right)^{\frac{N}{2}} \exp \left(- \dfrac{1}{2 \sigma^2} \sum\limits_{n = 1}^{N}(y_n - \mu)^2 \right) \\
	& \hspace{0.5cm} \times \frac{\sqrt{\lambda_0}}{\sqrt{2 \pi}} \frac{\beta_0^{\alpha_0}}{\Gamma(\alpha_0)}\left(\frac{1}{\sigma^{2}}\right)^{\alpha_0 + \frac{3}{2}} \exp \left(-\frac{2 \beta_0 +\lambda_0(\mu -\mu_0)^{2}}{2 \sigma^{2}}\right) \\
	& \propto \left(\frac{1}{\sigma^{2}}\right)^{\alpha_0 + \frac{N + 3}{2}} \exp \left(\underbrace{-\frac{2 \beta_0 +\lambda_0(\mu -\mu_0)^{2} + \sum\limits_{n = 1}^{N}(y_n - \mu)^2}{2 \sigma^{2}}}_{\mathcal{A}}\right).
\end{align*}

Let us introduce \Cref{lem:ch5} below to decompose the term $\mathcal{A}$ as desired:
\begin{lemma}
\label{lem:ch5}
Assume a set $\boldsymbol{x_{\textcolor{colN}{1}}}, \dots, \boldsymbol{x_{\sampleN}} \in \mathbb{R}^{q}$, and note $\bar{\boldsymbol{x}} = \dfrac{1}{\sampleN} \sum\limits_{\samplen = 1}^{\sampleN}{\boldsymbol{x_{\samplen}}}$ the associated average vector.
For any $\boldsymbol{\mu} \in \mathbb{R}^{q}$: 
\begin{align*}
\displaystyle
	 \sum\limits_{\samplen = 1}^{\sampleN}{(\boldsymbol{x_{\samplen}} - \boldsymbol{\mu})(\boldsymbol{x_{\samplen}} - \boldsymbol{\mu})^{\intercal}}
	 &=  \sampleN (\bar{\boldsymbol{x}} - \boldsymbol{\mu})(\bar{\boldsymbol{x}} - \boldsymbol{\mu})^{\intercal} + \sum\limits_{\samplen = 1}^{\sampleN}{(\boldsymbol{x_{\samplen}} - \bar{\boldsymbol{x}})(\boldsymbol{x_{\samplen}} - \bar{\boldsymbol{x}})^{\intercal}}.
\end{align*}
\end{lemma}

\begin{proof}
	\begin{align*}
	\displaystyle
	 \sum\limits_{\samplen = 1}^{{\sampleN}}{(\boldsymbol{x_{\samplen}} - \boldsymbol{\mu})(\boldsymbol{x_{\samplen}} - \boldsymbol{\mu})^{\intercal}}
	 &= \sum\limits_{\samplen = 1}^{{\sampleN}}{\boldsymbol{x_{\samplen}}\boldsymbol{x_{\samplen}}^{\intercal} + \boldsymbol{\mu}\boldsymbol{\mu}^{\intercal} - 2 \boldsymbol{x_{\samplen}} \boldsymbol{\mu}^{\intercal}} \\
	 &= {\sampleN}\boldsymbol{\mu}\boldsymbol{\mu}^{\intercal} - 2 {\sampleN} \bar{\boldsymbol{x}}\boldsymbol{\mu}^{\intercal} + \sum\limits_{\samplen = 1}^{{\sampleN}}{\boldsymbol{x_{\samplen}}\boldsymbol{x_{\samplen}}^{\intercal}} \\
	 &= {\sampleN}\boldsymbol{\mu}\boldsymbol{\mu}^{\intercal} + {\sampleN}\bar{\boldsymbol{x}}\bar{\boldsymbol{x}}^{\intercal} + {\sampleN}\bar{\boldsymbol{x}}\bar{\boldsymbol{x}}^{\intercal} - 2 {\sampleN} \bar{\boldsymbol{x}}\bar{\boldsymbol{x}}^{\intercal} - 2 {\sampleN} \bar{\boldsymbol{x}} \boldsymbol{\mu}^{\intercal} + \sum\limits_{\samplen = 1}^{{\sampleN}}{\boldsymbol{x_{\samplen}}\boldsymbol{x_{\samplen}}^{\intercal}} \\
	 &= {\sampleN} \left(\bar{\boldsymbol{x}}\bar{\boldsymbol{x}}^{\intercal} - \boldsymbol{\mu}\boldsymbol{\mu}^{\intercal} - 2 \bar{\boldsymbol{x}} \boldsymbol{\mu}^{\intercal} \right)  + \sum\limits_{\samplen = 1}^{{\sampleN}}{\boldsymbol{x_{\samplen}}\boldsymbol{x_{\samplen}}^{\intercal} + \bar{\boldsymbol{x}}\bar{\boldsymbol{x}}^{\intercal} - 2 \boldsymbol{x_{\samplen}} \bar{\boldsymbol{x}}^{\intercal}} \\
	 &= {\sampleN} \left(\bar{\boldsymbol{x}} - \boldsymbol{\mu} \right)\left(\bar{\boldsymbol{x}} - \boldsymbol{\mu} \right)^{\intercal}  + \sum\limits_{\samplen = 1}^{{\sampleN}}{(\boldsymbol{x_{\samplen}} - \bar{\boldsymbol{x}})(\boldsymbol{x_{\samplen}} - \bar{\boldsymbol{x}})^{\intercal}}.
\end{align*}
\end{proof}

Applying this result in our context for $q=1$, we obtain:
\begin{align*}
\displaystyle
	\mathcal{A} 
	&= -\frac{1}{2 \sigma^{2}} \left( 2 \beta_0 +\lambda_0(\mu -\mu_0)^{2} + {\sampleN}(\bar{y} - \mu)^2 + \sum\limits_{\samplen = 1}^{{\sampleN}}(y_{\samplen} - \bar{y})^2 \right) \\
	&= -\frac{1}{2 \sigma^{2}} \left( 2 \beta_0 + \sum\limits_{\samplen = 1}^{{\sampleN}}(y_{\samplen} - \bar{y})^2 + (\lambda_0 + {\sampleN}) \mu^2 - 2 \mu ({\sampleN} \bar{y} + \lambda_0 \mu_0) + {\sampleN} \bar{y}^2 + \lambda_0 \mu_0^2 \right) \\
	&= -\frac{1}{2 \sigma^{2}} \Bigg( 2 \beta_0 + \sum\limits_{\samplen = 1}^{{\sampleN}}(y_{\samplen} - \bar{y})^2  +  {\sampleN} \bar{y}^2 + \lambda_0 \mu_0^2 \\ 
	& \hspace{0.5cm} + (\lambda_0 + {\sampleN}) \left[ \mu^2 - 2 \mu \dfrac{{\sampleN} \bar{y} + \lambda_0 \mu_0}{\lambda_0 + {\sampleN}} + \left( \dfrac{{\sampleN} \bar{y} + \lambda_0 \mu_0}{\lambda_0 + {\sampleN}} \right)^2 - \left( \dfrac{{\sampleN} \bar{y} + \lambda_0 \mu_0}{\lambda_0 + {\sampleN}} \right)^2  \right] \Bigg) \\
	&= -\frac{1}{2 \sigma^{2}} \Bigg( 2 \beta_0 + \sum\limits_{\samplen = 1}^{{\sampleN}}(y_{\samplen} - \bar{y})^2  + {\sampleN} \bar{y}^2 + \lambda_0 \mu_0^2  - \dfrac{({\sampleN} \bar{y} + \lambda_0 \mu_0)^2}{\lambda_0 + {\sampleN}} \\
	&\hspace{0.5cm} + (\lambda_0 + {\sampleN}) \left( \mu - \dfrac{{\sampleN} \bar{y} + \lambda_0 \mu_0}{\lambda_0 + {\sampleN}}\right)^2 \Bigg) \\ 
	&= -\frac{1}{2 \sigma^{2}} \Bigg( 2 \beta_0 + \sum\limits_{\samplen = 1}^{{\sampleN}}(y_{\samplen} - \bar{y})^2 + \dfrac{(\lambda_0 + {\sampleN}) ({\sampleN} \bar{y}^2 + \lambda_0 \mu_0^2) - {\sampleN}^2 \bar{y}^2 - \lambda_0^2 \mu_0^2 + 2 {\sampleN} \bar{y} \lambda_0 \mu_0}{\lambda_0 + {\sampleN}} \\
	&\hspace{0.5cm} + (\lambda_0 + {\sampleN}) \left( \mu - \dfrac{{\sampleN} \bar{y} + \lambda_0 \mu_0}{\lambda_0 + {\sampleN}}\right)^2 \Bigg) \\ 
	&= -\frac{1}{2 \sigma^{2}} \Bigg( 2 \beta_0 + \sum\limits_{\samplen = 1}^{{\sampleN}}(y_{\samplen} - \bar{y})^2 + \dfrac{\lambda_0 {\sampleN}}{\lambda_0 + {\sampleN}} (\bar{y} - \mu_0)^2 + (\lambda_0 + {\sampleN}) \left( \mu - \dfrac{{\sampleN} \bar{y} + \lambda_0 \mu_0}{\lambda_0 + {\sampleN}}\right)^2 \Bigg).
\end{align*}

\subsection{Proof of General Bayesian framework for evaluating mean differences}
\label{sec:model_multi:proof}
\begin{proof}
	For the sake of clarity, let us omit the $\groupK$ groups here and first consider a general case with $\yb_{\groupk} = \yb \in \mathbb{R}^{\peptP}$. Moreover, let us focus on only one imputed dataset and maintain the notation $\tilde{\yb}_1^{(\drawd)}, \dots, \tilde{\yb}_{\sampleN}^{(\drawd)} = \yb_1, \dots, \yb_{\sampleN}$ for convenience. From the hypotheses of the model, we can derive $\mathcal{L}$, the posterior $\log$-PDF over $\left(\mub, \Sigmab \right)$, following the same idea as for the univariate case presented in \Cref{sec:model_uni}:
	
    \begin{align*}
    \displaystyle
    \mathcal{L}
    &= \log p(\mub, \Sigmab \mid \yb_1, \dots, \yb_{\sampleN}) \\
    &= \log \underbrace{p(\yb_1, \dots, \yb_{\sampleN} \mid \mub, \Sigmab)}_{\mathcal{N}(\mub,  \Sigmab)} + \log \underbrace{p(\mub, \Sigmab)}_{\mathcal{NW}^{-1}(\mub_0, \lambda_0,  \Sigmab_0, \nu_0)} + C_1 \\
    &= - \frac{{\sampleN}}{2} \log \vert \Sigmab \vert - \frac{1}{2} \left(\sum_{\samplen = 1}^{{\sampleN}} (\yb_{\samplen} - \mub)^{\intercal} \Sigmab^{-1} (\yb_{\samplen} - \mub)  \right)\\
    & \hspace{0.5cm}  - \frac{\nu_0 + \peptP + 2}{2} \log \vert \Sigmab \vert - \frac{1}{2} \left( \tr{\Sigmab_0 \Sigmab^{-1}} - \dfrac{\lambda_0}{2} (\mub - \mub_0)^{\intercal} \Sigmab^{-1}(\mub - \mub_0) \right) + C_2 \\
    &= -\frac{1}{2} \Bigg[ \left( \nu_0 + \peptP + 2 + \sampleN \right)\log \vert \Sigmab \vert + \tr{\Sigmab_0 \Sigmab^{-1}}  \\
    & \hspace{0.5cm} +  \sum_{\samplen = 1}^{\sampleN} \tr{(\yb_{\samplen} - \mub)^\mathrm{T} \Sigmab^{-1} (\yb_{\samplen} - \mub) } + \tr{\lambda_0 (\mub - \mub_0)^{\intercal}\Sigmab^{-1}(\mub - \mub_0)} \Bigg] + C_2 \\
    &= -\frac{1}{2} \Bigg[ \left( \nu_0 + \peptP + 2 + N \right) \log \vert \Sigmab \vert + \textrm{tr} \Bigg( \Sigmab^{-1} \Big\{ \Sigmab_0 + \lambda_0 (\mub - \mub_0)(\mub - \mub_0)^{\intercal}   \\
    & \hspace{0.5cm} + \underbrace{ \sampleN (\bar{\yb} - \mub)(\bar{\yb} - \mub)^{\intercal} +  \sum_{\samplen = 1}^{\sampleN} (\yb_{\samplen} - \bar{\yb})(\yb_{\samplen} - \bar{\yb})^{\intercal}}_{\text{Lemma 1}} \Big\} \Bigg) \Bigg] + C_2 \\
    &= -\frac{1}{2} \Bigg[ \left( \nu_0 + P + 2 + \sampleN \right) \log \vert \Sigmab \vert + \textrm{tr} \Bigg( \Sigmab^{-1} \Big\{ \Sigmab_0 +  \sum_{\samplen = 1}^{\sampleN} (\yb_{\samplen} - \bar{\yb})(\yb_{\samplen} - \bar{\yb})^{\intercal}  \\
    & \hspace{0.5cm} + (\sampleN + \lambda_0) \mub \mub^{\intercal} - \mub \left( \sampleN \bar{\yb}^{\intercal} + \lambda_0 \mub_0^{\intercal} \right)  - (\lambda_0 \mub_0 + \sampleN \bar{\yb}) \mub^{\intercal} + \lambda_0 \mub_0\mub_0^{\intercal} + \sampleN \bar{\yb}\bar{\yb}^{\intercal} \Big\} \Bigg) \Bigg] + C_2 \\
    &= -\frac{1}{2} \Bigg[ \left( \nu_0 + \peptP + 2 + \sampleN \right) \log \vert \Sigmab \vert  \\
    & \hspace{0.5cm} + \textrm{tr} \Bigg( \Sigmab^{-1} \Big\{ \Sigmab_0 +  \sum_{\samplen = 1}^{\sampleN} (\yb_{\samplen} - \bar{\yb})(\yb_{\samplen} - \bar{\yb})^{\intercal}  +  \dfrac{\sampleN \lambda_0}{\sampleN + \lambda_0} (\bar{\yb} - \mub_0)(\bar{\yb} - \mub_0)^{\intercal} \\
    & \hspace{0.5cm} + \left( \sampleN + \lambda_0 \right) \left( \mub - \dfrac{\sampleN \bar{\yb} + \lambda_0 \mub_0}{\sampleN + \lambda_0} \right)\left( \mub - \dfrac{N \bar{\yb} + \lambda_0 \mub_0}{\sampleN + \lambda_0} \right)^{\intercal} \Big\} \Bigg) \Bigg] + C_2 \\
    &= -\frac{1}{2} \Bigg[ \left( \nu_{\sampleN} + \peptP + 2 \right) \log \vert \Sigmab \vert + \tr{ \Sigmab^{-1} \Sigmab_{\sampleN}} + \lambda_{\sampleN} \left( \mub - \mub_{\sampleN}\right)^{\intercal} \Sigmab^{-1} \left( \mub - \mub_{\sampleN}\right) \Bigg] + C_2.
    \end{align*}

    By identification, we recognise the log-PDF that characterises the Gaussian-inverse-Wishart distribution $\mathcal{NIW}^{-1}(\mub_{\sampleN},\lambda_{\sampleN},\Sigmab_{\sampleN}, \nu_{\sampleN})$ with:
    \begin{itemize}
    	\item $\mub_{\sampleN} = \dfrac{\sampleN \bar{\yb} + \lambda_0 \mub_0}{\sampleN + \lambda_0}$,
    	\item $\lambda_{\sampleN} = \lambda_0 + \sampleN$,
    	\item $\Sigmab_{\sampleN} = \Sigmab_0 + \sum\limits_{\samplen = 1}^{\sampleN}(\yb_{\sampleN} - \bar{\yb})(\yb_{\sampleN} - \bar{\yb})^{\intercal} + \dfrac{\lambda_0 \sampleN}{(\lambda_0 + \sampleN)} (\bar{\yb} - \mub_0)(\bar{\yb} - \mub_0)^{\intercal} $,
    	\item $\nu_{\sampleN} = \nu_0 + \sampleN$.
    \end{itemize}

    Once more, we can integrate over $\Sigmab$ to compute the mean's marginal posterior distribution by identifying the PDF of the inverse-Wishart distribution $\mathcal{W}^{-1}\Big( \Sigmab_{\sampleN} +  \lambda_{\sampleN} \left( \mub - \mub_{\sampleN} \right) \left( \mub - \mub_{\sampleN} \right)^{\intercal}, \\ \nu_{\sampleN} + 1 \Big)$ and by reorganising the terms:
    \begin{align*}
    \displaystyle
    	p(\mub \mid \yb) 
    	&= \int p(\mub, \Sigmab \mid \yb) \dif \Sigmab \\
    	&= \frac{\lambda_{\sampleN}^{\frac{\peptP}{2}}|\boldsymbol{\Sigma_{\sampleN}}|^{\frac{\nu_{\sampleN}}{2}}}{(2 \pi)^{\frac{\peptP}{2}} 2^{\frac{\peptP \nu_{\sampleN}}{2}} \Gamma_{\peptP}\left(\frac{\nu_{\sampleN}}{2}\right)}  \\ 
    	& \hspace{0.5cm} \times \int \vert \boldsymbol{\Sigma} \vert^{-\frac{\nu_{\sampleN}+\peptP+2}{2}} \exp \left(- \dfrac{1}{2} \left( \tr{\Sigmab_{\sampleN} \Sigmab^{-1}} - \dfrac{\lambda_{\sampleN} }{2} \left( \mub - \mub_{\sampleN}\right)^{\intercal} \Sigmab^{-1} \left( \mub - \mub_{\sampleN}\right) \right) \right) \dif \Sigmab \\ 
    	&= \frac{\lambda_{\sampleN}^{\frac{\peptP}{2}}|\boldsymbol{\Sigma_{\sampleN}}|^{\frac{\nu_{\sampleN}}{2}}}{(2 \pi)^{\frac{\peptP}{2}} 2^{\frac{\peptP \nu_{\sampleN}}{2}} \Gamma_{\peptP}\left(\frac{\nu_{\sampleN} }{2}\right)} \times \frac{2^{ \frac{\peptP (\nu_{\sampleN} + 1)}{2}} \Gamma_{\peptP}\left(\frac{\nu_{\sampleN} + 1}{2}\right)}{ \vert \Sigmab_{\sampleN} +  \lambda_{\sampleN} \left( \mub - \mub_{\sampleN} \right) \left( \mub - \mub_{\sampleN} \right)^{\intercal} \vert^{\frac{\nu_{\sampleN} + 1}{2}}} \times 1\\ 
    	&= \dfrac{ \pi^{\peptp(\peptp-1) / 4} \prod\limits_{\peptp = 0}^{\peptP-1} \Gamma \left(\frac{\nu_{\sampleN} + 1 - \peptp}{2}\right)}{ \pi^{\peptP(\peptP-1) / 4} \prod\limits_{\peptp = 1}^{\peptP} \Gamma \left(\frac{\nu_{\sampleN} + 1 - \peptp}{2}\right)} \times \dfrac{\lambda_{\sampleN}^{\frac{P}{2}}  }{\pi^{\frac{P}{2}} } \\
    	& \hspace{0.5cm} \times \underbrace{\dfrac{ \vert\Sigmab_{\sampleN} \vert^{\frac{\nu_{\sampleN}}{2}}}{ \vert \Sigmab_{\sampleN} \vert^{\frac{\nu_{\sampleN} + 1}{2}}} \times \left( 1 + \lambda_{\sampleN} \left( \mub - \mub_{\sampleN} \right)^{\intercal} \Sigmab_{\sampleN}^{-1} \left( \mub - \mub_{\sampleN} \right) \right)^{- \frac{\nu_{\sampleN} + 1}{2}} }_{\text{Matrix determinant lemma}} \\ 
    	&= \dfrac{ \Gamma \left(\frac{\nu_{\sampleN} + 1}{2}\right)}{\Gamma \left(\frac{\nu_{\sampleN} + 1 - \peptP}{2}\right)} \times \dfrac{ \left[\lambda_{\sampleN} (\nu_{\sampleN} - \peptP + 1)\right]^{\frac{P}{2}}  }{\left[\pi(\nu_{\sampleN} - \peptP + 1)\right]^{\frac{P}{2}} \vert \Sigmab_{\sampleN} \vert^{\frac{1}{2}} } \\
    	&\hspace{0.5cm} \times \left( 1 + \dfrac{\lambda_{\sampleN} (\nu_{\sampleN} - \peptP + 1)}{(\nu_{\sampleN} - \peptP + 1)} \left( \mub - \mub_{\sampleN} \right)^{\intercal} \Sigmab_{\sampleN}^{-1} \left( \mub - \mub_{\sampleN} \right) \right)^{- \frac{\nu_{\sampleN} + 1}{2}} \\ 
    	&= \dfrac{\Gamma \left(\frac{(\nu_{\sampleN} - \peptP + 1) + \peptP}{2}\right)}{\Gamma\left(\frac{\nu_{\sampleN} - \peptP + 1}{2}\right) \left[\pi(\nu_{\sampleN} - \peptP + 1)\right]^{\frac{P}{2}} \vert \dfrac{\Sigmab_{\sampleN} }{\lambda_{\sampleN} (\nu_{\sampleN} - \peptP + 1)} \vert^{\frac{1}{2}}} \\ 
    	& \hspace{0.5cm}  \times \left( 1 + \dfrac{1}{\nu_{\sampleN} - \peptP + 1} \left( \mub - \mub_{\sampleN} \right)^{\intercal} \left( \dfrac{ \Sigmab_{\sampleN}}{\lambda_{\sampleN} (\nu_{\sampleN} - \peptP + 1)} \right)^{-1} \left( \mub - \mub_{\sampleN} \right) \right)^{- \frac{(\nu_{\sampleN} - \peptP + 1) + \peptP}{2}}.
    \end{align*}
    The above expression corresponds to the PDF of a multivariate $t$-distribution $\mathcal{T}_{\nu}\left(\mub_{\sampleN}, \hat{\Sigmab} \right)$, with:
    \begin{itemize}
    	\item $\nu = \nu_{\sampleN}  - \peptP + 1$,
    	\item $\hat{\Sigmab} = \dfrac{\Sigmab_{\sampleN}}{ \lambda_{\sampleN} (\nu_{\sampleN} - \peptP + 1) }$.
    \end{itemize}
    
    Therefore, we demonstrated that for each group and imputed dataset, the complete-data posterior distribution over $\mub_{\groupk}$ is a multivariate $t$-distribution. 
    Thus, following Rubin's rules for multiple imputation \citep{littleStatisticalAnalysisMissing2019} for a small number of imputation draws $\drawD$, we can propose an approximation to the true posterior distribution (that is only conditioned over observed values):
    \begin{align*}
    	p\left( \mub_{\groupk} \mid \yb_{\groupk}^{(0)} \right) 
    	&= \int p\left( \mub_{\groupk} \mid \yb_{\groupk}^{(0)}, \yb_{\groupk}^{(1)}\right) p\left( \yb_{\groupk}^{(1)} \mid \yb_{\groupk}^{(0)} \right) \dif \yb_{\groupk}^{(1)}  \\
    	& \simeq \dfrac{1}{\drawD} \sum\limits_{\peptp = 1}^{\drawD}  p \left( \mub_{\groupk} \mid \yb_{\groupk}^{(0)}, \tilde{\yb}_{\groupk}^{(1), \drawd} \right)
    \end{align*}
    Leading to the desired results when evaluating the previously derived posterior distribution on each multiple-imputed dataset.
\end{proof} 

\newpage

\section{Sensitivity analysis for hyperparameters}

In this section, we aim to explore a wide range of prior values for each free
hyperparameter (thus excluding prior means, which are always set equal across groups, leading to invariant effects on posterior differences). 
The tested values in the following experiments are deliberately more extreme than those typically assumed in proteomics studies, to evaluate robustness to strong misspecifications.
Sensitivity is measured using two metrics, Negative Log-Likelihood (NLL) and Credible Interval Coverage ($CIC_{95}$), to quantify both sharpness and uncertainty calibration of the posterior distributions computed with different prior values on the same synthetic datasets. 
For NLL, the lower the value, the better, whereas $CIC_{95}$ values should be as close as possible to the theoretical threshold $95\%$.

In all experiments, we generated 1000 synthetic datasets using our simulation scheme with a unique group/condition. 
Unless otherwise stated, 5 samples are observed for each peptide.
The posterior distributions are derived analytically for all the synthetic datasets, using default hyperparameter values, except for the tested one, whose range of values is explicitly stated in the corresponding figures. 
The true mean of each peptide is used to compute NLL and $CIC_{95}$ metrics, and results are averaged over the 1000 synthetic datasets.

\subsection{Sensitivity of $\alpha_0$, $\beta_0$, and $\lambda_0$}

\begin{figure}[ht]
 	\makebox[\textwidth][c]{
    \includegraphics[width = \textwidth]{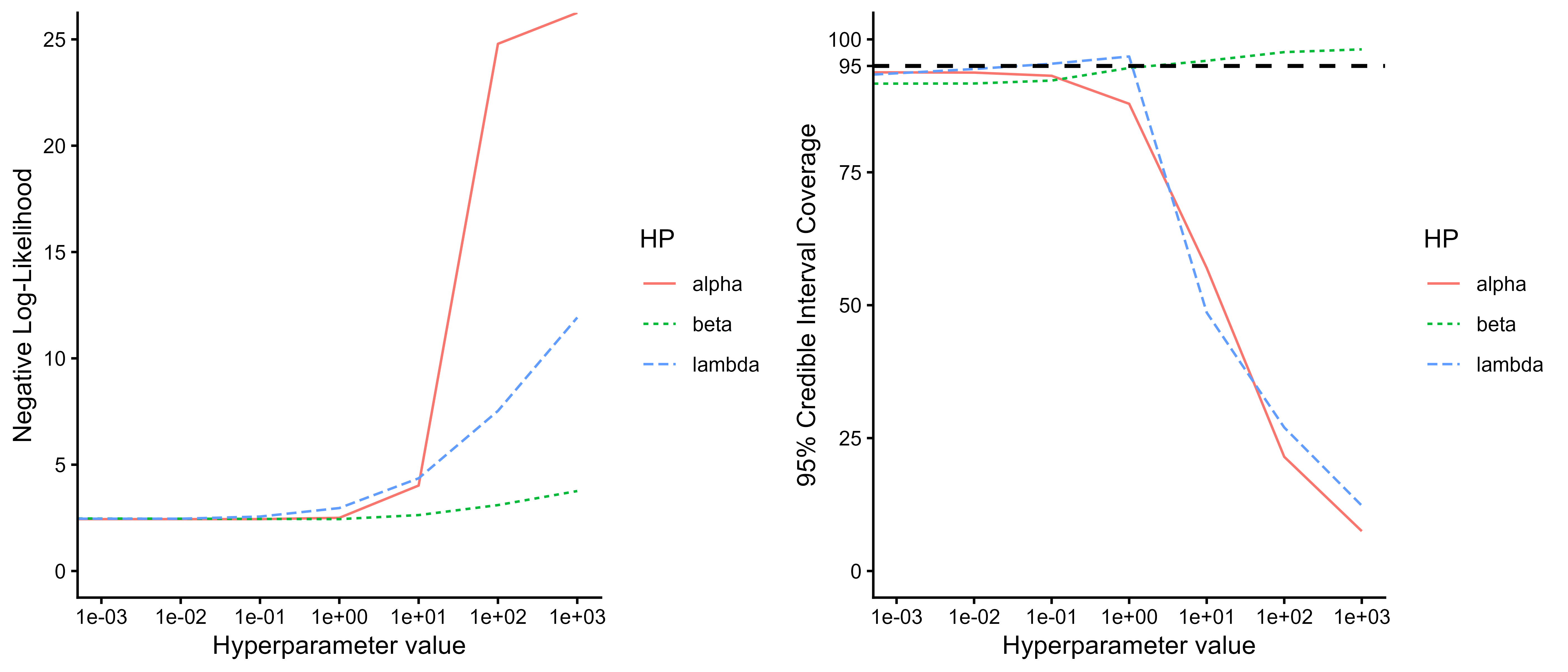}}
    \caption{Sensitivity analysis for the hyperparameters $\alpha_0$, $\beta_0$, and $\lambda_0$ in a univariate ProteoBayes setting. For each curve, one hyperparameter varies along the x-axis while the others are fixed to default values. \textbf{Left:} Negative Log-Likelihood of the posterior distribution evaluated on the true mean value. \textbf{Right}: $95\%$ Credible Interval Coverage for the true mean value. All results are averaged over 1000 runs of the simulation scheme.}
    \label{fig:sensitivity_uni}
\end{figure}

The sensitivity of hyperparameter priors in univariate ProteoBayes is reported in \Cref{fig:sensitivity_uni}. 
We can observe that both $\alpha_0$ and  $\lambda_0$ seem to deteriorate posteriors when set to excessively high values.
This result is expected, as $\lambda_0$ controls the confidence one places in the prior mean, which can dramatically bias the posterior mean if $\mu_0$ is far from the data. 
Similarly, $\alpha_0$ is often thought of as the prior degrees of freedom, and intuitively acts as the number of \emph{pseudo-observations} that may dominate data if set too high. 
Overall, we can see that all hyperparameters seem robust when maintained at reasonable values for such models, typically below 1, roughly meaning that the prior should not influence the posterior more than one data point would.  

\subsection{Sensitivity of $\Sigma_0$}

\begin{figure}[h!]
 	\makebox[\textwidth][c]{
    \includegraphics[width = \textwidth]{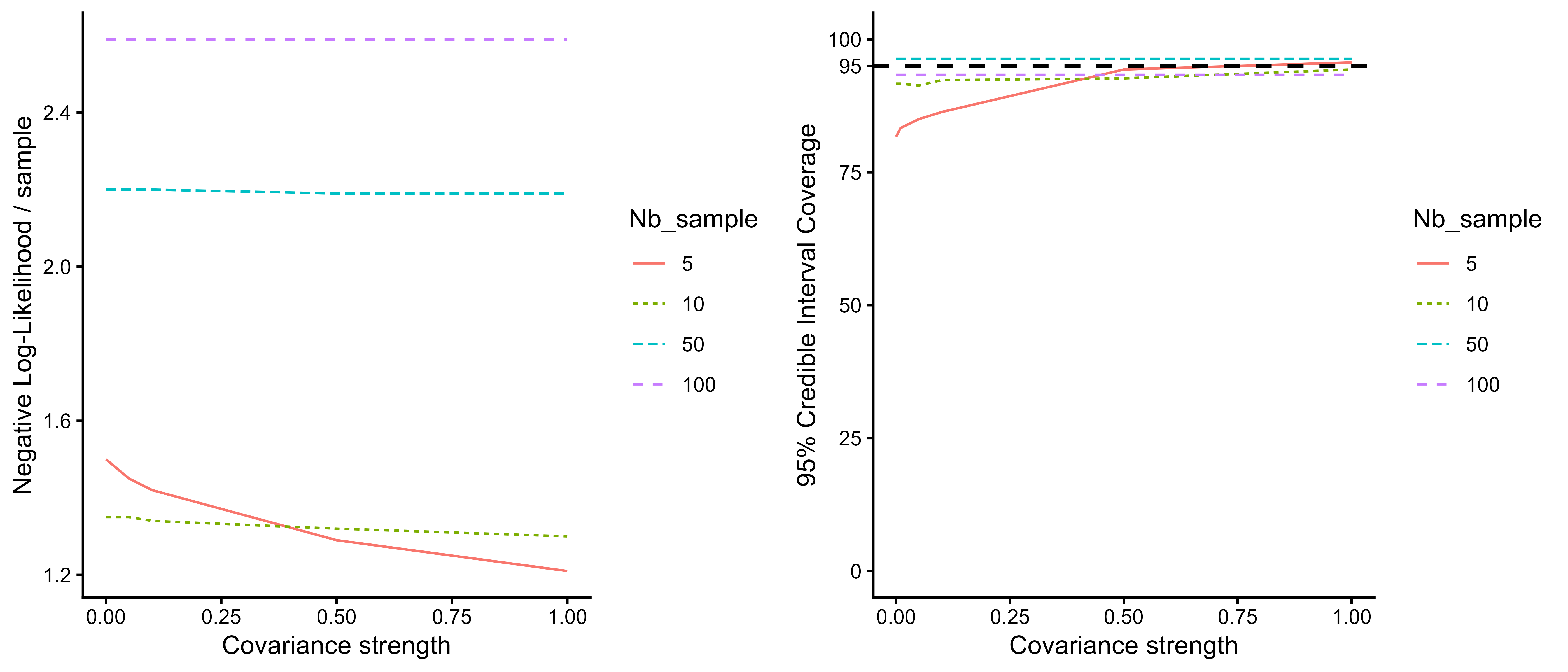}}
    \caption{Sensitivity analysis for the hyperparameter $\Sigma_0$ in a multivariate ProteoBayes setting. For each curve, posterior distributions are computed from datasets with the corresponding sample sizes. The off-diagonal terms of the prior matrix $\Sigma_0$ are scaled by a factor that varies along the x-axis. \textbf{Left:} Negative Log-Likelihood of the posterior distribution evaluated on the true mean value, divided by the number of samples. \textbf{Right}: $95\%$ Credible Interval Coverage for the true mean value.  All results are averaged over 1000 runs of the simulation scheme.}
    \label{fig:sensitivity_multi_matrix}
\end{figure}

In \Cref{fig:sensitivity_multi_matrix}, we displayed the sensitivity of the posteriors to misspecification of the scale matrix $\Sigma_0$. 
To measure different orders of misspecification, we generated data using the $\Sigma_{ref}$ matrix defined in Section 3.1 \emph{Synthetic datasets}. 
Then, the off-diagonal terms controlling correlations between peptides are scaled by a factor that we call the \emph{covariance strength}, which varies along the x-axis. 
We can see that, in most cases, the posteriors are remarkably robust to misspecification of $\Sigma_0$, except for very small sample sizes, which slightly deteriorate as we approach a diagonal matrix.

\subsection{Sensitivity of $\nu_0$}

\begin{figure}[ht]
 	\makebox[\textwidth][c]{
    \includegraphics[width = \textwidth]{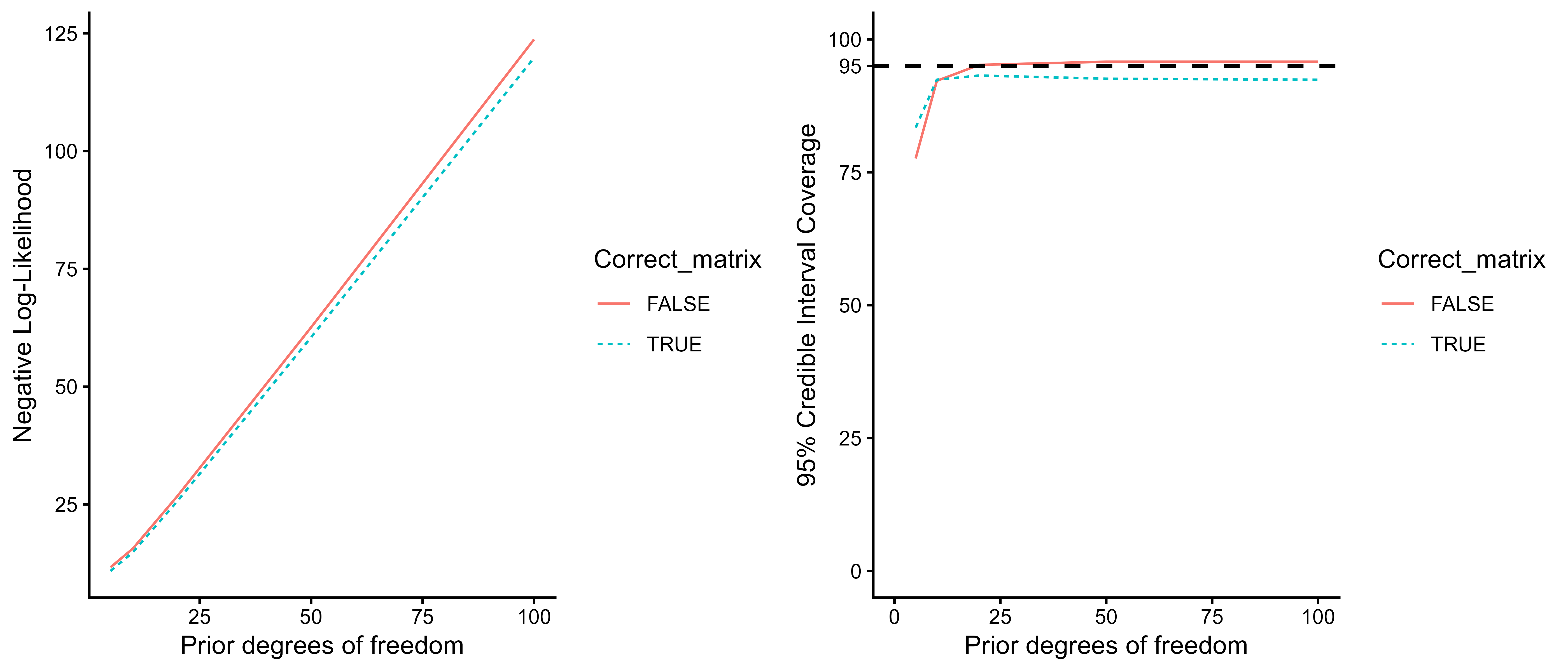}}
    \caption{Sensitivity analysis for the hyperparameter $\nu_0$ in a multivariate ProteoBayes setting. The prior degrees of freedom vary along the x-axis. The distinct curves correspond to a prior specification of the scale matrix $\Sigma_0$ either matching the true covariance structure of the data (TRUE) or an incorrect diagonal matrix (FALSE). \textbf{Left:} Negative Log-Likelihood of the posterior distribution evaluated on the true mean value. \textbf{Right}: $95\%$ Credible Interval Coverage for the true mean value.  All results are averaged over 1000 runs of the simulation scheme.}
    \label{fig:sensitivity_multi_df}
\end{figure}

In \Cref{fig:sensitivity_multi_df}, we evaluated the robustness of $\nu_0$, the prior degrees of freedom. 
As for its univariate counterpart $\alpha_0$, we observe that the NLL deteriorates as we increase its value, which justifies our choice to keep it close to the theoretically lowest value, since the minimal constraint is $\nu_0 \geq P - N_K - 1$.
Uncertainty remains consistently well-calibrated once we are above the stability threshold described above.
We also notice that the effect of $\nu_0$ is reasonably insensitive to the correct specification of $\Sigma_0$. 

\subsection{Joint effect of $\alpha_0$ and $\beta_0$}

\begin{figure}[ht]
 	\makebox[\textwidth][c]{\includegraphics[width = 0.77\textwidth]{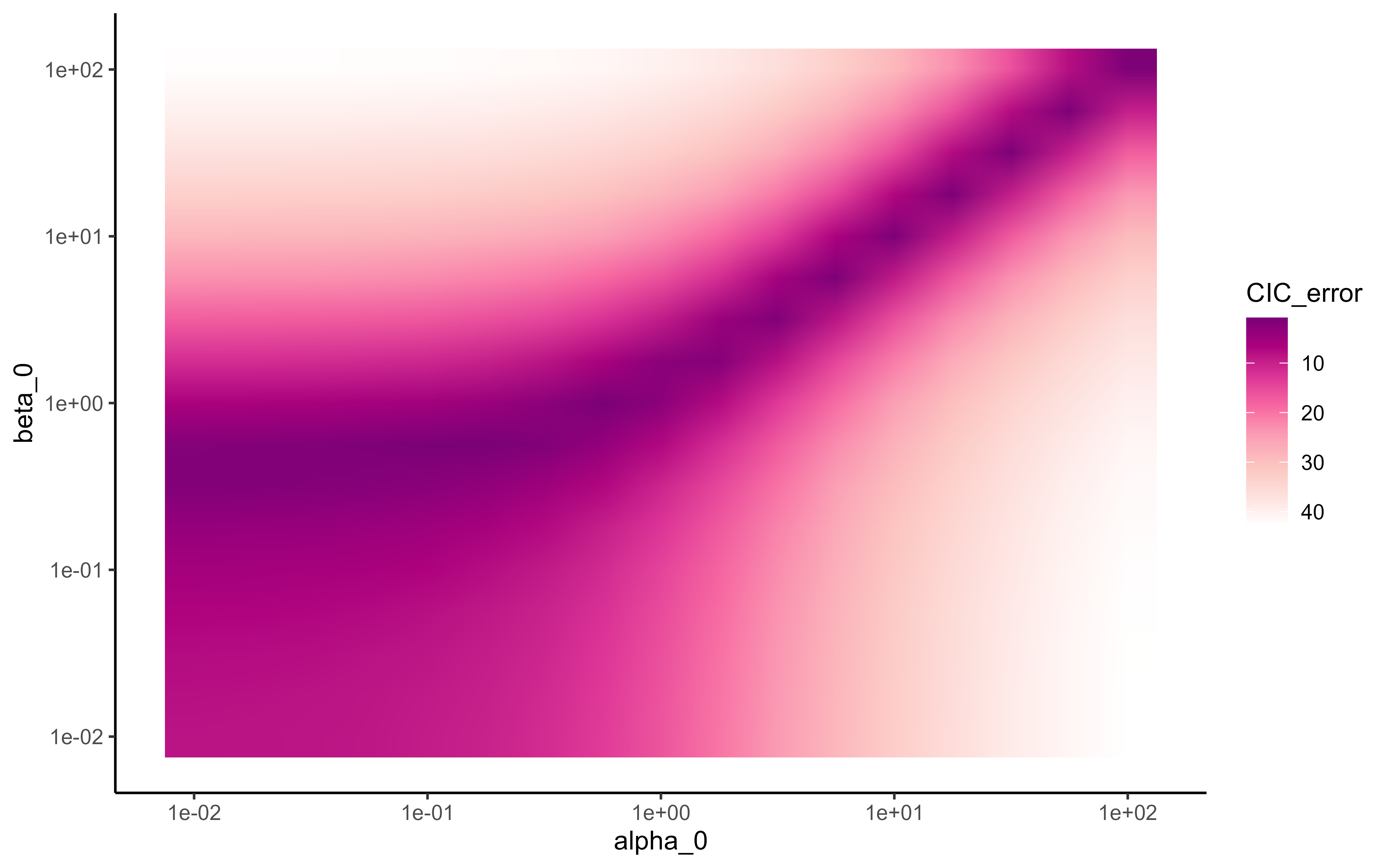}}
    \caption{Heatmap of errors for the values of the Credible Interval Coverage with respect to $\alpha_0$ and $\beta_0$ values. Empirical errors are computed over 1000 runs on synthetic data according to the simulated scheme with a fixed value of $\lambda_0 = 10^{-10}$.}
    \label{fig:alpha_beta}
\end{figure}

In \Cref{fig:alpha_beta}, we provide an additional insight into the joint effect of the inverse-gamma hyperparameters. 
This heatmap of errors in uncertainty quantification is computed using wide ranges of prior values for both $\alpha_0$ and $\beta_0$. We observe that low errors are obtained when both are of the same order of magnitude.
In particular, although previous \Cref{fig:sensitivity_uni} exhibited pathological regimes for high values of $\alpha_0$, we observe that this behaviour disappears when $\beta_0$ increases at the same rate.
The values retained as default for the experiments correspond to the $\{\alpha_0 = 0.01, \beta_0 = 0.3\}$ pair, which minimises calibration errors across 1000 simulation runs, although this figure illustrates that many other choices would have been equally valid. 

\subsection{Uncertainty calibration for different thresholds}

\begin{figure}[h!]
 	\makebox[\textwidth][c]{\includegraphics[width = 0.77\textwidth]{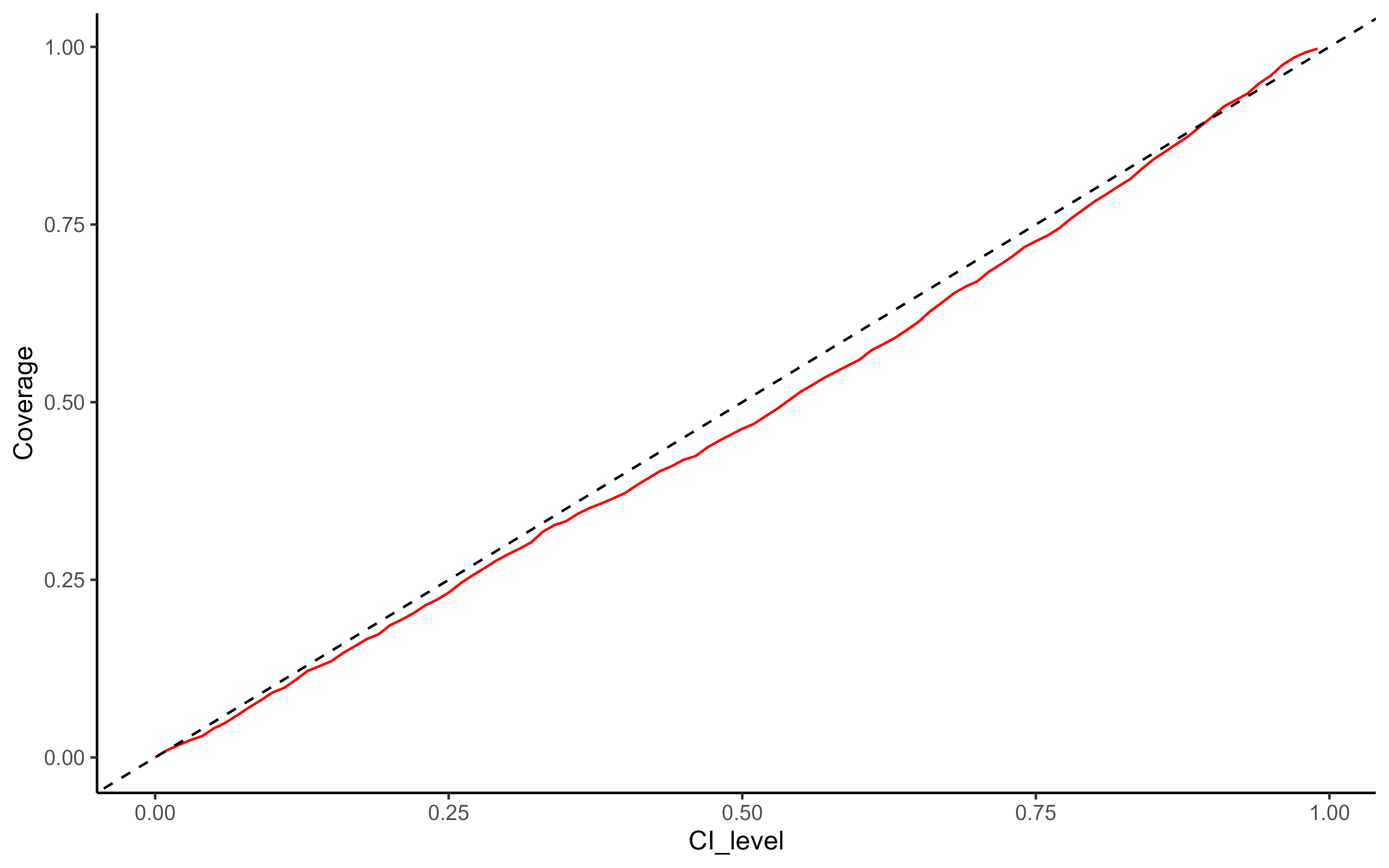}}
    \caption{Empirical validation of the Credible Interval Coverage (CIC) for all threshold probabilities between 0 and 1. The dashed line represents the theoretical level of the Credible Interval. The red line corresponds to the empirical coverage computed from synthetic data using the simulated scheme.}
    \label{fig:cic_validation}
\end{figure}

To confirm that uncertainty calibration is measured correctly, we depict in \Cref{fig:cic_validation} the empirical calibration of the Credible Interval Coverage (CIC) across all probability levels (although $0.95$ is used as the default in all experiments). 
The empirical plain red line overlapping the theoretical dashed line confirms that, for the default set of prior hyperparameters, our credible intervals remains conform to the expected values.

\newpage
\section{Real-data experiments: additional results}

\begin{table}[h]
\makebox[\textwidth][c]{\begin{tabular}{|c|c|c|ccc|ccc|}
\hline
\multirow{2}{*}{\textbf{Type}} &
  \multirow{2}{*}{\textbf{\begin{tabular}[c]{@{}c@{}}Vs.\\50 fmol\end{tabular}}} &
  \multirow{2}{*}{\textbf{\begin{tabular}[c]{@{}c@{}}Nb of\\ peptides\end{tabular}}} &
  \multicolumn{3}{c|}{\textbf{Mean difference}} &
  \multicolumn{3}{c|}{\textbf{ProteoBayes}} \\ \cline{4-9} 
 &
   &
   &
  \multicolumn{1}{c|}{\textbf{True}} &
  \multicolumn{1}{c|}{\textbf{limma}} &
  \textbf{ProteoBayes} &
  \multicolumn{1}{c|}{\textbf{CI$_{95}$ width}} &
  \multicolumn{1}{c|}{\textbf{RMSE}} &
  \textbf{CIC$_{95}$} \\ \hline
\multirow{9}{*}{\rotatebox[origin=c]{90}{\textbf{UPS}}} &
  10 amol &
  101 &
  \multicolumn{1}{c|}{12.29} &
  \multicolumn{1}{c|}{5.49 (1.99)} &
  5.49 (1.99) &
  \multicolumn{1}{c|}{11.47 (8.07)} &
  \multicolumn{1}{c|}{5.58 (3.91)} &
  55.45 (49.95) \\ 
 &
  50 amol &
  94 &
  \multicolumn{1}{c|}{9.97} &
  \multicolumn{1}{c|}{5.64 (1.96)} &
  5.64 (1.96) &
  \multicolumn{1}{c|}{10.52 (8.20)} &
  \multicolumn{1}{c|}{3.15 (2.68)} &
  55.32 (49.98) \\ 
 &
  100 amol &
  108 &
  \multicolumn{1}{c|}{8.97} &
  \multicolumn{1}{c|}{5.88 (1.96)} &
  5.88 (1.96) &
  \multicolumn{1}{c|}{10.71 (8.17)} &
  \multicolumn{1}{c|}{2.19 (2.28)} &
  62.04 (48.76) \\ 
 &
  250 amol &
  181 &
  \multicolumn{1}{c|}{7.64} &
  \multicolumn{1}{c|}{5.89 (1.81)} &
  5.89 (1.81) &
  \multicolumn{1}{c|}{9.17 (8.04)} &
  \multicolumn{1}{c|}{1.45 (2.17)} &
  86.19 (34.60) \\ 
 &
  500 amol &
  252 &
  \multicolumn{1}{c|}{6.64} &
  \multicolumn{1}{c|}{5.86 (1.29)} &
  5.86 (1.29) &
  \multicolumn{1}{c|}{7.95 (7.85)} &
  \multicolumn{1}{c|}{0.76 (1.20)} &
  99.21 (8.89) \\ 
 &
  1 fmol &
  351 &
  \multicolumn{1}{c|}{5.64} &
  \multicolumn{1}{c|}{5.20 (1.06)} &
  5.20 (1.06) &
  \multicolumn{1}{c|}{6.20 (7.20)} &
  \multicolumn{1}{c|}{0.64 (1.08)} &
  92.88 (25.76) \\ 
 &
  5 fmol &
  545 &
  \multicolumn{1}{c|}{3.32} &
  \multicolumn{1}{c|}{3.25 (0.56)} &
  3.25 (0.56) &
  \multicolumn{1}{c|}{3.62 (5.47)} &
  \multicolumn{1}{c|}{0.66 (1.11)} &
  88.81 (31.56) \\ 
 &
  10 fmol &
  623 &
  \multicolumn{1}{c|}{2.32} &
  \multicolumn{1}{c|}{2.26 (0.56)} &
  2.26 (0.56) &
  \multicolumn{1}{c|}{2.79 (4.47)} &
  \multicolumn{1}{c|}{0.71 (1.34)} &
  88.76 (31.61) \\ 
 &
  25 fmol &
  680 &
  \multicolumn{1}{c|}{1} &
  \multicolumn{1}{c|}{0.99 (0.39)} &
  0.99 (0.39) &
  \multicolumn{1}{c|}{2.02 (3.25)} &
  \multicolumn{1}{c|}{0.69 (1.40)} &
  88.38 (32.07) \\ \hline
\multirow{9}{*}{\rotatebox[origin=c]{90}{\textbf{YEAST}}} &
  10 amol &
  19739 &
  \multicolumn{1}{c|}{0} &
  \multicolumn{1}{c|}{0.12 (0.41)} &
  0.12 (0.41) &
  \multicolumn{1}{c|}{2.89 (4.62)} &
  \multicolumn{1}{c|}{0.23 (0.53)} &
  99.75 (4.98) \\ 
 &
  50 amol &
  19776 &
  \multicolumn{1}{c|}{0} &
  \multicolumn{1}{c|}{0.13 (0.42)} &
  0.13 (0.42) &
  \multicolumn{1}{c|}{2.74 (4.41)} &
  \multicolumn{1}{c|}{0.24 (0.69)} &
  99.69 (5.59) \\
 &
  100 amol &
  19749 &
  \multicolumn{1}{c|}{0} &
  \multicolumn{1}{c|}{0.14 (0.40)} &
  0.14 (0.40) &
  \multicolumn{1}{c|}{2.72 (4.39)} &
  \multicolumn{1}{c|}{0.22 (0.61)} &
  99.78 (4.66) \\  
 &
  250 amol &
  19770 &
  \multicolumn{1}{c|}{0} &
  \multicolumn{1}{c|}{0.14 (0.42)} &
  0.14 (0.42) &
  \multicolumn{1}{c|}{2.76 (4.46)} &
  \multicolumn{1}{c|}{0.23 (0.64)} &
  99.76 (4.92) \\ 
 &
  500 amol &
  19852 &
  \multicolumn{1}{c|}{0} &
  \multicolumn{1}{c|}{0.16 (0.42)} &
  0.16 (0.42) &
  \multicolumn{1}{c|}{2.74 (4.40)} &
  \multicolumn{1}{c|}{0.23 (0.65)} &
  99.83 (4.07) \\ 
 &
  1 fmol &
  19783 &
  \multicolumn{1}{c|}{0} &
  \multicolumn{1}{c|}{0.16 (0.41)} &
  0.16 (0.41) &
  \multicolumn{1}{c|}{2.72 (4.38)} &
  \multicolumn{1}{c|}{0.23 (0.57)} &
  99.81 (4.38) \\ 
 &
  5 fmol &
  19768 &
  \multicolumn{1}{c|}{0} &
  \multicolumn{1}{c|}{0.14 (0.40)} &
  0.14 (0.40) &
  \multicolumn{1}{c|}{2.73 (4.40)} &
  \multicolumn{1}{c|}{0.23 (0.59)} &
  99.76 (4.92) \\ 
 &
  10 fmol &
  19790 &
  \multicolumn{1}{c|}{0} &
  \multicolumn{1}{c|}{0.13 (0.38)} &
  0.13 (0.38) &
  \multicolumn{1}{c|}{2.72 (4.40)} &
  \multicolumn{1}{c|}{0.24 (0.64)} &
  99.66 (5.81) \\  
 &
  25 fmol &
  19632 &
  \multicolumn{1}{c|}{0} &
  \multicolumn{1}{c|}{0.06 (0.35)} &
  0.06 (0.35) &
  \multicolumn{1}{c|}{2.83 (4.55)} &
  \multicolumn{1}{c|}{0.25 (0.64)} &
  99.67 (5.70) \\ \hline
\end{tabular}}
\caption{Results table for the univariate differential analysis of the Bouyssie2020 dataset. All results are averaged over all peptides in each group and reported using the format \textit{Mean (Sd)}.}
\label{tab:Bouyssie2020}
\end{table}

\begin{table}[h]
\makebox[\textwidth][c]{\begin{tabular}{|c|c|c|ccc|ccc|}
\hline
\multirow{2}{*}{\textbf{Type}} &
  \multirow{2}{*}{\textbf{\begin{tabular}[c]{@{}c@{}}Vs.\\7.54 amol\end{tabular}}} &
  \multirow{2}{*}{\textbf{\begin{tabular}[c]{@{}c@{}}Nb of\\ peptides\end{tabular}}} &
  \multicolumn{3}{c|}{\textbf{Mean difference}} &
  \multicolumn{3}{c|}{\textbf{ProteoBayes}} \\ \cline{4-9} 
 &
   &
   &
  \multicolumn{1}{c|}{\textbf{True}} &
  \multicolumn{1}{c|}{\textbf{limma}} &
  \textbf{ProteoBayes} &
  \multicolumn{1}{c|}{\textbf{CI$_{95}$ width}} &
  \multicolumn{1}{c|}{\textbf{RMSE}} &
  \textbf{CIC$_{95}$} \\ \hline
\multirow{4}{*}{\rotatebox[origin=c]{90}{\textbf{UPS}}} &
  0.75 amol &
  382 &
  \multicolumn{1}{c|}{3.33} &
  \multicolumn{1}{c|}{2.81 (1.77)} &
  2.81 (1.77) &
  \multicolumn{1}{c|}{3.34 (4.46)} &
  \multicolumn{1}{c|}{0.88 (1.62)} &
  91.10 (28.51) \\ 
 &
  0.83 amol &
  382 &
  \multicolumn{1}{c|}{3.18} &
  \multicolumn{1}{c|}{2.82 (1.69)} &
  2.82 (1.69) &
  \multicolumn{1}{c|}{3.45 (4.75)} &
  \multicolumn{1}{c|}{0.86 (1.50)} &
  91.62 (27.74) \\  
 &
  1.07 amol &
  382 &
  \multicolumn{1}{c|}{2.82} &
  \multicolumn{1}{c|}{2.56 (1.51)} &
  2.56 (1.51) &
  \multicolumn{1}{c|}{3.20 (4.66)} &
  \multicolumn{1}{c|}{0.71 (1.25)} &
  94.50 (22.82) \\ 
 &
  2.04 amol &
  390 &
  \multicolumn{1}{c|}{1.89} &
  \multicolumn{1}{c|}{1.74 (1.34)} &
  1.74 (1.34) &
  \multicolumn{1}{c|}{2.63 (3.97)} &
  \multicolumn{1}{c|}{0.65 (1.04)} &
  93.85 (24.06) \\ \hline
\multirow{4}{*}{\rotatebox[origin=c]{90}{\textbf{MOUSE}}} &
  0.75 amol &
  95599 &
  \multicolumn{1}{c|}{0} &
  \multicolumn{1}{c|}{0.03 (0.78)} &
  0.03 (0.78) &
  \multicolumn{1}{c|}{1.82 (2.41)} &
  \multicolumn{1}{c|}{0.46 (1.27)} &
  97.74 (14.86) \\ 
 &
  0.83 amol &
  95591 &
  \multicolumn{1}{c|}{0} &
  \multicolumn{1}{c|}{0.02 (0.78)} &
  0.02 (0.78) &
  \multicolumn{1}{c|}{1.83 (2.46)} &
  \multicolumn{1}{c|}{0.45 (1.15)} &
  97.77 (14.76) \\ 
 &
  1.07 amol &
  95588 &
  \multicolumn{1}{c|}{0} &
  \multicolumn{1}{c|}{0.02 (0.77)} &
  0.02 (0.77) &
  \multicolumn{1}{c|}{1.83 (2.46)} &
  \multicolumn{1}{c|}{0.45 (1.21)} &
  98.00 (14.00) \\ 
 &
  2.04 amol &
  95553 &
  \multicolumn{1}{c|}{0} &
  \multicolumn{1}{c|}{0.01 (0.77)} &
  0.01 (0.77) &
  \multicolumn{1}{c|}{1.90 (2.54)} &
  \multicolumn{1}{c|}{0.46 (1.17)} &
  97.96 (14.14) \\ \hline
\end{tabular}}
\caption{Results table for the univariate differential analysis of the Huang2020 dataset. All results are averaged over all peptides in each group and reported using the format \textit{Mean (Sd)}.}
\label{tab:Huang2020}
\end{table}

\begin{table}[h]
\makebox[\textwidth][c]{\begin{tabular}{|c|c|c|ccc|ccc|}
\hline
\multirow{2}{*}{\textbf{Type}} &
  \multirow{2}{*}{\textbf{\begin{tabular}[c]{@{}c@{}}Vs.\\10 fmol\end{tabular}}} &
  \multirow{2}{*}{\textbf{\begin{tabular}[c]{@{}c@{}}Nb of\\ peptides\end{tabular}}} &
  \multicolumn{3}{c|}{\textbf{Mean difference}} &
  \multicolumn{3}{c|}{\textbf{ProteoBayes}} \\ \cline{4-9} 
 &
   &
   &
  \multicolumn{1}{c|}{\textbf{True}} &
  \multicolumn{1}{c|}{\textbf{limma}} &
  \textbf{ProteoBayes} &
  \multicolumn{1}{c|}{\textbf{CI$_{95}$ width}} &
  \multicolumn{1}{c|}{\textbf{RMSE}} &
  \textbf{CIC$_{95}$} \\ \hline
\multirow{6}{*}{\rotatebox[origin=c]{90}{\textbf{UPS}}} &
  0.05 fmol &
  205 &
  \multicolumn{1}{c|}{7.64} &
  \multicolumn{1}{c|}{4.21 (2.41)} &
  4.21 (2.41) &
  \multicolumn{1}{c|}{7.64 (7.64)} &
  \multicolumn{1}{c|}{3.20 (3.36)} &
  49.76 (50.12) \\ 
 &
  0.25 fmol &
  350 &
  \multicolumn{1}{c|}{5.32} &
  \multicolumn{1}{c|}{4.59 (0.90)} &
  4.59 (0.90) &
  \multicolumn{1}{c|}{6.44 (7.12)} &
  \multicolumn{1}{c|}{0.72 (1.29)} &
  96.29 (18.94) \\ 
 &
  0.5 fmol &
  459 &
  \multicolumn{1}{c|}{4.32} &
  \multicolumn{1}{c|}{3.52 (0.87)} &
  3.52 (0.87) &
  \multicolumn{1}{c|}{4.71 (5.97)} &
  \multicolumn{1}{c|}{0.65 (0.89)} &
  94.99 (21.84) \\ 
 &
  1.25 fmol &
  539 &
  \multicolumn{1}{c|}{3} &
  \multicolumn{1}{c|}{3.06 (0.72)} &
  3.06 (0.72) &
  \multicolumn{1}{c|}{4.82 (5.75)} &
  \multicolumn{1}{c|}{0.72 (0.99)} &
  91.47 (27.97) \\ 
 &
  2.5 fmol &
  608 &
  \multicolumn{1}{c|}{2} &
  \multicolumn{1}{c|}{1.7 (0.49)} &
  1.7 (0.49) &
  \multicolumn{1}{c|}{3.35 (4.45)} &
  \multicolumn{1}{c|}{0.58 (0.92)} &
  92.76 (25.93) \\ 
 &
  5 fmol &
  618 &
  \multicolumn{1}{c|}{1} &
  \multicolumn{1}{c|}{1.43 (0.57)} &
  1.43 (0.57) &
  \multicolumn{1}{c|}{3.69 (4.78)} &
  \multicolumn{1}{c|}{0.88 (1.25)} &
  86.89 (33.77) \\ \hline
\multirow{6}{*}{\rotatebox[origin=c]{90}{\textbf{ARATH}}} &
  0.05 fmol &
  15874 &
  \multicolumn{1}{c|}{0} &
  \multicolumn{1}{c|}{0.03 (0.60)} &
  0.03 (0.60) &
  \multicolumn{1}{c|}{3.25 (4.37)} &
  \multicolumn{1}{c|}{0.37 (0.77)} &
  99.21 (8.84) \\ 
 &
  0.25 fmol &
  15879 &
  \multicolumn{1}{c|}{0} &
  \multicolumn{1}{c|}{0.06 (0.58)} &
  0.06 (0.58) &
  \multicolumn{1}{c|}{3.12 (4.25)} &
  \multicolumn{1}{c|}{0.35 (0.79)} &
  99.31 (8.26) \\ 
 &
  0.5 fmol &
  15989 &
  \multicolumn{1}{c|}{0} &
  \multicolumn{1}{c|}{0.07 (0.56)} &
  0.07 (0.56) &
  \multicolumn{1}{c|}{3.15 (4.25)} &
  \multicolumn{1}{c|}{0.33 (0.93)} &
  99.49 (7.10) \\ 
 &
  1.25 fmol &
  16397 &
  \multicolumn{1}{c|}{0} &
  \multicolumn{1}{c|}{0.08 (0.61)} &
  0.08 (0.61) &
  \multicolumn{1}{c|}{3.74 (4.62)} &
  \multicolumn{1}{c|}{0.45 (0.90)} &
  98.33 (12.82) \\  
 &
  2.5 fmol &
  16253 &
  \multicolumn{1}{c|}{0} &
  \multicolumn{1}{c|}{0.04 (0.46)} &
  0.04 (0.46) &
  \multicolumn{1}{c|}{3.45 (4.61)} &
  \multicolumn{1}{c|}{0.28 (0.80)} &
  99.73 (5.20) \\ 
 &
  5 fmol &
  16228 &
  \multicolumn{1}{c|}{0} &
  \multicolumn{1}{c|}{0.03 (0.51)} &
  0.03 (0.51) &
  \multicolumn{1}{c|}{3.88 (4.95)} &
  \multicolumn{1}{c|}{0.48 (0.88)} &
  98.24 (13.14) \\ \hline
\end{tabular}}
\caption{Results table for the univariate differential analysis of the Chion2022 dataset. All results are averaged over all peptides in each group and reported using the format \textit{Mean (Sd)}.}
\label{tab:Chion2022}
\end{table}

\newpage
\section{Simulation experiment on MNAR data}

\begin{table}[h]
\centering
\makebox[\textwidth][c]{\begin{tabular}{c|c|clccc|c|c|}
\cline{2-9}
\multirow{2}{*}{\textbf{}}                   & \multirow{2}{*}{\textbf{\begin{tabular}[c]{@{}c@{}}MNAR \\ Missing \\ ratio\end{tabular}}} & \multicolumn{5}{c|}{\textbf{ProteoBayes}}     & \multirow{2}{*}{\textbf{\begin{tabular}[c]{@{}c@{}}limma \\ p\_value\end{tabular}}} & \multirow{2}{*}{\textbf{\begin{tabular}[c]{@{}c@{}} proDA \\ p\_value\end{tabular}}} \\ &  & \textbf{\begin{tabular}[c]{@{}c@{}}Mean \\ difference\end{tabular}} & \textbf{RMSE}                    & $\mathbf{CIC_{95}}$                & \textbf{\begin{tabular}[c]{@{}c@{}}$\mathbf{CI_{95}}$ \\ width\end{tabular}} & \textbf{TVD} &   & \\ \hline
\multicolumn{1}{|c|}{\textbf{Complete data}} & 0\%                                                                                       & \multicolumn{1}{c|}{1.01 (0.39)}                                    & \multicolumn{1}{l|}{0.32 (0.37}  & \multicolumn{1}{c|}{95.00 (21.83)} & \multicolumn{1}{c|}{1.18 (0.26)}                                             & 0.19 (0.21)                                                          & 0.08 (0.14)                                                                         & 0.13 (0.18)  \\ \hline
\multicolumn{1}{|c|}{\textbf{}}              & 20\%                                                                                      & \multicolumn{1}{c|}{0.98 (0.50)}                                    & \multicolumn{1}{l|}{0.50 (0.56)} & \multicolumn{1}{c|}{77.33 (41.94)} & \multicolumn{1}{c|}{1.07 (0.28)}                                             & 0.18 (0.22)                                                          & \multicolumn{1}{l|}{0.13 (0.23)}                                                    & 0.12 (0.20)                                                                                                                                                       \\
\multicolumn{1}{|c|}{\textbf{No imputation}} & 50\%                                                                                      & \multicolumn{1}{c|}{0.96 (0.43)}                                    & \multicolumn{1}{l|}{0.80 (0.78)} & \multicolumn{1}{c|}{52.67 (50.01)} & \multicolumn{1}{c|}{1.11 (0.30)}                                             & 0.21 (0.20)                                                          & \multicolumn{1}{l|}{0.13 (0.23)}                                                    & 0.17 (0.25) \\
\multicolumn{1}{|c|}{}                       & 80\%                                                                                      & \multicolumn{1}{c|}{1.05 (0.56)}                                    & \multicolumn{1}{l|}{1.31 (1.14)} & \multicolumn{1}{c|}{88.28 (32.23)} & \multicolumn{1}{c|}{1.78 (0.45)}                                             & 0.43 (0.23)                                                          & \multicolumn{1}{l|}{0.15 (0.20)}                                                    & 1.00 (0.00)  \\ \hline
\multicolumn{1}{|c|}{\textbf{}}              & 20\%                                                                                      & \multicolumn{1}{c|}{1.08 (0.45)}                                    & \multicolumn{1}{l|}{0.48 (0.51)} & \multicolumn{1}{c|}{67.67 (46.85)} & \multicolumn{1}{c|}{0.85 (0.22)}                                             & 0.10 (0.17)                                                          & 0.05 (0.14)                                                                         & 0.07 (0.16)  \\
\multicolumn{1}{|c|}{\textbf{Imputation}}    & 50\%                                                                                      & \multicolumn{1}{c|}{1.02 (0.51)}                                    & \multicolumn{1}{l|}{0.82 (0.76)} & \multicolumn{1}{c|}{11.67 (32.16)} & \multicolumn{1}{c|}{0.56 (0.16)}                                             & 0.06 (0.13)                                                          & 0.04 (0.12)                                                                         & 0.03 (0.11)   \\
\multicolumn{1}{|c|}{}                       & 80\%                                                                                      & \multicolumn{1}{c|}{1.06 (0.65)}                                    & \multicolumn{1}{l|}{1.45 (1.15)} & \multicolumn{1}{c|}{1.67 (12.82)}  & \multicolumn{1}{c|}{0.36 (0.07)}                                             & 0.05 (0.11)                                                          & 0.03 (0.13)                                                                         & 0.03 (0.12)  \\ \hline
\end{tabular}}
\caption{Performance metrics of ProteoBayes, limma, and proDA, for different scenarios of missing data ratios. Observations with the lowest observed values were removed, and imputation is performed by replacing missing values with the average of the observed ones. All results are averaged over 1000 repetitions of the experiments with 10 samples per peptide and reported using the format Mean (Sd).}
\label{tab:tab_MNAR}
\end{table}

\end{document}